\documentclass[usegraphicx,usenatbib,useAMS]{mn2e} 
\usepackage{mathabx}
\newcommand\Eqn[1]     {Eq.\,(\ref{#1})}
\newcommand\Eqns[2]    {Eqs\,(\ref{#1}) and~(\ref{#2})}

\newcommand\Fig[1]     {Figure\,{\ref{#1}}}

\newcommand\fig[1]     {Figure\,{\ref{#1}}}
\newcommand\figs[2]    {Figures\,\ref{#1} and~{\ref{#2}}}

\newcommand\nn         {\nonumber}

\def\mnras{{Mon.~ Not.~ R.~ Astron.~ Soc.~}}
\def\aj{{Astronomical Journal}}

\def\prd{{Phys.~ Rev.~ D.~}}

\def\apj{{Astrophys.~ J.~}}
\def\apjs{{Astrophys.~ J.~ Suppl.~}}
\def\apjl{{Astrophys.~ J.~ Lett.~}}

\def\nat{{Nature (London)~}}

\def\mnras{{MNRAS}}

\def\prd{{PRD}}

\def\apj{{ApJ}}
\def\apjs{{ApJS}}
\def\apjl{{ApJL}}
\def\aap{{A\&A}}
\def\nat{{Nature}}

\def\physrep{{Phys.~ Rep.~}}

\newcommand{\be}{\begin{equation}}
\newcommand{\ee}{\end{equation}}
\newcommand{\ba}{\begin{eqnarray}}
\newcommand{\ea}{\end{eqnarray}}

\newcommand{\gt}{\geq}
\newcommand{\mx}{\mbox}
\newcommand{\bm}{\boldmath }
\def\gt{\geq}

\def\g{{\rm g}}

\def\btheta{{\mx {\bm $\theta$}}}
\def\bphi{{\mx {\bm $\phi$}}}

\def\bl{{\mx{\bm $l$}}}
\def\bk{{\mx{\bm $k$}}}

\def\bx{{\mx{\bm $x$}}}
\def\by{{\mx{\bm $y$}}}
\def\bz{{\mx{\bm $z$}}}

\def\deg2{\rm deg^2}
\def\arcmin2{\rm arcmin^2}

\def\br{{\bf r}}
\def\bR{{\bf R}}
\def\nn{{\nonumber}}

\def\s8{{\sigma_8}}

\def\Msol{h^{-1}M_{\odot}}
\def\kpc{\, h^{-1}{\rm kpc}}
\def\Mpc{\, h^{-1}{\rm Mpc}}
\def\Mpccube{\, h^{-3} \, {\rm Mpc}^3}
\def\Gpc{\, h^{-1}{\rm Gpc}}

\def\nbar{\bar{n}}

\title[Galaxy-galaxy lensing and clustering in the Millennium-XXL simulation]
{An exploration of galaxy-galaxy lensing and galaxy clustering in the Millennium-XXL simulation }

\author[{\it Marian~et~al.~2014}] 
  {Laura~Marian$^{1,2}$\thanks{l.marian@sussex.ac.uk},
  Robert~E.~Smith$^{1,3}$  \& Raul~E.~Angulo$^{4}$ \\ 
  $^1$ Department of Physics and Astronomy, University of Sussex, Brighton BN1 9QH, UK\\
  $^2$ Argelander-Institute for Astronomy, Auf dem H\"ugel 71, D-53121
  Bonn, Germany\\ 
 $^3$ Max-Planck Institute for Astrophysics,
  Karl-Schwarzschild-Str. 1, Garching, D-85748, Germany \\ 
  $^4$ Centro de Estudios de Fisica del Cosmos de
  Aragon, Plaza San Juan 1, Planta-2, 44001, Teruel, Spain}

\begin{document}

\maketitle

%%%%%%%%%%%%%%%%%%%%%%%%%%%%%%%%%%%%%%%%%%%%%%%%%%%%%%%

\begin{abstract}
The combination of galaxy-galaxy lensing and galaxy clustering data
has the potential to simultaneously constrain both the cosmological
and galaxy formation models. In this paper we perform a comprehensive
exploration of these signals and their covariances through a
combination of analytic and numerical approaches. First, we derive
analytic expressions for the projected galaxy correlation function and
stacked tangential shear profile and their respective covariances,
which include Gaussian and discreteness noise terms. Secondly, we
measure these quantities from mock galaxy catalogues obtained from the
Millennium-XXL simulation and semi-analytic models of galaxy
formation. We find that on large scales ($R>10\Mpc$), the galaxy bias
is roughly linear and deterministic. On smaller scales ($R\lesssim 5
\Mpc $) the bias is a complicated function of scale and luminosity,
determined by the different spatial distribution and abundance of
satellite galaxies present when different magnitude cuts are applied,
as well as by the mass dependence of the host haloes on magnitude. Our
theoretical model for the covariances provides a reasonably good
description of the measured ones on small and large scales. However,
on intermediate scales $(1<R<10\Mpc)$, the predicted errors are
$\sim$2--3 times smaller, suggesting that the inclusion of
higher-order, non-Gaussian terms in the covariance will be required
for further improvements. Importantly, both our theoretical and
numerical methods show that the galaxy-galaxy lensing and clustering
signals have a non-zero cross-covariance matrix with significant
bin-to-bin correlations. Future surveys aiming to combine these probes
must take this into account in order to obtain unbiased and realistic
constraints.
\end{abstract}

%%%%%%%%%%%%%%%%%%%%%%%%%%%%%%%%%%%%%%%%%%%%%%%%%%%%%%%


%%%%%%%%%%%%%%%%%%%%%%%%%%%%%%%%%%%%%%%%%%%%%%%%%%%%%%%
\section{Introduction}\label{SI}
%%%%%%%%%%%%%%%%%%%%%%%%%%%%%%%%%%%%%%%%%%%%%%%%%%%%%%%

Since the first pioneering attempt to measure the galaxy-galaxy
lensing (hereafter GGL) signal by \citet{Tysonetal1984}, there have
been significant technological developments in deep and wide-field
astronomy, which have lead to the emergence of GGL as one of the most
promising probes for simultaneously constraining both the
cosmological and galaxy formation models.

The first robust detection of the GGL signal was made by
\citet{Brainerdetal1996} using 90~arcmin$^2$ of imaging data from the
5m Palomar telescope. They showed that whilst the tangential shear
profile around individual galaxies was too weak to be measured, the
stacked signal around all lens galaxies could be detected with high
signal-to-noise. Since then there has been a rapid explosion in the
field: the addition of the Wide-Field Camera to the Hubble Space
Telescope enabled a number of key GGL studies
\citep{Griffithsetal1996, Hudsonetal1998, Dell'Antonioetal1996,
  Leauthaudetal2012}.  The improvement of ground-based facilities such
as the Canada-France-Hawaii Telescope has also led to significant
developments \citep{Wilsonetal2001, Hoekstraetal2004, Parkeretal2007,
  vanUitertetal2011, Hudsonetal2015, Velanderetal2014}.  However,
perhaps the most prolific work in this area comes from the analysis of
the data from the Sloan Digital Sky Survey (hereafter SDSS)
\citep{McKayetal2001, Guziketal2001, Guziketal2002, Sheldonetal2004,
  Hirataetal2004, Mandelbaumetal2005, Mandelbaumetal2006a,
  Mandelbaumetal2006b, Mandelbaumetal2006c, Johnstonetal2007,
  Sheldonetal2009a, Sheldonetal2009b, Nakajimaetal2012,
  Mandelbaumetal2013}.  All these works have revealed that the GGL
signal is a complex function depending on a number of galaxy
properties, such as luminosity, colour, spectral type etc. The key
importance of GGL is that it enables one to make a direct link from
galaxy properties to the underlying dark matter distribution. Indeed,
these works have also constrained the mass, density profiles,
ellipticity of the dark matter haloes hosting the lens galaxies.

One of the first to realize that cosmic shear could help to
simultaneously constrain galaxy formation and cosmology, through
directly measuring the bias, was \citet{Schneider1998}. Schneider's
approach of using aperture mass filters was implemented by
\citet{Hoekstraetal2002} who directly measured galaxy bias,
establishing that it was a complicated function of scale.  This
approach was further theoretically developed for the Halo Occupation
Distribution (hereafter HOD) framework by \citet{Guziketal2001} and
later \citet{Seljaketal2005} and \citet{Yooetal2006}. More recent work
has been performed by \citet{Cacciatoetal2009,Cacciatoetal2013} who
have combined the results from GGL and galaxy clustering (hereafter
GC) studies, along with measurements of the galaxy luminosity function
(hereafter GLF) from SDSS to constrain the parameters of the
Conditional Luminosity Function (hereafter CLF) model -- which fully
specifies the link between a given dark matter halo and the galaxies
it hosts, albeit with assumptions about the functional form of the CLF
\citep{Yangetal2003,vandenBoschetal2013}. An interesting result to
emerge from this work, was that if one did not include the GGL
measurements in the analysis, then equally good fits to the CLF model
parameters could be obtained for either WMAP1 or WMAP3 cosmology
\citep{Spergeletal2003, Spergeletal2007}. Including the GGL data broke
this degeneracy and identified the WMAP3 parameters as the preferred
cosmological model. See also the very recent works of
\cite{Miyatakeetal2013, Moreetal2014}.

Whilst there has been significant progress in attempting to understand
and interpret the GGL signal \citep[see also][]{Baldaufetal2010,
  Saghihaetal2012}, our understanding of how to perform a robust
likelihood analysis with such data sets has been lacking. For example,
in the recent development of the CLF framework
\citep{vandenBoschetal2013,Cacciatoetal2013,Moreetal2013}, the GGL and
GLF measurements were taken to have diagonal covariance matrices, and
the GC covariance matrix was obtained from jackknife
estimation. Moreover, these probes were assumed to have {\em zero}
cross-covariance. This is clearly a simplification that future large
data sets analyses should improve on. A better analysis of the errors
was performed by \citet{Leauthaudetal2012} who used numerical
simulations to estimate the covariance matrices of the GGL and GC
measurements, while \citet{Mandelbaumetal2013} used the jackknife
approach. The very recent works of \cite{Miyatakeetal2013,
  Moreetal2014} also had an improved error analysis. However, again in
their analyses the cross-covariance of the two measurements was
assumed to be negligible.

If upcoming surveys such as the Dark Energy Survey (DES), the
Javalambre-Physics of the Accelerated Universe Astrophysical Survey
(J-PAS), Euclid, and the Large Synoptic Survey Telescope (LSST) are to
optimally constrain the cosmological model, then it is inevitable that
they must also jointly constrain the model of galaxy formation. The
best way to do this will be to combine the GGL, GC and GLF
measurements. This will require not only accurate models for the
signals themselves, but also accurate modelling of the covariance and
cross-covariance matrices of these probes.

In this paper we develop an analytical framework to compute both the
covariance and cross-covariance of GC and GGL. Our work builds upon
the analysis of the earlier work of \citet{Jeongetal2009} for GGL and
that of \citet{SmithMarian2014,SmithMarian2015} for the GC signal.
We then use the semi-analytic galaxy catalogues and dark matter
distribution from the Millennium-XXL (hereafter MXXL) simulation
\citep{Anguloetal2012} to directly measure these observables and their
associated auto- and cross-covariances, for several bins in
luminosity. Unlike the CLF approach, semi-analytic models (hereafter
SAM) make no direct assumption on how galaxies populate dark matter
haloes. Instead, they attempt to model the relevant physical processes
for galaxy formation and evolution, and how these are affected by
environment and assembly history. Thus, the nonlinearity and
stochasticity of galaxy bias are predictions, not assumptions in our
study.

The paper is structured as follows. In \S\ref{SII} we present the
necessary theoretical expressions for modelling the stacked tangential
shear profiles and projected galaxy clustering signals. In
\S\ref{SIII} we present expressions for their associated auto- and
cross-covariances. In \S\ref{SIV} we provide an overview of the
MXXL-simulation and the SAM galaxy catalogues that we use. In
\S\ref{SV} we present the measurements of the GGL and GC signals for a
set of luminosity bins, and compare them with the predictions from the
theory. In \S\ref{SVI} we present our results for the GC and GGL
covariance and cross-covariance matrices.  In \S\ref{SVII} we
summarize our findings and draw our conclusions.

%%%%%%%%%%%%%%%%%%%%%%%%%%%%%%%%%%%%%%%%%%%%%%%%%%%%%%%
%%%%%%%%%%%%%%%%%%%%%%%%%%%%%%%%%%%%%%%%%%%%%%%%%%%%%%%
\section{Theoretical predictions for the GGL and GC signals}
\label{SII}
%%%%%%%%%%%%%%%%%%%%%%%%%%%%%%%%%%%%%%%%%%%%%%%%%%%%%%%
%begin with some definitions here
In this section we present theoretical expressions for the stacked
tangential shear signal of a population of galaxy lenses and the
signal for the projected galaxy correlation function.

%%%%%%%%%%%%%%%%%%%%%%%%%%%%%%%%%%%%%%%%%%%%%%%%%%%%%%%
%%%%%%%%%%%%%%%%%%%%%%%%%%%%%%%%%%%%%%%%%%%%%%%%%%%%%%%
\subsection{Overview of required lensing ingredients}
\label{SII.1}
%%%%%%%%%%%%%%%%%%%%%%%%%%%%%%%%%%%%%%%%%%%%%%%%%%%%%%%
%%%%%%%%%%%%%%%%%%%%%%%%%%%%%%%%%%%%%%%%%%%%%%%%%%%%%%%

In the flat-sky approximation, the complex shear is written in terms
of the convergence as \citep{BartelmannSchneider2001}
%%%%%%%%%%%%%%%%%%%%%%%%%%%%%%%%%%%%%%%%%%%%%%%%%%%%%%%
\be
\gamma(\btheta)=\frac{1}{\pi}\int_{\mathbf{R}^2} d^2\theta' 
\kappa(\btheta')\mathcal{D}(\btheta-\btheta')\ , \label{eq:Dreal}
\ee
where the lensing kernel is defined to be,
\be 
\mathcal{D}(\btheta)\equiv(\theta_y^2-\theta_x^2- 2 i \theta_x \theta_y)/|\btheta|^4,
\ee
%%%%%%%%%%%%%%%%%%%%%%%%%%%%%%%%%%%%%%%%%%%%%%%%%%%%%%%
with $\theta_x$ and $\theta_y$ being the Cartesian components of the
Euclidean vector $\btheta=(\theta_x,\theta_y)$. These equations can be
written in Fourier space as:
%%%%%%%%%%%%%%%%%%%%%%%%%%%%%%%%%%%%%%%%%%%%%%%%%%%%%%%
\be
\gamma(\bl)=\frac{1}{\pi}\kappa(\bl) \mathcal{D}(\bl), \hspace{0.2cm}{\rm with}\hspace{0.2cm}
\mathcal{D}(\bl)=\pi \frac{l_x^2-l_y^2 + 2 i l_x l_y}{|\bl|^2},
\label{eq:DFourier}
\ee
%%%%%%%%%%%%%%%%%%%%%%%%%%%%%%%%%%%%%%%%%%%%%%%%%%%%%%%
where $l_x$ and $l_y$ are the Cartesian components of the vector
$\bl$; $\gamma(\bl)$ and $\mathcal{D}(\bl)$ and are the Fourier
transforms of $\gamma(\btheta)$ and $\mathcal{D}(\btheta)$
respectively. From Eq.~(\ref{eq:DFourier}), the complex shear can also
be written using the polar coordinates of the Fourier vector $\bl$ as:
%%%%%%%%%%%%%%%%%%%%%%%%%%%%%%%%%%%%%%%%%%%%%%%%%%%%%%%
\be
\gamma(\bl)=\gamma_1(\bl) + i \gamma_2(\bl)=\kappa(\bl)[\cos(2\phi_{\bl}) + i \sin(2\phi_{\bl})].
\label{eq:shear_Fourier}
\ee
%%%%%%%%%%%%%%%%%%%%%%%%%%%%%%%%%%%%%%%%%%%%%%%%%%%%%%%
The tangential shear at position $\btheta$ with respect to position
$\btheta_0$ (where $\btheta$ and $\btheta_0$ are defined with respect
to the same origin), is:
%%%%%%%%%%%%%%%%%%%%%%%%%%%%%%%%%%%%%%%%%%%%%%%%%%%%%%%
\be
\gamma_t(\btheta ; \btheta_0)=-\gamma_1(\btheta)\cos(2\phi_{\btheta-\btheta_0})
-\gamma_2(\btheta)\sin(2\phi_{\btheta-\btheta_0}),
\label{eq:def_tan_shear}
\ee
%%%%%%%%%%%%%%%%%%%%%%%%%%%%%%%%%%%%%%%%%%%%%%%%%%%%%%%
where $\phi_{\btheta-\btheta_0}$ is the polar angle of the relative
position vector $\btheta-\btheta_0$. Combining the last two equations,
one can write:
%%%%%%%%%%%%%%%%%%%%%%%%%%%%%%%%%%%%%%%%%%%%%%%%%%%%%%%
\be 
\gamma_t(\btheta ; \btheta_0)=-\int \frac{d^2 l}{(2\pi)^2}
\kappa(\bl)\cos[2(\phi_{\btheta-\btheta_0}-\phi_{\bl})] \,e^{i\, \bl \cdot \btheta} \ .
\label{eq:tan_shear}
\ee
%%%%%%%%%%%%%%%%%%%%%%%%%%%%%%%%%%%%%%%%%%%%%%%%%%%%%%%
Finally, we define the azimuthal average of the tangential shear as:
%%%%%%%%%%%%%%%%%%%%%%%%%%%%%%%%%%%%%%%%%%%%%%%%%%%%%%%
\be 
\gamma_{t,a}(\theta; \btheta_0)=\int_0^{2\pi} \frac{d\phi_{\btheta-\btheta_0}}
{2\pi}\gamma_t(\btheta; \btheta_0).
\label{eq:aa}
\ee
%%%%%%%%%%%%%%%%%%%%%%%%%%%%%%%%%%%%%%%%%%%%%%%%%%%%%%%
With the help of the very useful Bessel relation
%%%%%%%%%%%%%%%%%%%%%%%%%%%%%%%%%%%%%%%%%%%%%%%%%%%%%%%
\be
J_{n}(x)=\int_{\alpha}^{2\pi-\alpha} \frac{d\phi}{2\pi}\, e^{i\, [n\phi-x\sin(\phi)]} \ ,
\label{eq:Bessel}
\ee
%%%%%%%%%%%%%%%%%%%%%%%%%%%%%%%%%%%%%%%%%%%%%%%%%%%%%%%
one obtains the following equation which will recur many times in this section:
%%%%%%%%%%%%%%%%%%%%%%%%%%%%%%%%%%%%%%%%%%%%%%%%%%%%%%%
\be
J_2(x)=-\int_0^{2\pi} \frac{d\phi}{2\pi}\cos[2(\phi-\phi')]e^{i\, x \cos(\phi-\phi')}.
\label{eq:Bessel2}
\ee
%%%%%%%%%%%%%%%%%%%%%%%%%%%%%%%%%%%%%%%%%%%%%%%%%%%%%%%
Therefore the azimuthal average of the tangential shear is:
%%%%%%%%%%%%%%%%%%%%%%%%%%%%%%%%%%%%%%%%%%%%%%%%%%%%%%%
\be 
{\gamma_{t,a}}(\theta; \btheta_0)=\int\frac{d^2 l}{(2\pi)^2}
\kappa(\bl)J_2(l\theta) e^{i\,\bl \cdot \btheta_0}. 
\ee

%%%%%%%%%%%%%%%%%%%%%%%%%%%%%%%%%%%%%%%%%%%%%%%%%%%%%%%
%%%%%%%%%%%%%%%%%%%%%%%%%%%%%%%%%%%%%%%%%%%%%%%%%%%%%%%
%%%%%%%%%%%%%%%%%%%%%%%%%%%%%%%%%%%%%%%%%%%%%%%%%%%%%%%
\subsection{Estimator for the stacked tangential shear}
\label{SII.2}
%%%%%%%%%%%%%%%%%%%%%%%%%%%%%%%%%%%%%%%%%%%%%%%%%%%%%%%
In galaxy-galaxy lensing, the signal of individual lenses is very
weak, so in order to improve the signal-to-noise of this probe, one
has to stack several lenses. Suppose there is a population of $N_g$
galaxy lenses at positions $\bx_i$ in a chosen coordinate
system. Suppose also the total area of the survey to be $\int
d^2x=\Omega_s$. The number density of these lenses can be written as a
sum over their positions:
%%%%%%%%%%%%%%%%%%%%%%%%%%%%%%%%%%%%%%%%%%%%%%%%%%%%%%%
\be
n_g(\bx)=\sum_{i=1}^{N_{g}}\delta_D(\bx-\bx_i)
\label{eq:gal_den} \ ,
\ee
%%%%%%%%%%%%%%%%%%%%%%%%%%%%%%%%%%%%%%%%%%%%%%%%%%%%%%%
where $\bx$ is a 2D vector in the survey area. Using the above
equation, an estimator for the tangential shear of such a lens
population at an arbitrary position $\btheta$ with respect to the
location of the galaxy centres $\bx_i$ is \citep{Jeongetal2009}:
%%%%%%%%%%%%%%%%%%%%%%%%%%%%%%%%%%%%%%%%%%%%%%%%%%%%%%%
\ba
\widehat{\gamma}_t^g(\btheta) & = & 
\frac{1}{N_g}\sum_{i=1}^{N_{g}}\gamma_t(\bx_i+\btheta ; \bx_i)\nn \\
& = &
\frac{1}{N_{g}}\int_{\Omega_s}d^2x\,n_g(\bx)\gamma_t(\bx+\btheta ; \bx)  \ .
\label{eq:shear_est}
\ea
%%%%%%%%%%%%%%%%%%%%%%%%%%%%%%%%%%%%%%%%%%%%%%%%%%%%%%%
We define the fluctuation in the number density of lenses as
$n_g(\bx)=\bar{n}_g[1+\delta_g(\bx)]$, where the mean angular density
of lens galaxies is $\bar{n}_g=N_g/\Omega_s$. At this point we also
introduce the definitions of the convergence and galaxy density auto-
and cross-power spectra, which shall be used throughout this paper:
%%%%%%%%%%%%%%%%%%%%%%%%%%%%%%%%%%%%%%%%%%%%%%%%%%%%%%%
\ba
\langle\kappa(\bl)\kappa(\bl')\rangle & = & 
(2\pi)^2\delta_D(\bl+\bl')\mathcal{C}_{\kappa
  \kappa}(l) \ ; \nonumber \\
 \langle\delta_g(\bl) \delta_g(\bl')\rangle & = & 
(2\pi)^2\delta_D(\bl+\bl')\mathcal{C}_{g g}(l) \ ;
\nonumber \\
\langle\delta_g(\bl)\kappa(\bl')\rangle & = & 
(2\pi)^2\delta_D(\bl+\bl')\mathcal{C}_{g\kappa}(l) \ .
\label{eq:power_defs}
\ea
%%%%%%%%%%%%%%%%%%%%%%%%%%%%%%%%%%%%%%%%%%%%%%%%%%%%%%%
Since the tangential shear with respect to an origin ${\bf 0}$
vanishes on average $\langle \gamma_t(\btheta | {\bf 0})\rangle=0$,
the ensemble average of the estimator in Eq.~(\ref{eq:shear_est}) is
%%%%%%%%%%%%%%%%%%%%%%%%%%%%%%%%%%%%%%%%%%%%%%%%%%%%%%%
\ba
\langle\widehat{\gamma}_t^g(\btheta)\rangle & = &\frac{1}{N_g}
\int_{\Omega_s}d^2x\,\langle n_g(\bx)\gamma_t(\bx+\btheta ; \bx)\rangle \nn \\
& = & \frac{1}{\Omega_s}\int_{\Omega_s}d^2x\,\langle 
\delta_g(\bx)\gamma_t(\bx+\btheta ; \bx)\rangle \nn \\
& = & \langle \delta_g({\bf 0})\gamma_t(\btheta ; {\bf 0})\rangle \nn \\
& = &  -\int \frac{d^2l}{(2\pi)^2}\mathcal{C}_{g\kappa}(l)
\cos[2(\phi_{\btheta}-\phi_{\bl})]e^{i \,\bl \cdot \btheta}.
\label{eq:shear_est_ave0}
\ea
%%%%%%%%%%%%%%%%%%%%%%%%%%%%%%%%%%%%%%%%%%%%%%%%%%%%%%%
In the above we have also used the homogeneity of the Universe, which
makes the ensemble average of two cosmological fields to be invariant
under translations. We shall take advantage extensively of this
property throughout this work. With Eq.~(\ref{eq:Bessel2}), we arrive
at the azimuthally-averaged expression for the ensemble average of the
stacked shear estimator:
%%%%%%%%%%%%%%%%%%%%%%%%%%%%%%%%%%%%%%%%%%%%%%%%%%%%%%%
\be 
\langle\widehat{\gamma}_{t,a}^g(\theta)\rangle = \int
\frac{d^2l}{(2\pi)^2}\mathcal{C}_{g\kappa}(l)J_2(l \theta).
\label{eq:shear_est_ave}
\ee
%%%%%%%%%%%%%%%%%%%%%%%%%%%%%%%%%%%%%%%%%%%%%%%%%%%%%%%
Our goal is to compare theory predictions with estimates from
simulations, so we must take into account that the measured tangential
shear is bin averaged and not just azimuthally averaged. We introduce
the bin area:
%%%%%%%%%%%%%%%%%%%%%%%%%%%%%%%%%%%%%%%%%%%%%%%%%%%%%%%
\be
A(\theta_i)\equiv\int_0^{2\pi} d\phi_\btheta \int_{\theta_{\rm min}^i}^{\theta^i_{\rm max}} 
d\theta \,\theta=\pi\left(\theta^{i\,2}_{\rm max}-\theta^{i\,2}_{\rm min}\right),
\label{eq:area_bin}
\ee
%%%%%%%%%%%%%%%%%%%%%%%%%%%%%%%%%%%%%%%%%%%%%%%%%%%%%%%
with $\theta^{i}_{\rm min}$ and $\theta^{i}_{\rm max}$ being the lower
and upper bounds of the radial bin $i$. The bin-averaged stacked
tangential shear estimator is defined by
%%%%%%%%%%%%%%%%%%%%%%%%%%%%%%%%%%%%%%%%%%%%%%%%%%%%%%%
\be
\langle\widehat{\widebar{\gamma}}_t^g(\theta_i)\rangle
 \equiv \int_{A(\theta_i)} \frac{d^2\theta}{A(\theta_i)}
\langle\widehat{\gamma}_{t}^g(\btheta)\rangle =\frac{2\pi}{A(\theta_i)}
\int_{\theta^i_{\rm min}}^{\theta^i_{\rm max}} d\theta\,\theta 
\langle\widehat{\gamma}^{g}_{t,a}(\theta)\rangle
\label{eq:bin_ave} 
\ee
%%%%%%%%%%%%%%%%%%%%%%%%%%%%%%%%%%%%%%%%%%%%%%%%%%%%%%%
Defining the bin-averaged Bessel function of order $n$ to be:
%%%%%%%%%%%%%%%%%%%%%%%%%%%%%%%%%%%%%%%%%%%%%%%%%%%%%%%
\be 
\widebar{J}_n(l \theta_i) \equiv \int_{A(\theta_i)} 
\frac{d^2\theta}{A(\theta_i)}J_n(l \theta),
\ee
%%%%%%%%%%%%%%%%%%%%%%%%%%%%%%%%%%%%%%%%%%%%%%%%%%%%%%%
and using Eq.~(\ref{eq:shear_est_ave}), we write the final expression
for the bin-averaged stacked tangential shear estimator
%%%%%%%%%%%%%%%%%%%%%%%%%%%%%%%%%%%%%%%%%%%%%%%%%%%%%%%
\be
\langle\widehat{\widebar{\gamma}}_t^g(\theta_i)\rangle = 
\int \frac{d^2l}{(2\pi)^2}\mathcal{C}_{g\kappa}(l)\widebar{J}_2(l \theta_i). 
\label{eq:shear_est_BA}
\ee
%%%%%%%%%%%%%%%%%%%%%%%%%%%%%%%%%%%%%%%%%%%%%%%%%%%%%%%
%%%%%%%%%%%%%%%%%%%%%%%%%%%%%%%%%%%%%%%%%%%%%%%%%%%%%%%
%%%%%%%%%%%%%%%%%%%%%%%%%%%%%%%%%%%%%%%%%%%%%%%%%%%%%%%
\subsection{The projected galaxy correlation function}
\label{SII.3}
%%%%%%%%%%%%%%%%%%%%%%%%%%%%%%%%%%%%%%%%%%%%%%%%%%%%%%%
%%%%%%%%%%%%%%%%%%%%%%%%%%%%%%%%%%%%%%%%%%%%%%%%%%%%%%%
We define the estimator for the projected galaxy correlation function
to be:
%%%%%%%%%%%%%%%%%%%%%%%%%%%%%%%%%%%%%%%%%%%%%%%%%%%%%%%
\be
\widehat{w}_{\rm gg}(R)\equiv \int_0^{2\pi}\frac{d\phi_\bR}{2\pi}\int_{-\chi_{\rm max}}^{\chi_{\rm max}} d\chi
\,\widehat{\xi}_{\rm gg}(\br),
\label{eq:def_wgg}
\ee
%%%%%%%%%%%%%%%%%%%%%%%%%%%%%%%%%%%%%%%%%%%%%%%%%%%%%%%
where $\widehat{\xi}_{\rm gg}$ is an estimator for the 3D galaxy
correlation function, $\chi_{\rm max}$ is the comoving projection
length, and the position vector $\br$ has the components
$\{R,\phi_\bR, \chi\}$ in cylindrical coordinates. The estimator for
the galaxy correlation function is discussed in section
\S\ref{app:gc-cov} of the appendix, here we just mention that it is
unbiased, i.e. $\langle\widehat{\xi}_{\rm gg}(\br)\rangle=\xi_{\rm
  gg}(\br)$. Note that in this study we shall ignore redshift-space
distortions, as for our chosen projection length $\chi_{\rm
  max}=100\,\Mpc$, their contribution to $\widehat{w}_{\rm gg}$ is of
$\sim 10\%$ on the largest scales that we consider and less otherwise
\citep{Baldaufetal2010}. The galaxy correlation function can be
written in terms of its Fourier transform, the galaxy power spectrum:
%%%%%%%%%%%%%%%%%%%%%%%%%%%%%%%%%%%%%%%%%%%%%%%%%%%%%%%
\ba
\xi_{\rm gg}(\br) & = & \int\frac{d^3 k}{(2\pi)^3} \mathcal{P}_{\rm gg}(k)\, 
e^{i\,\bk \cdot \br} \nn \\
& = & \int_0^{\infty} \frac{d k_{\perp}}{(2\pi)^3} k_\perp 
\int_{\infty}^{\infty} d k_z \mathcal{P}_{\rm gg}(k)
\nn \\ & & \times \,
e^{i\,k_z \chi} \int_0^{2\pi} d\phi_\bk \, e^{i\,k_\perp R \cos(\phi_\bR-\phi_\bk)},
\label{eq:Pgg}
\ea
%%%%%%%%%%%%%%%%%%%%%%%%%%%%%%%%%%%%%%%%%%%%%%%%%%%%%%%
where the second line follows from expressing the wavevector $\bk$ in
cylindrical coordinates, with components $\{k_\perp,\phi_\bk, k_z\}$.
$k_z$ is the component along the line-of-sight and the magnitude $k$
is defined in the standard way $k=\sqrt{k_\perp^2+k_z^2}$. The
expectation of the projected galaxy correlation function estimator may
therefore be written
%%%%%%%%%%%%%%%%%%%%%%%%%%%%%%%%%%%%%%%%%%%%%%%%%%%%%%%
\ba
\langle\widehat{w}_{\rm gg}(R)\rangle \hspace{-0.3cm} & = &\hspace{-0.4cm} 
\int_0^{\infty}\hspace{-0.2cm} \frac{dk_\perp}{(2\pi)^2}
k_\perp J_0(k_\perp R) \hspace{-0.1cm} \int_{-\infty}^{\infty} \hspace{-0.4cm} 
dk_z \mathcal{P}_{\rm gg}(k) \int_{-\chi_{\rm max}}^{\chi_{\rm max}}\hspace{-0.3cm} 
d\chi\, e^{i\,k_z \chi} \label{eq:wgg_ave-tmp} \\
 \hspace{-0.3cm} & = &  \hspace{-0.3cm} 4\chi_{\rm max}  \hspace{-0.1cm}\int_0^{\infty}
 \hspace{-0.2cm}\frac{d k_\perp}{(2\pi)^2} k_\perp J_0(k_\perp R)
\int_0^{\infty}  \hspace{-0.3cm} d k_z \mathcal{P}_{\rm gg}(k) j_0(k_z\chi_{\rm max}), \nn\\
\label{eq:wgg_ave}
\ea
%%%%%%%%%%%%%%%%%%%%%%%%%%%%%%%%%%%%%%%%%%%%%%%%%%%%%%%
where $j_0$ is the zero$^{th}$ order spherical Bessel function.  Note that
we can also obtain an expression for the ensemble average of our
projected correlation function estimator in spherical coordinates,
choosing a particular frame where $\br\parallel{\bf e_z}$
%%%%%%%%%%%%%%%%%%%%%%%%%%%%%%%%%%%%%%%%%%%%%%%%%%%%%%%
\ba
\langle\widehat{w}_{\rm gg}(R)\rangle \hspace{-0.3cm} & = & \hspace{-0.3cm}
2\int_0^{\infty}\hspace{-0.2cm}\frac{dk}{(2\pi)^2} k^2 \mathcal{P}_{\rm gg}(k)
\int_{-\chi_{\rm max}}^{\chi_{\rm max}}\hspace{-0.4cm} d\chi \, j_0(kr) \nn \\
\hspace{-0.3cm} & = & \hspace{-0.3cm} 4\int_0^{\infty}\hspace{-0.2cm}
\frac{dk}{(2\pi)^2} k^2 \mathcal{P}_{\rm gg}(k)
\int_0^{\chi_{\rm max}}\hspace{-0.4cm} d\chi \, j_0(k \sqrt{R^2+\chi^2}),
\label{eq:wgg_ave_SC}
\ea
%%%%%%%%%%%%%%%%%%%%%%%%%%%%%%%%%%%%%%%%%%%%%%%%%%%%%%%
where in the last equality we took advantage of the fact that the
result did not depend on our particular choice of frame, and switched
back to cylindrical coordinates. Whilst Eqs.~(\ref{eq:wgg_ave})
and~(\ref{eq:wgg_ave_SC}) are expected to yield the same result, the
evaluation of the latter should more accurate since it involves a
single Bessel function integral.

Finally, we may apply the Limber approximation to simplify
Eq.~(\ref{eq:wgg_ave}). In this approximation it is only modes that
are transverse to the line-of-sight which contribute to the power
spectrum integral, and so the second integral in \Eqn{eq:wgg_ave-tmp})
becomes,
%%%%%%%%%%%%%%%%%%%%%%%%%%%%%%%%%%%%%%%%%%%%%%%%%%%%%%%
\ba
\int_{-\infty}^{\infty} dk_z\mathcal{P}_{\rm gg}(\sqrt{k_\perp^2 + k_z^2})\int_{-\chi_{\rm max}}^{\chi_{\rm max}}
d \chi \,e^{i\,k_z \chi} \nn \\
& & \hspace{-6cm}\approx \mathcal{P}_{\rm gg}(k_\perp)\int_{-\chi_{\rm max}}^{\chi_{\rm max}} d\chi
\int_{-\infty}^{\infty} dk_z \,e^{i\,k_z \chi} = 2\pi \mathcal{P}_{\rm gg}(k_\perp) .
\label{eq:Limber}
\ea
%%%%%%%%%%%%%%%%%%%%%%%%%%%%%%%%%%%%%%%%%%%%%%%%%%%%%%%
Using this relation in Eq.~(\ref{eq:wgg_ave-tmp}) we find that the
Limber-approximated ensemble average of the projected galaxy
correlation function estimator is therefore\footnote{Note that this
  result can also be obtained if in Eq.~(\ref{eq:wgg_ave}) one takes
  $\chi_{\rm max}\rightarrow \infty$.}:
%%%%%%%%%%%%%%%%%%%%%%%%%%%%%%%%%%%%%%%%%%%%%%%%%%%%%%%
\be
\langle\widehat{w}_{\rm gg}^L(R)\rangle = \int_0^{\infty} \frac{dk_\perp}{2\pi}
k_\perp \mathcal{P}_{\rm gg}(k_\perp) J_0(k_\perp R).
\label{eq:wgg_ave_Limber}
\ee
%%%%%%%%%%%%%%%%%%%%%%%%%%%%%%%%%%%%%%%%%%%%%%%%%%%%%%%
%%%%%%%%%%%%%%%%%%%%%%%%%%%%%%%%%%%%%%%%%%%%%%%%%%%%%%%
%%%%%%%%%%%%%%%%%%%%%%%%%%%%%%%%%%%%%%%%%%%%%%%%%%%%%%%
\section{Signal covariance matrices}
\label{SIII}
%%%%%%%%%%%%%%%%%%%%%%%%%%%%%%%%%%%%%%%%%%%%%%%%%%%%%%%
%%%%%%%%%%%%%%%%%%%%%%%%%%%%%%%%%%%%%%%%%%%%%%%%%%%%%%%
In this section we compute the auto- and cross-covariance matrices of
the stacked tangential shear signal and the projected galaxy
correlation function.

%%%%%%%%%%%%%%%%%%%%%%%%%%%%%%%%%%%%%%%%%%%%%%%%%%%%%%%
\subsection{The covariance matrix of the stacked tangential shear estimator}
\label{SIII.1}
%%%%%%%%%%%%%%%%%%%%%%%%%%%%%%%%%%%%%%%%%%%%%%%%%%%%%%%
The definition of the covariance of the estimator in
Eq.~(\ref{eq:shear_est}) is
%%%%%%%%%%%%%%%%%%%%%%%%%%%%%%%%%%%%%%%%%%%%%%%%%%%%%%%
\be
{\rm Cov}[\widehat{\gamma}_t^{g}](\btheta_1, \btheta_2)=
\langle \widehat{\gamma}_t^g(\btheta_1)
\widehat{\gamma}_t^g(\btheta_2)\rangle-\langle\widehat{\gamma}_t^g(\btheta_1)\rangle \langle
\widehat{\gamma}_t^g(\btheta_2)\rangle
\label{eq:cov_shear_def}
\ee
%%%%%%%%%%%%%%%%%%%%%%%%%%%%%%%%%%%%%%%%%%%%%%%%%%%%%%%
The bin-averaged estimate of the GGL covariance is defined according
to Eq.~(\ref{eq:bin_ave}) as:
%%%%%%%%%%%%%%%%%%%%%%%%%%%%%%%%%%%%%%%%%%%%%%%%%%%%%%%
\be 
{\rm Cov}[\widehat{\widebar{\gamma}}_t^{g}](\theta_i, \theta_j)
\equiv 
\int_{A_i}\int_{A_j} \frac{d^2\theta_1}{A_i} \frac{d^2\theta_2}{A_j}
{\rm Cov}[\widehat{\gamma}_t^{g}](\btheta_1, \btheta_2) .
\ee
%%%%%%%%%%%%%%%%%%%%%%%%%%%%%%%%%%%%%%%%%%%%%%%%%%%%%%%
In appendix~\ref{app:ggl-cov} we provide the complete details of the
derivation of the covariance of the stacked tangential shear
profiles. The main result is (c.f.~\Eqn{eq:cov_shear3}):
%%%%%%%%%%%%%%%%%%%%%%%%%%%%%%%%%%%%%%%%%%%%%%%%%%%%%%%
\ba
{\rm Cov}[\widehat{\widebar{\gamma}}_t^{g}](\theta_i, \theta_j) & = & 
\frac{1}{\Omega_s} \int_0^{\infty} \frac{d^2 l}{(2\pi)^2}\, 
\widebar{J}_2(l\theta_i) \widebar{J}_2(l\theta_j) \times \nn \\
& & \hspace{-1.6cm}\left\{ \mathcal{C}_{g\kappa}^2(l) + \left[\mathcal{C}_{\kappa
      \kappa}(l) + \frac{\sigma_{\gamma}^2}{2 \bar{n}_s}\right] 
  \left[\mathcal{C}_{\rm gg}(l) +\frac{1}{\bar{n}_g} \right]
 \right\},
\label{eq:cov_shear4}
\ea
%%%%%%%%%%%%%%%%%%%%%%%%%%%%%%%%%%%%%%%%%%%%%%%%%%%%%%%
where $\sigma_{\gamma}^2/2 = \sigma_{\gamma_1}^2 =
\sigma_{\gamma_2}^2$ is the variance per shear component in the
measurement of one source galaxy, and $\bar{n}_s$ is the mean angular
density of the source galaxies.
%%%%%%%%%%%%%%%%%%%%%%%%%%%%%%%%%%%%%%%%%%%%%%%%%%%%%%%
%%%%%%%%%%%%%%%%%%%%%%%%%%%%%%%%%%%%%%%%%%%%%%%%%%%%%%%
\subsection{The covariance matrix of the projected galaxy correlation function estimator}
\label{SIII.2}
%%%%%%%%%%%%%%%%%%%%%%%%%%%%%%%%%%%%%%%%%%%%%%%%%%%%%%%
The azimuthally-averaged covariance of the projected correlation
function estimator can be written as a projection of the covariance of
the estimator for the 3D galaxy correlation function, which we denote
$\widehat{\xi}_{\rm gg}$. Hence,
%%%%%%%%%%%%%%%%%%%%%%%%%%%%%%%%%%%%%%%%%%%%%%%%%%%%%%%
\ba {\rm Cov}[\widehat{w}_{\rm gg}](R_1, R_2) \hspace{-0.3cm}& = &\hspace{-0.4cm}
\int_0^{2\pi}\hspace{-0.1cm}\frac{d\phi_{\bR_1}}{2\pi}\hspace{-0.1cm} \frac{d\phi_{\bR_2}}{2\pi}
\int_{-\chi_{\rm max}}^{\chi_{\rm max}} \hspace{-0.7cm} d\chi_1 d\chi_2\, 
{\rm Cov}[\widehat{\xi}_{\rm gg}](\br_1, \br_2).
\nn
\ea
%%%%%%%%%%%%%%%%%%%%%%%%%%%%%%%%%%%%%%%%%%%%%%%%%%%%%%%
In the appendix~\ref{app:gc-cov} we provide complete details of the
derivation of the covariance matrix of $\widehat{\xi}_{\rm gg}$ and
the final result is given by \Eqn{eq:cov_xigg_AA}. On combining the
equation above and \Eqn{eq:cov_xigg_AA}, we arrive at the expression
for the azimuthally-averaged covariance of the projected galaxy
correlation function estimator:
%%%%%%%%%%%%%%%%%%%%%%%%%%%%%%%%%%%%%%%%%%%%%%%%%%%%%%%
\ba {\rm Cov}[\widehat{w}_{\rm gg}](R_1, R_2) & & \hspace{-0.7cm} =
\frac{1}{V_s}\left\{8 \chi^2_{\rm
  max}\int_0^{\infty} \hspace{-0.2cm}\frac{dk_\perp} {(2\pi)^2}
k_\perp J_0(k_\perp R_1)J_0(k_\perp R_2) \right. \nn \\ &
& \hspace{-3.5cm} \left. \times \int_{-\infty}^{\infty}
\hspace{-0.2cm}dk_z j_0^2(k_z \chi_{\rm max}) \mathcal{P}_{\rm gg}
(\sqrt{k_\perp^2+k_z^2}) \left[\mathcal{P}_{\rm
    gg}(\sqrt{k_\perp^2+k_z^2}) +\frac{2(1+\alpha)}{\bar{n}_g}\right]
\right.\nn \\ & & \hspace{-3.3cm} \left. \,+
\,\frac{2}{\bar{n}_g^2}\left[w_{\rm gg}(R_1) + 2\chi_{\rm max}
  (1+\alpha)^2\right] \frac{\delta_D(R_1-R_2)}{2\pi R_1} \right\}
\,,\hspace{-4cm}
\label{eq:cov_wgg}
\ea
%%%%%%%%%%%%%%%%%%%%%%%%%%%%%%%%%%%%%%%%%%%%%%%%%%%%%%%
where $\alpha$ is a constant quantifying how dense the random
catalogue used to estimate the correlation function is relative to the
galaxy data (see \S\ref{app:gc-cov} for more details). Note that in
the above $\bar{n}_g$ is the mean galaxy volume density, whereas in
\S\ref{SIII.1} the same notation is used for the mean angular density of
lens galaxies. As in the case of the ensemble-averaged projected
galaxy correlation function estimator, here too we apply the Limber
approximation to simplify the above result. The Limber-approximated
covariance is given by:
%%%%%%%%%%%%%%%%%%%%%%%%%%%%%%%%%%%%%%%%%%%%%%%%%%%%%%%
\ba {\rm Cov^L}[\widehat{w}_{\rm gg}](R_1, R_2) & & \hspace{-0.7cm}=
\frac{1}{V_s}\left\{4\chi_{\rm max}\int_0^{\infty} \hspace{-0.1cm}
\frac{dk_\perp}{2\pi}\,k_\perp J_0(k_\perp R_1) J_0(k_\perp R_2)
\right.\nn \\ & & \left.\hspace{-3cm}\, \times\, \mathcal{P}_{\rm
  gg}(k_\perp) \left[\mathcal{P}_{\rm gg}(k_\perp)
  +\frac{2(1+\alpha)}{\bar{n}_g}\right]\right. \nn \\ & &
\left.\hspace{-3cm}\, + \,\frac{2}{\bar{n}_g^2}\left[w_{\rm
    gg}^L(R_1)+ 2\chi_{\rm max}
  (1+\alpha)^2\right]\frac{\delta_D(R_1-R_2)}{2\pi R_1}\right\}.
\label{eq:cov_wgg_L}
\ea
%%%%%%%%%%%%%%%%%%%%%%%%%%%%%%%%%%%%%%%%%%%%%%%%%%%%%%%
Note that the survey volume can be expressed as $V_s=2\chi_{\rm max}
A_s$, where $A_s$ denotes the transverse area of the survey. If
$\chi_{\rm max}\rightarrow \infty$, then we also have
$V_s\rightarrow\infty$ and the Limber-approximated covariance becomes:
%%%%%%%%%%%%%%%%%%%%%%%%%%%%%%%%%%%%%%%%%%%%%%%%%%%%%%%
\ba
\hspace{-0.3cm}\lim_{\chi_{\rm
    max}\rightarrow\infty} \hspace{-0.3cm}{\rm Cov^L}[\widehat{w}_{\rm
    gg}](R_1, R_2) & & \hspace{-0.6cm} = \frac{2}{A_s} \left\{
\int_0^{\infty} \hspace{-0.1cm}\frac{dk_\perp}{2\pi}\,k_\perp
J_0(k_\perp R_1) J_0(k_\perp R_2)\right. \nn \\ & &
\left.\hspace{-4cm}\,\times\, \mathcal{P}_{\rm gg}(k_\perp)
\left[\mathcal{P}_{\rm gg}(k_\perp)
  +\frac{2(1+\alpha)}{\bar{n}_g}\right] +
\frac{(1+\alpha)^2}{\bar{n}_g^2}\frac{\delta_D(R_1-R_2)}{2\pi R_1}
\right\}.  \nonumber \ea
%%%%%%%%%%%%%%%%%%%%%%%%%%%%%%%%%%%%%%%%%%%%%%%%%%%%%%%
Therefore, the Limber covariance is well-behaved in the limit where
$\chi_{\rm max}\rightarrow \infty$. Finally, using the definitions in
Eqs.~(\ref{eq:area_bin}) and~(\ref{eq:bin_ave}), we write the
expression for the bin-averaged, Limber-approximated covariance of the
projected galaxy correlation function estimator:
%%%%%%%%%%%%%%%%%%%%%%%%%%%%%%%%%%%%%%%%%%%%%%%%%%%%%%%
\ba {\rm Cov^L}[\widehat{\widebar{w}}_{\rm gg}](R_i, R_j) &
& \hspace{-0.6cm} = \frac{2}{V_s} \left\{ 2\chi_{\rm max}
\int_0^{\infty}\hspace{-0.1cm} \frac{d^2k_\perp}{(2\pi)^2}
\widebar{J}_0(k_\perp R_i)\widebar{J}_0(k_\perp R_j)\right. \nn \\ & &
\left. \hspace{-0.65cm}\,\times\, \mathcal{P}_{\rm gg}(k_\perp)
\left[\mathcal{P}_{\rm gg}(k_\perp)
  +\frac{2(1+\alpha)}{\bar{n}_g}\right]\right. \nn \\ & &
\left. \hspace{-0.65cm}\, +\, \frac{\delta^K_{ij}}
     {\bar{n}_g^2}\left[\widebar{w}_{\rm gg}^L(R_i) + 2\chi_{\rm max}
       (1+\alpha)^2\right]\right\} .
\label{eq:cov_wgg_L_BA}
\ea
%%%%%%%%%%%%%%%%%%%%%%%%%%%%%%%%%%%%%%%%%%%%%%%%%%%%%%%
%%%%%%%%%%%%%%%%%%%%%%%%%%%%%%%%%%%%%%%%%%%%%%%%%%%%%%%
%%%%%%%%%%%%%%%%%%%%%%%%%%%%%%%%%%%%%%%%%%%%%%%%%%%%%%%
%%%%%%%%%%%%%%%%%%%%%%%%%%%%%%%%%%%%%%%%%%%%%%%%%%%%%%%
\subsection{The cross-covariance of the stacked tangential 
shear and the projected galaxy correlation function estimators}
\label{SIII.3}
%%%%%%%%%%%%%%%%%%%%%%%%%%%%%%%%%%%%%%%%%%%%%%%%%%%%%%%
In this section our goal is to compute the cross-covariance of the
estimators for the stacked tangential shear and projected galaxy
correlation functions, defined by Eqs.~(\ref{eq:shear_est})
and~(\ref{eq:def_wgg}). To this avail, we follow the same procedure
as before, and define the cross-covariance as:
%%%%%%%%%%%%%%%%%%%%%%%%%%%%%%%%%%%%%%%%%%%%%%%%%%%%%%%
\ba {\rm Cov}[\widehat{w}_{\rm gg}(\bR_1),
  \widehat{\gamma}^g_t(\bR_2)] & \equiv &  \langle\widehat{w}_{\rm
  gg}(\bR_1) \widehat{\gamma}^g_t(\bR_2)\rangle \nn \\ & &
-\langle\widehat{w}_{\rm gg}(\bR_1)\rangle\,\langle
\widehat{\gamma}^g_t(\bR_2)\rangle \ ,
\label{eq:def_cov_cross}
\ea
%%%%%%%%%%%%%%%%%%%%%%%%%%%%%%%%%%%%%%%%%%%%%%%%%%%%%%%
where $\bR_1$ and $\bR_2$ are two 2D position vectors. To simplify
this calculation, we shall use the angular correlation function
instead of the projected correlation function. This is justified by
the fact that our lenses are in a thin redshift slice, in which case
the two correlation functions are equivalent. In
appendix~\ref{app:ggl-gc-cov} we provide complete details of the
derivation of the cross-covariance matrix.

The final expression is given by (c.f.~\Eqn{eq:cross_cov_angular}):
%%%%%%%%%%%%%%%%%%%%%%%%%%%%%%%%%%%%%%%%%%%%%%%%%%%%%%%
\ba
{\rm Cov}[\widehat{w}_{\rm gg}(\btheta_1), \widehat{\gamma}^g_t(\btheta_2)] 
\hspace{-0.3cm} & = & \hspace{-0.3cm}-\frac{1}{\Omega_s}\int\hspace{-0.1cm} \frac{d^2l}{(2\pi)^2} 
\mathcal{C}_{g\kappa}(l)\hspace{-0.1cm}\left[\mathcal{C}_{\rm gg}(l) 
+\frac{1}{\bar{n}_g}\hspace{-0.05cm}\right]  \nn \\
& & \hspace{-2.3cm} \times \,\cos[2(\phi_{\btheta_2}-\phi_{\bl})] 
\left[e^{i\,\bl \cdot (\btheta_2-\btheta_1)} +  e^{i\,\bl \cdot (\btheta_2+\btheta_1)}\right]. 
\nonumber
\ea
%%%%%%%%%%%%%%%%%%%%%%%%%%%%%%%%%%%%%%%%%%%%%%%%%%%%%%%
Just like before, we are interested in the cross-covariance matrix of
the bin-averaged measurements. In a similar fashion to the analysis of
the previous sections, we see that under binning the above equation
becomes: %&  & \hspace{-0.7cm} & & \hspace{-0.55cm} \times\:
%%%%%%%%%%%%%%%%%%%%%%%%%%%%%%%%%%%%%%%%%%%%%%%%%%%%%%%
\ba
{\rm Cov}[\widehat{\widebar{w}}_{\rm gg}(\theta_i), \widehat{\widebar{\gamma}}^g_t(\theta_j)] 
\hspace{-0.2cm} & = & \hspace{-0.2cm} \frac{2}{\Omega_s}
\int \frac{d^2l}{(2\pi)^2} 
\widebar{J}_0(l \theta_i)\widebar{J}_2(l \theta_j)  \nn\\ 
\hspace{-0.2cm} & \times & \hspace{-0.2cm} 
\mathcal{C}_{g\kappa}(l)\left[\mathcal{C}_{\rm gg}(l) +\frac{1}{\bar{n}_g}\right].
\label{eq:cross_cov_BA}
\ea
%%%%%%%%%%%%%%%%%%%%%%%%%%%%%%%%%%%%%%%%%%%%%%%%%%%%%%%
Eq.~(\ref{eq:cross_cov_BA}) represents the bin-averaged
cross-covariance of the estimator for the angular galaxy correlation
function and the stacked tangential shear estimator. As expected, it
has no dimensions.

%%%%%%%%%%%%%%%%%%%%%%%%%%%%%%%%%%%%%%%%%%%%%%%%%%%%%%%
\begin{figure}
\centering {
\includegraphics[scale=0.45]{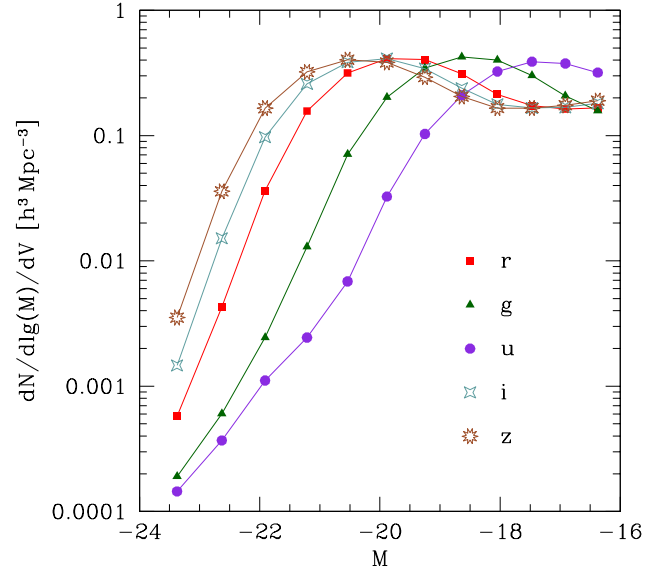}
\caption{Luminosity function of semi-analytic galaxies in the MXXL
  simulation, in the five SDSS bands $r$, $g$, $u$, $i$, $z$. Note
  that artifacts produced by the finite mass resolution of the
  simulation are evident for M $\gt$ $-20$ -- $-19$.}
\label{fig:lum_func}}
\end{figure}

%%%%%%%%%%%%%%%%%%%%%%%%%%%%%%%%%%%%%%%%%%%%%%%%%%%%%%%

\section{The MXXL simulation}
\label{SIV}

%%%%%%%%%%%%%%%%%%%%%%%%%%%%%%%%%%%%%%%%%%%%%%%%%%%%%%%
%%%%%%%%%%%%%%%%%%%%%%%%%%%%%%%%%%%%%%%%%%%%%%%%%%%%%%%
\begin{figure*}
\centering {
\includegraphics[scale=0.62]{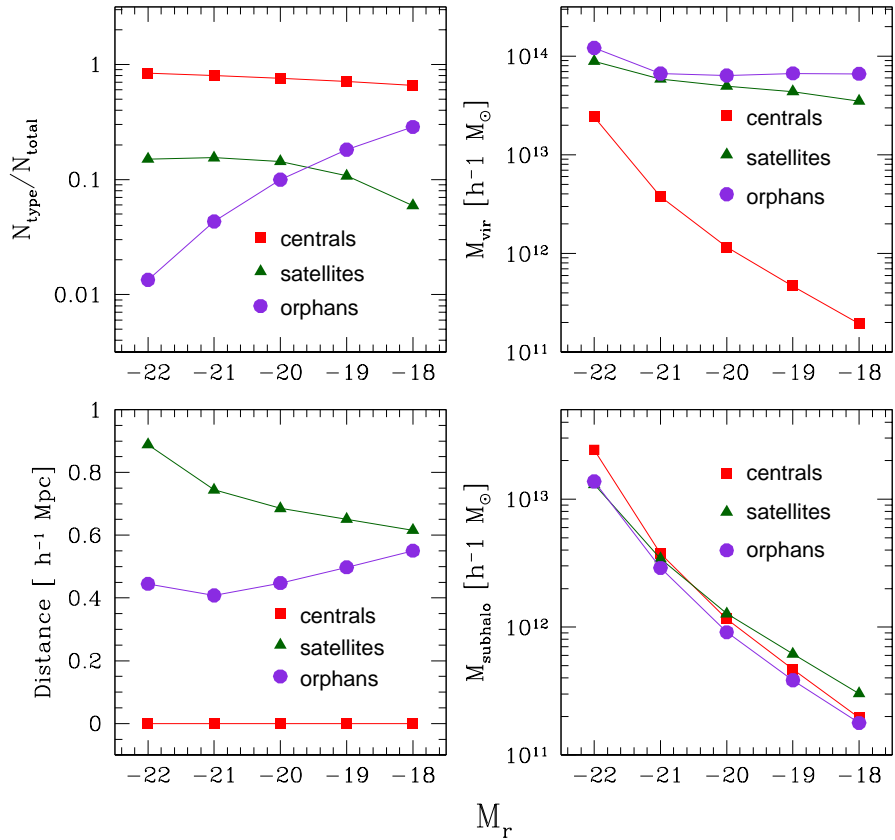}
\caption{Galaxy properties as a function of absolute red-band
  magnitude. The galaxies are at redshift 0.24. {\it Upper left
    panel:} The ratio of the number of galaxies of each type to the
  total number of galaxies in the respective magnitude bin. {\it Upper
    right panel:} The virial mass of the host halo. {\it Lower left
    panel:} The average distance of galaxies to the central
  galaxy. {\it Lower right panel:} The mass of the subhaloes where the
  galaxies formed.}
\label{fig:gal_prop}}
\end{figure*}
%%%%%%%%%%%%%%%%%%%%%%%%%%%%%%%%%%%%%%%%%%%%%%%%%%%%%%%

The MXXL is the largest simulation in the Millennium series, with a
volume of $V=[3\,\Gpc]^3$ and $6720^3$ dark matter particles of mass
$m_p=6.9\times10^9\,\Msol$. The cosmological model corresponds to a
flat $\Lambda$CDM universe with: the matter density parameter
$\Omega_m=0.25$; the dimensionless Hubble parameter $h=0.73$; the
amplitude of matter fluctuations $\sigma_8=0.9$; the primordial
spectral index $n_s=1$; and a constant dark energy equation of state
with $w=-1$. For a complete description of the MXXL we refer the
reader to \cite{Anguloetal2012}.

Halo and subhalo catalogues were stored for $63$ snapshots. The
smallest object in these catalogues has a mass $\sim 1.4\times
10^{11}\,\Msol$. Merger trees were built by identifying for every
subhalo in each snapshot the most likely descendant in the next
snapshot. The trees were then used to build a galaxy catalogue with
the SAM galaxy formation code {\tt L-Galaxies}
\citep{Springeletal2005}.

The {\tt L-Galaxies} code corresponds to a set of differential
equations that couple with the above-mentioned merger trees and that
encode the key physical mechanisms for galaxy formation. Processes
such as gas cooling, star formation, feedback from SN and AGN, galaxy
mergers, black hole formation and growth, and generation of metals are
all implemented in a self-consistent manner. We refer the interested
reader to \citet{Guoetal2011, Henriquesetal2012} and references
therein for specific details on the method, and to
\citet{Anguloetal2014} for details on the implementation in the MXXL
simulation. Here, we just highlight that the galaxy population of a
given halo does not depend on its mass alone, as commonly assumed in
many models, but also on the details of the halo assembly history and
environment. For each galaxy, the full star formation history is
stored, and when coupled with population synthesis models and an
assumed initial stellar mass function, it allows us to compute the
expected luminosity for each of the five SDSS filters.
%%%%%%%%%%%%%%%%%%%%%%%%%%%%%%%%%%%%%%%%%%%%%%%%%%%%%%%

In Figure~\ref{fig:lum_func} we present the luminosity function for
the five SDSS bands. We find that all of the luminosity functions show
qualitatively similar behaviour: a steep fall-off at bright magnitudes
and a turn-over followed by a power-law-like tail at intermediate and
faint magnitudes. We also see that for a given magnitude band there
are greater abundances of galaxies at red wavelengths than at
blue. This is qualitatively consistent with observational results from
the SDSS \citep{Blantonetal2003}. For the faintest magnitudes we
notice artifacts produced by the finite mass resolution of the MXXL
simulation. We elaborate on this next.

In the upper left panel of Figure~\ref{fig:gal_prop} we present the
relative abundance of central, satellite and orphan galaxies as a
function of their red-band absolute magnitude. `Central' galaxies
reside at the centres of the main halo(subhalo), and are therefore the
main galaxies of the FoF haloes. `Satellite' galaxies inhabit
satellite subhaloes within the FoF haloes. `Orphan' galaxies are
satellite galaxies whose dark matter subhalo has been stripped down
below the resolution limit of the simulation. The figure clearly shows
that the brightest galaxies ($M_r\leq -18$) are mostly centrals. The
satellites with a resolved dark matter subhalo are sub-dominant for
all magnitude bins, but dominate among satellite galaxies with $M_r <
-20$. These features depend on the mass resolution of the simulation
(see Figure~\ref{fig:gal_prop_Bruno} for an analogous figure
constructed from the higher-resolution Millennium simulation).  With a
much higher mass resolution, central galaxies would dominate at any
luminosity and there would be no orphan galaxies.

The upper right panel of \Fig{fig:gal_prop} shows how the mass of the
host haloes evolves with the galaxy luminosity. Independent of the
brightness of satellite and orphan galaxies, the haloes hosting them are
quite massive ($M_{\rm vir}\in [3\times 10^{13}, 10^{14}]\,\Msol$),
whereas the haloes inhabited by central galaxies display a substantial
decrease in mass with decreasing luminosity. This drop in halo mass
spans more than two orders of magnitude, starting at $M\sim 2\times
10^{13}\,\Msol$. 

The bottom left panel of \Fig{fig:gal_prop} shows the average distance
of the galaxies from the central galaxy, as a function of
magnitude. By definition the distance of central galaxies is
zero. Satellite galaxies are on average most distant from the halo
centre: the brightest have a separation of $\sim0.9\,\Mpc$, and the
faintest of about $0.6\,\Mpc$.  On the other hand, orphan galaxies are
on average within $[0.4, 0.6]\,\Mpc$ from the halo centre, which is a
consequence of tidal forces being stronger closer to the halo centre
where tidal disruption and mass loss happen.

Finally, the bottom right panel of \Fig{fig:gal_prop} presents the
evolution of the host subhalo mass with luminosity. For centrals
this is in fact the same as shown in the upper right panel; the other
two types follow the same trend as the centrals. Note that the subhalo
mass associated with the orphan galaxies is defined to be the mass of
the last subhalo tracked before it fell below the mass resolution of
the simulation. We can see that there are no strong differences among
different types which is a consequence of the fact that the $r$-band
magnitude mostly traces the total amount of mass in stars, which in turn
depends primarily on the total amount of gas available for star formation
and thus on the mass of the host dark matter structure. 

In this paper we shall use only the red band, since most of the GGL
and GC studies to date have focused on this band.  We shall only
consider galaxies with $M_r < -19$, split into four absolute magnitude
bins, with each bin spanning a single unit of magnitude, except for
the brightest bin for which we take all galaxies with $M_r<-22$.  We
have ensured that above the chosen limit $M_r < -19$ our results are
qualitatively insensitive to the finite mass resolution of the MXXL by
explicitly comparing the GGL and GC signals with those derived from
the higher-resolution Millennium simulation.

%%%%%%%%%%%%%%%%%%%%%%%%%%%%%%%%%%%%%%%%%%%%%%%%%%%%%%%

\section{Results I: clustering and lensing measurements in MXXL}
\label{SV}

%%%%%%%%%%%%%%%%%%%%%%%%%%%%%%%%%%%%%%%%%%%%%%%%%%%%%%%
\subsection{Methodology for estimating the projected correlation functions}
\label{SV.1}
%%%%%%%%%%%%%%%%%%%%%%%%%%%%%%%%%%%%%%%%%%%%%%%%%%%%%%%
%%%%%%%%%%%%%%%%%%%%%%%%%%%%%%%%%%%%%%%%%%%%%%%%%%%%%%%
In order to estimate the GGL and GC signals and their covariance from
the MXXL data, we divide the simulation box into 216 subcubes of
volume $V_{\rm sub}=[500 \Mpc]^3$.  Each subcube therefore contains
$\approx 1.4\times 10^9$ dark matter particles, as well as
galaxies from the catalogues described in section \S\ref{SIV}. For
our analysis we assume both the Born and Limber approximations, which
allow us to perform all of the computations at the fixed redshift
$z=0.24$. For each of the subcubes, we measure the projected
matter-matter, galaxy-matter, and galaxy-galaxy correlation
functions. In order to handle the huge data volume, we have developed
a fast k-D tree code in {\tt C++} with {\tt MPI} parallelization. Our
algorithm is similar to that described by \citet{Mooreetal2001} and
\citet{Jarvisetal2004}. However, rather than invoking an approximate
scheme for binning the pair counts as is done in these algorithms, we
place every particle exactly into the correct radial bin. We have
carefully tested that our code obeys the pair counting scaling
$DD\propto\, t^{3/2}$ and that it reproduces exactly the answer obtained
from a brute-force pair summation code.

For the particular problem of computing the correlation functions in
the MXXL simulation, we count pairs in logarithmic bins of the
transverse distance $R$, and linear bins of the line-of-sight distance
$\chi$. Since the subcubes do not have periodic boundary conditions,
we also cross-correlate the data with a random catalogue to account
for boundary effects on the pair counts. With the pair counts for
the $(R, \chi)$ bins, the 3D correlation function can be estimated by
using the unbiased and minimum-variance (in the limit of
no-clustering) estimator of \citet{LandySzalay1993}:
\be
\hat{\xi}=(D_1D_2-D_1R-D_2R+RR)/RR \ ,
\label{eq:def_LS}
\ee
where $D_1$ and $D_2$ represent the first and second data catalogues,
and $R$ is the random catalogue. $D_1D_2$, $D_1R$, $D_2R$, and $RR$
represent the respective pair counts. For auto-correlations,
$D_1=D_2$, while for cross-correlations e.g. of galaxies and matter,
they differ. Note that this estimator perfectly matches the one we
defined in Eqs.~(\ref{eq:def_est_ngal}) and~(\ref{eq:est_xigg_v1}),
since we took the weights there to be constant. The ratio of the
number of data particles to the number of particles from the random
catalogue represents the $\alpha$ from Eq.~(\ref{eq:def_est_ngal}).
%%%%%%%%%%%%%%%%%%%%%%%%%%%%%%%%%%%%%%%%%%%%%%%%%%%%%%%
\begin{figure*}
\centering {
\includegraphics[scale=0.75]{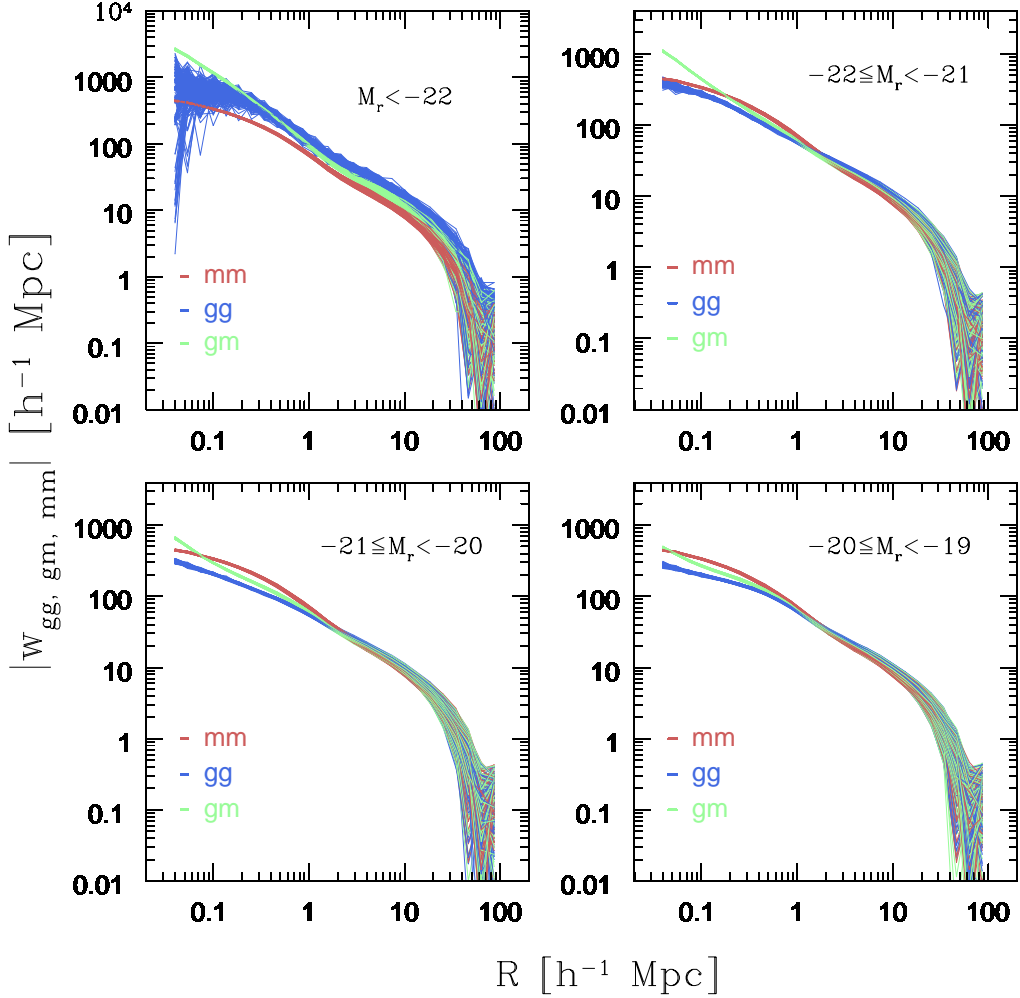}
\caption{Projected correlation functions measured for all 216 subcubes
  of the MXXL at redshift 0.24. The red, blue, and green lines
  represent the matter-matter, galaxy-galaxy, and galaxy-matter
  correlation functions respectively. In each panel the galaxies are
  binned according to their $r$-band absolute magnitudes. }
\label{fig:wp_subcubes}}
\end{figure*}
%%%%%%%%%%%%%%%%%%%%%%%%%%%%%%%%%%%%%%%%%%%%%%%%%%%%%%%
Some details of how we estimate these correlations are as follows. In
order to obey computing time constraints, we limit the random
catalogues to $10^6$ particles. To maintain a value as low as possible
for $\alpha$, we subsample both the matter and the galaxy data. The
number of subsamples is $32$, and for each of them we generate a
random catalogue. The rate of sampling for matter is $1/4000$, which
gives us about $350,\!000$ dark matter particles per subsample, while
for each luminosity bin we randomly select no more than $150,\!000$
galaxies. These values correspond to $\alpha \sim 0.3$ for matter, and
$\alpha\sim 0.15$ for galaxies in each luminosity bin. The only
exception is the first luminosity bin, containing the brightest
galaxies, which has only about $20,\!000$ galaxies per subcube, and is
therefore not subsampled. The matter particles and galaxies from every
subsampled subcube are correlated with the random catalogue generated
for that subsample. Thus for all the subcubes, the i{\it th} subsample
is correlated with the i{\it th} random catalogue. We have checked
that given the number of data particles, the number of random
particles is sufficiently large not to yield significant errors in the
measured projected correlations.

We found that it was crucial to correct the projected correlation
functions for the integral constraint, otherwise the results exhibited
a strong dependence on the projection length $\chi_{\rm max}$. This
owes to the fact that the total density of objects in each subcube is
not guaranteed to reach the universal mean.

We implement the integral constraint in the estimates of the 3D galaxy
correlation function in a similar fashion to the procedure described
in \citet{LandySzalay1993}: To begin, we define the `geometric
factor':
%%%%%%%%%%%%%%%%%%%%%%%%%%%%%%%%%%%%%%%%%%%%%%%%%%%%%%%
\be
G_p(R_i, \chi_j)\equiv\frac{RR_{i j}}{\sum_{i,j} RR_{i j}\,dV_{i j}} \ ,
\label{eq:GF}
\ee
%%%%%%%%%%%%%%%%%%%%%%%%%%%%%%%%%%%%%%%%%%%%%%%%%%%%%%%
where $RR_{i j}$ is the number of random pair counts for the bin
$(R_i, \chi_j)$, and $dV_{i j}$ is the cylindrical volume of the
respective bin, i.e. $dV_{ij}=\pi(R_{i+1}^2-R_i^2)\Delta\chi_j$, with
$\Delta\chi_j\equiv \chi_{j+1}-\chi_j$. Thus $G_p(R_i, \chi_j)$ is the
number of pairs possible in the bin $ij$ relative to the total number
of pairs in the survey volume. It is normalized to unity over the
survey volume, i.e. $\sum_{i, j} G_p(R_i, \chi_j)dV_{ij}=1$. The
integral constraint is defined by the equation:
%%%%%%%%%%%%%%%%%%%%%%%%%%%%%%%%%%%%%%%%%%%%%%%%%%%%%%%
\be
\sum_{i,j} (1+\widehat{\xi}_{i j})G_p(R_i,\chi_j)\, dV_{i j}\equiv 1+\xi_{V_s},
\ee
%%%%%%%%%%%%%%%%%%%%%%%%%%%%%%%%%%%%%%%%%%%%%%%%%%%%%%%
where $\widehat{\xi}_{i j}$ is the estimator introduced in
Eq.~(\ref{eq:def_LS}) at the respective bin. The
integral-constraint-corrected estimator for the projected galaxy
correlation function is given by:
%%%%%%%%%%%%%%%%%%%%%%%%%%%%%%%%%%%%%%%%%%%%%%%%%%%%%%%
\be
\widehat{w}_{\rm corr}(R_i)=\sum_{j=1}^{N_\chi}
\left[\frac{1+\widehat{\xi}_{i j}}{1+\xi_{V_s}}-1\right]\Delta\chi_j\,,
\label{eq:corr_LS}
\ee
%%%%%%%%%%%%%%%%%%%%%%%%%%%%%%%%%%%%%%%%%%%%%%%%%%%%%%%
where $N_\chi$ is the number of bins in $\chi$. We apply this to all
of the measurements.

Finally, we mention the choice of bins: we have 25 logarithmic bins in
$R$, spanning the interval $[0.04, 54]\,\Mpc$, and 10 linear bins in
$\chi$, with a chosen $\chi_{\rm max}$ of $100\,\Mpc$.  We checked
that the number of line-of-sight bins is sufficiently large to obtain
an accurate estimation of the projected correlation function.

To summarize: the projected correlation functions are estimated
through the following steps: i) use the tree code to evaluate the 3D
pair counts for each $(R, \chi)$ bin, and for every subsample; ii)
build the Landy-Szalay estimator for the 3D correlation function from
the pair counts and determine the integral constraint factor; iii) add
the line-of-sight bins to obtain the projected correlation functions,
i.e. compute $\widehat{w}_{\rm corr}$ following
\Eqn{eq:corr_LS}; iv) calculate the average of the subsamples.
%%%%%%%%%%%%%%%%%%%%%%%%%%%%%%%%%%%%%%%%%%%%%%%%%%%%%%%
%%%%%%%%%%%%%%%%%%%%%%%%%%%%%%%%%%%%%%%%%%%%%%%%%%%%%%%
\subsection{Galaxy and matter correlation functions}
\label{SV.2}
%%%%%%%%%%%%%%%%%%%%%%%%%%%%%%%%%%%%%%%%%%%%%%%%%%%%%%%
%%%%%%%%%%%%%%%%%%%%%%%%%%%%%%%%%%%%%%%%%%%%%%%%%%%%%%%
Figure~\ref{fig:wp_subcubes} presents the projected matter-matter,
galaxy-matter, and galaxy-galaxy correlation functions -- red, green,
blue solid lines respectively -- from the 216 subcubes of the
MXXL. Each line corresponds to the estimate from one subcube, and each
panel to a magnitude bin, starting with the brightest galaxies in the
upper left corner, and down to the faintest in the lower right
corner. The measurements have some scatter for the two brightest
magnitude bins, but are relatively tight otherwise. This plot also
provides a check that none of the subcubes displays any anomalous
behaviour.
%%%%%%%%%%%%%%%%%%%%%%%%%%%%%%%%%%%%%%%%%%%%%%%%%%%%%%%
\subsection{Galaxy bias and cross-correlation coefficient}
\label{SV.3}
%%%%%%%%%%%%%%%%%%%%%%%%%%%%%%%%%%%%%%%%%%%%%%%%%%%%%%%
%%%%%%%%%%%%%%%%%%%%%%%%%%%%%%%%%%%%%%%%%%%%%%%%%%%%%%%
We may obtain the galaxy bias parameter either from the galaxy-galaxy
or galaxy-matter projected correlation functions through:
%%%%%%%%%%%%%%%%%%%%%%%%%%%%%%%%%%%%%%%%%%%%%%%%%%%%%%%
\ba
b_{\rm gg}(R) &  = & \sqrt{\frac{w_{\rm gg}(R)}{w_{\rm mm}(R)}} \ ; \label{eq:bgg} \\
b_{\rm gm}(R) &  = & \frac{w_{\rm gm}(R)}{w_{\rm mm}(R)}\ . \label{eq:bgm}
\ea
%%%%%%%%%%%%%%%%%%%%%%%%%%%%%%%%%%%%%%%%%%%%%%%%%%%%%%%
%%%%%%%%%%%%%%%%%%%%%%%%%%%%%%%%%%%%%%%%%%%%%%%%%%%%%%%
\begin{figure*}
\vspace{0.5cm} \centering {
  \includegraphics[scale=0.7]{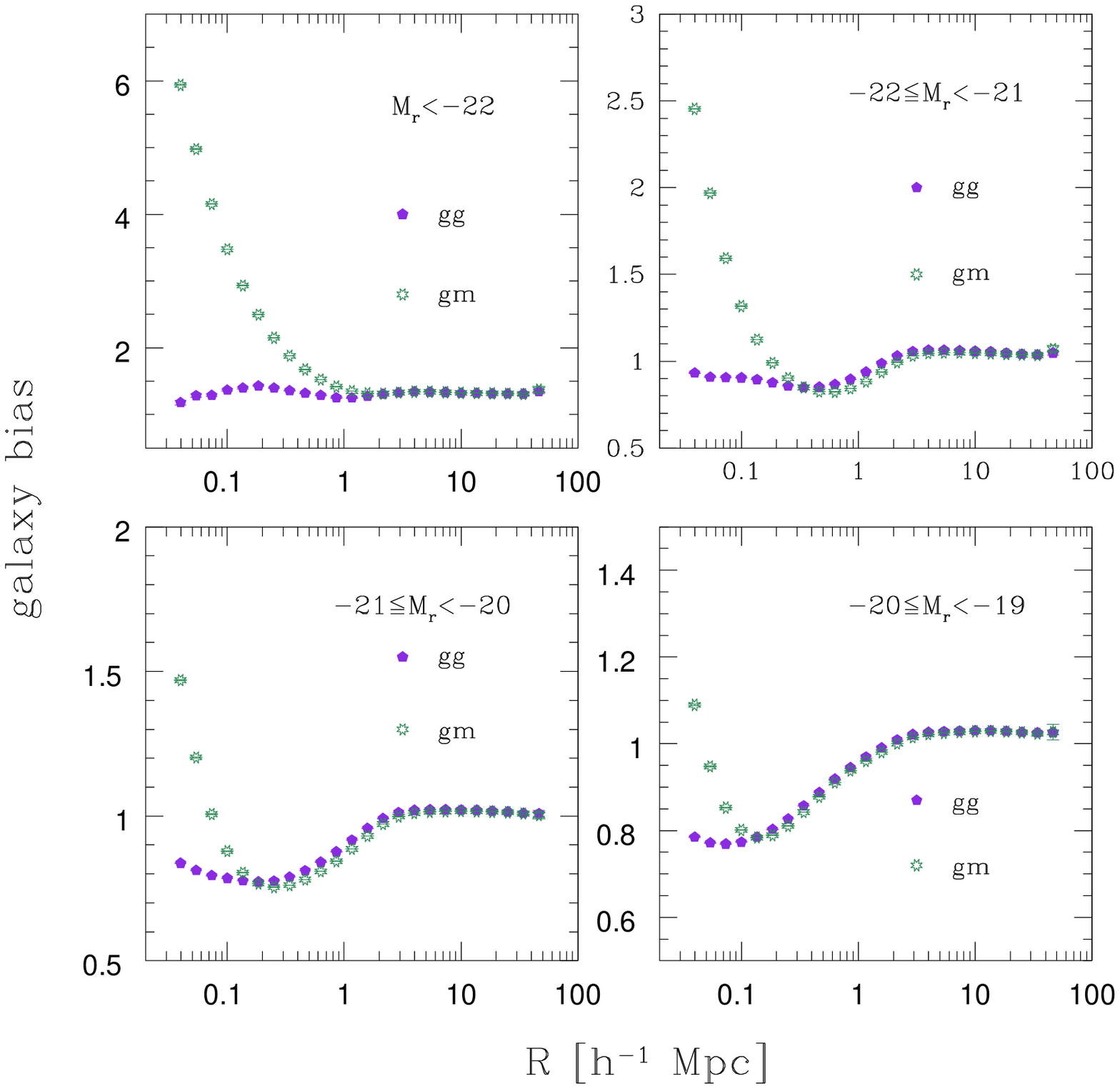}
\caption{The average galaxy bias for the various absolute-magnitude
  bins.  The violet solid pentagons depict the bias computed as in
  Eq.~(\ref{eq:bgg}), while the green symbols show the bias estimated
  as in Eq.~(\ref{eq:bgm}). The error bars correspond to errors on the
  mean of the 216 subcubes. }
\label{fig:bias_ave}}
%%%%%%%%%%%%%%%%%%%%%%%%%%%%%%%%%%%%%%%%%%%%%%%%%%%%%%%
\vspace{0.5cm}
%%%%%%%%%%%%%%%%%%%%%%%%%%%%%%%%%%%%%%%%%%%%%%%%%%%%%%%
\centering {
\includegraphics[scale=0.46]{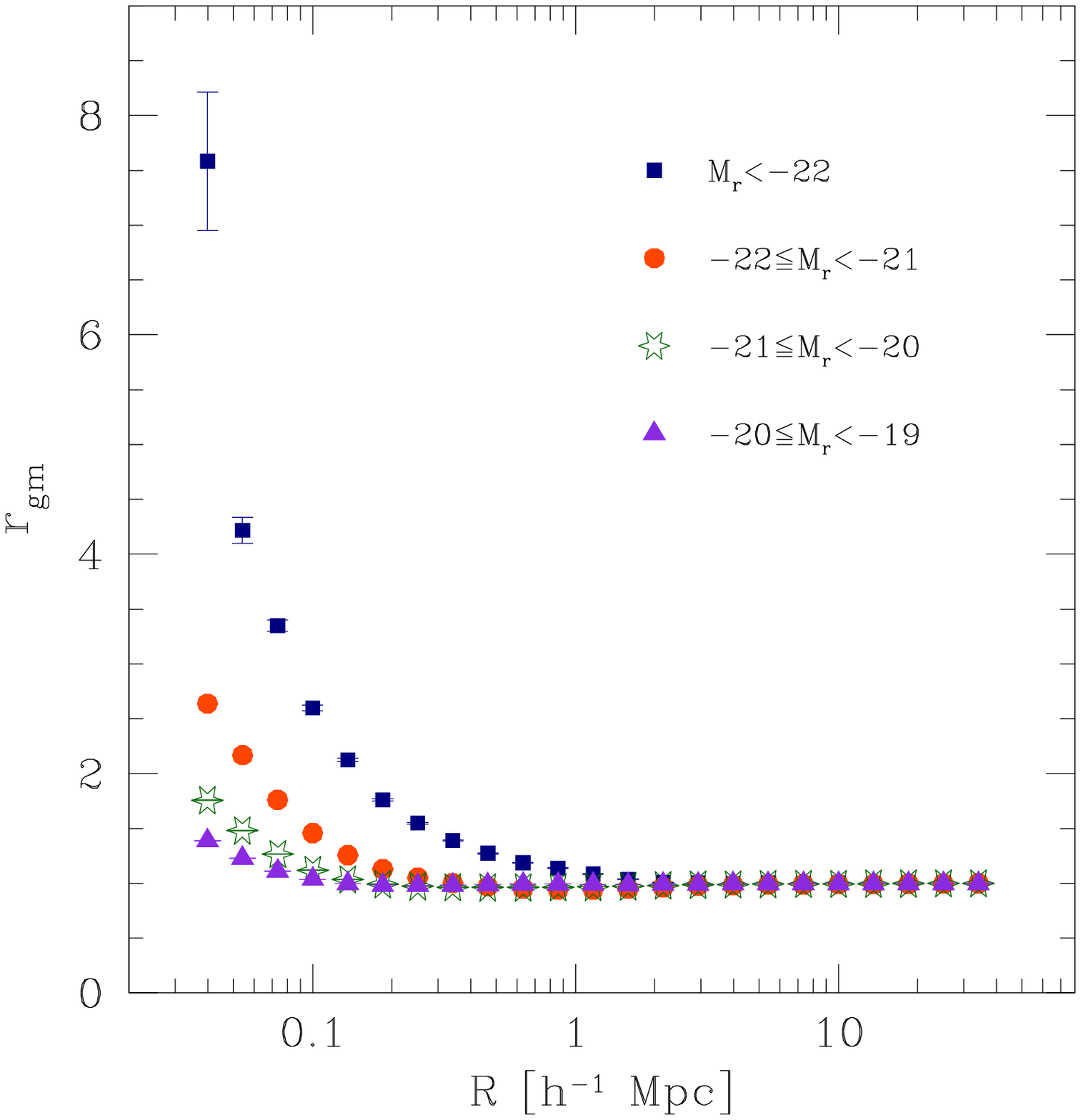}
\includegraphics[scale=0.46]{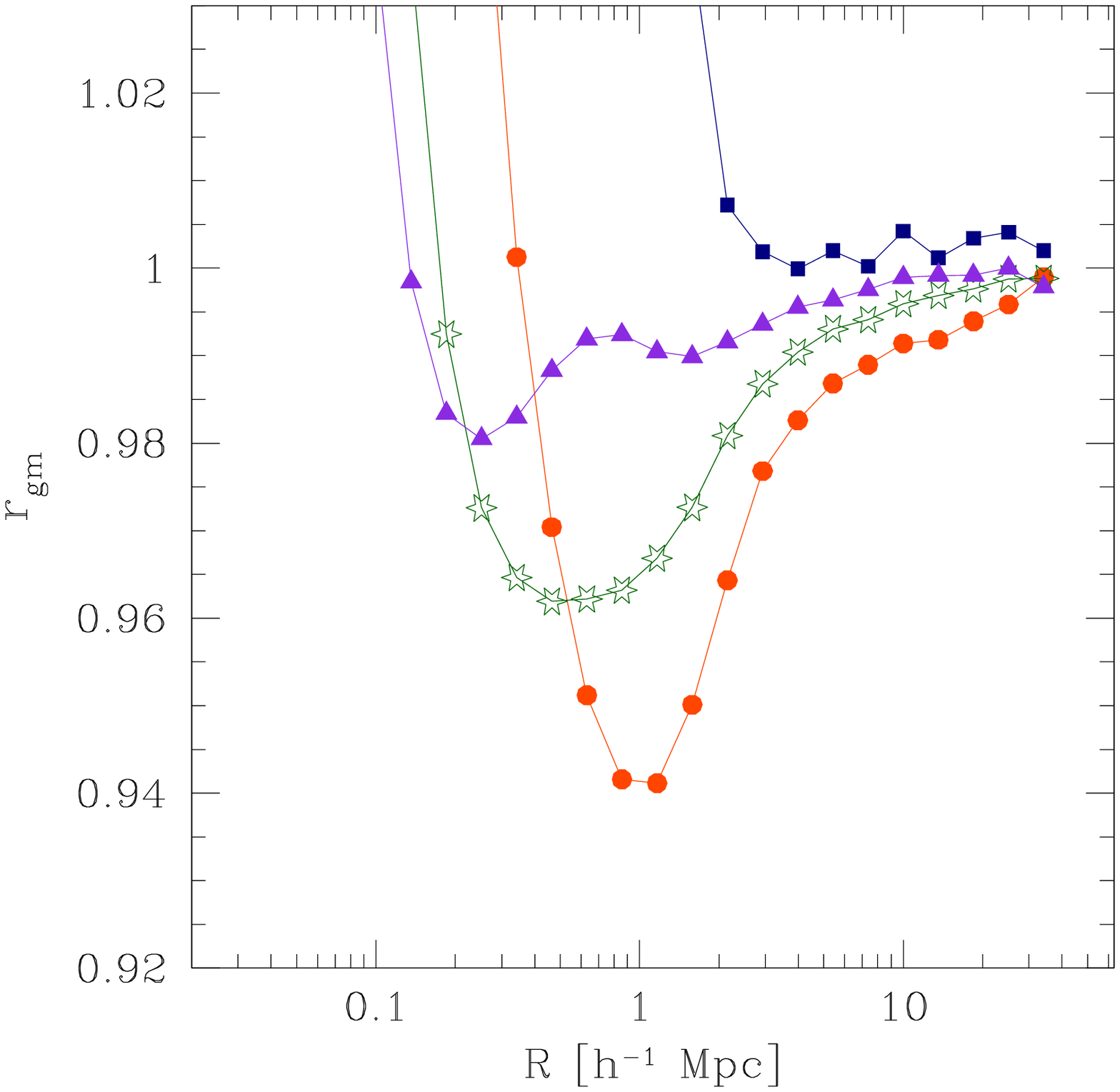}
\caption{The average cross-correlation coefficient,
  e.g. Eq.~(\ref{eq:corr_coef}) as a function of transverse radius.
  {\it Left panel:} The full range.{\it Right panel:} Zoom-in. }
\label{fig:corr_coef}}
\end{figure*}
%%%%%%%%%%%%%%%%%%%%%%%%%%%%%%%%%%%%%%%%%%%%%%%%%%%%%%%
Figure~\ref{fig:bias_ave} presents the two bias estimates for all
magnitude bins and averaged over all 216 subcubes of the MXXL
simulation. There are several points worth noticing in these
figures. Firstly, on large scales, we see that both $b_{\rm gg}$ and
$b_{\rm gm}$ appear to be constant and qualitatively consistent with
one another. We also note that the bias is relatively similar for the
fainter bins, but it increases sharply for the brightest
galaxies. This luminosity dependence of the bias in SAM models is
consistent with earlier studies \citep[see][and references
  therein]{Smith2012}, and can be understood from
Figure~\ref{fig:gal_prop}. There we see that it is only for the
brightest magnitude bin that the host haloes of both the central and
satellite galaxies are very massive. For the fainter bins, the mass of
the host haloes decreases with luminosity in the case of central
galaxies, while remaining relatively constant in the case of the
satellites. Thus the bias of the latter is boosted. This explanation
is consistent with the picture where an individual galaxy inherits the
bias of the halo hosting it.

On smaller scales, we notice that $b_{\rm gm} > b_{\rm gg}$: the scale
where this transition occurs decreases with increasing absolute
magnitude, ranging from $R\sim 1\,\Mpc$ for the brightest galaxies to
$R\sim 0.07\,\Mpc$ for the faintest bin.  The fact that $b_{\rm gm} >
b_{\rm gg}$ can be qualitatively understood from the following
reasons. The 3D galaxy-galaxy correlations in SAM models obey an
exclusion condition, which is that individual galaxies cannot come
closer than the separation of the sum of their individual subhalo
virial radii. On scales smaller than this, the correlation function
drops to -1. Whereas for the galaxy-matter cross-correlation function
no such exclusion is present and one simply probes the density profile
of matter around galaxies -- which is known to be cuspy
\citep{HayashiWhite2008}. However, in projection these effects are
less significant, but nevertheless still operate and lead to the shape
shown in \Fig{fig:bias_ave}. The transition scale varies with
magnitude because the halo mass of central galaxies decreases with
increasing absolute magnitude (c.f.~Figure~\ref{fig:gal_prop}).  

The cross-correlation coefficient can be defined as:
%%%%%%%%%%%%%%%%%%%%%%%%%%%%%%%%%%%%%%%%%%%%%%%%%%%%%%%
\be
r_{\rm gm}(R)\equiv\frac{w_{\rm gm}(R)}{\sqrt{w_{\rm gg}(R)\,w_{\rm mm}(R)}}
 \equiv  \frac{b_{\rm gm}(R)}{b_{\rm gg}(R)}.
\label{eq:corr_coef}
\ee
%%%%%%%%%%%%%%%%%%%%%%%%%%%%%%%%%%%%%%%%%%%%%%%%%%%%%%%
Note that this is not the same as is usually understood in statistics,
where the cross-correlation coefficient of two variables $X$ and $Y$
is constrained to be $|r|\le 1$. \Eqn{eq:corr_coef} is defined in
terms of correlation functions, and hence provided $R\ne0$ it is not
required to obey the condition $|r|\le1$. Indeed if either the
galaxy-galaxy or matter-matter correlation functions cross zero, which
they most certainly do, then $r$ is formally divergent. Nevertheless,
the diagnostic properties of $r$ are key: if the galaxy bias were
linear and deterministic then $r=1$, and measurements of either
$w_{\rm gm}$ or $w_{\rm gg}$ may be directly related to the underlying
matter distribution, modulo an amplitude factor. However, any
departure from unity indicates that the bias is either nonlinear or
stochastic, or both \citep[for more discussion
  see][]{DekelLahav1999,SeljakWarren2004}.

Figure~\ref{fig:corr_coef} shows the cross-correlation coefficient as
a function of the transverse scale $R$. We find that on large scales,
the correlation coefficient approaches unity for the four magnitude
bins that we have considered. This implies that, at least for the SAM
galaxies in MXXL, the large-scale bias is linear and deterministic and
that it describes both the galaxy-galaxy and galaxy-matter correlation
functions. On small scales we see that the correlation coefficient
decreases sharply and then shoots up above unity. This is consistent
with the nonlinear scale-dependent bias presented in
\Fig{fig:bias_ave}. Note that the small-scale clustering of galaxies
is very sensitive to the treatment of dynamical friction of orphan
galaxies, so we do not wish to over-interpret the results of
Figures~\ref{fig:bias_ave} and~\ref{fig:corr_coef} for $R< 100\,\kpc$.
%%%%%%%%%%%%%%%%%%%%%%%%%%%%%%%%%%%%%%%%%%%%%%%%%%%%%%%
\begin{figure*}
\centering { \includegraphics[scale=0.52]{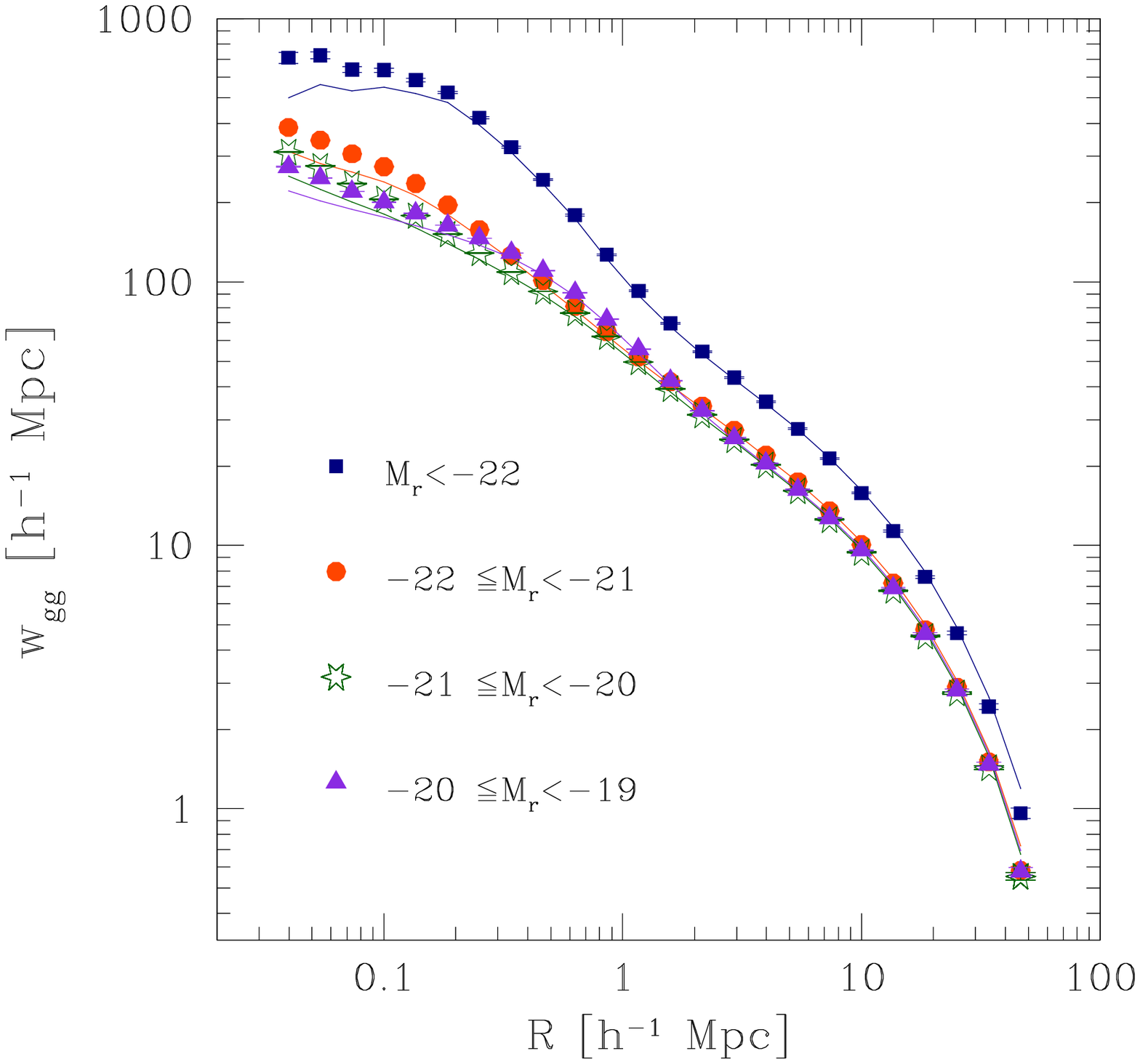}
  \includegraphics[scale=0.52]{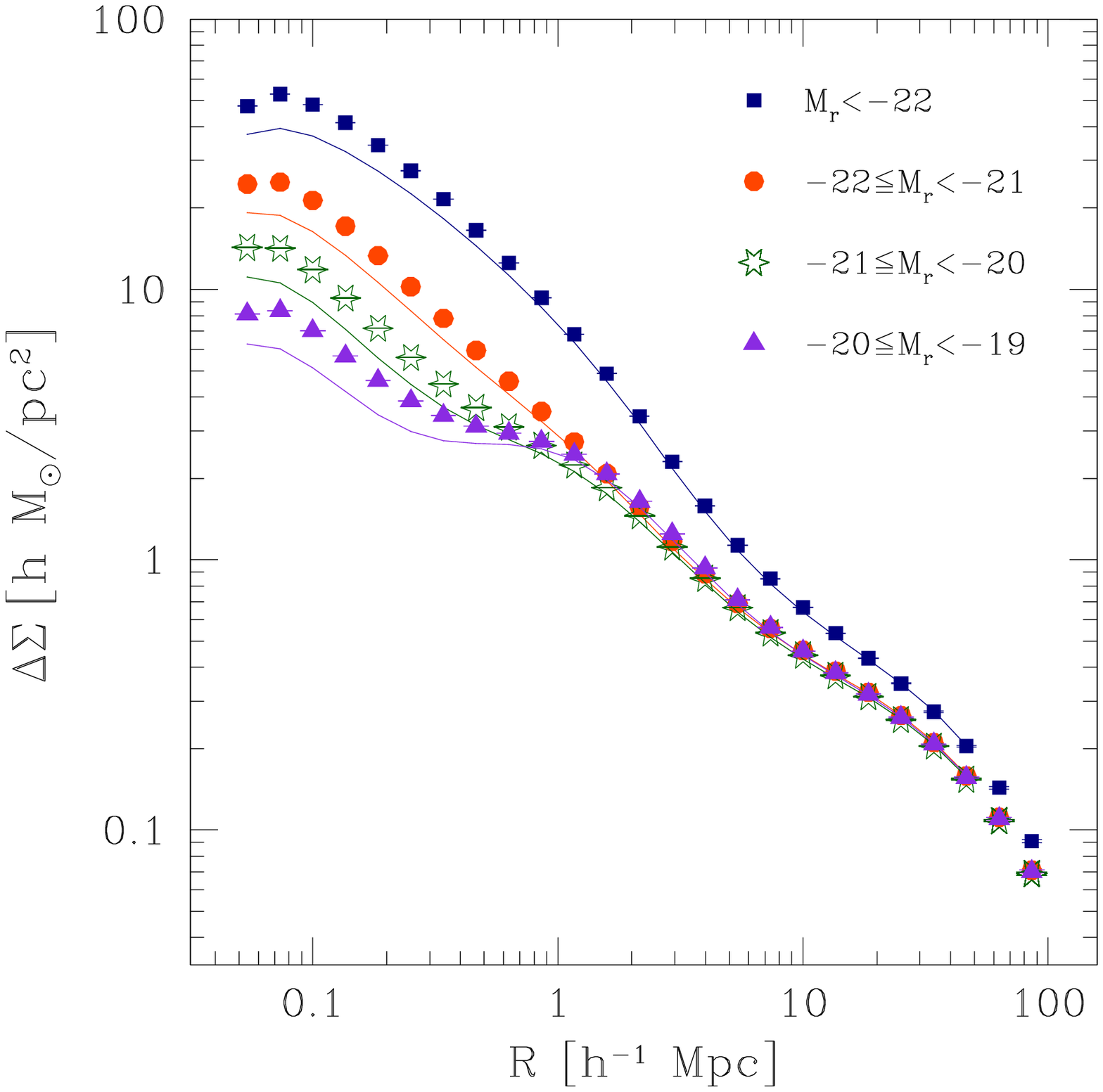}
\caption{{\it Left panel:} The projected galaxy correlation function
  for various magnitude bins. The symbols denote the average of the
  216 subcubes, the error bars are on the mean. The lines represent
  the theoretical predictions of Eq.~(\ref{eq:wgg_ave_Limber}).
{\it Right panel:} The excess surface density. Symbols and lines are
  just like in the left panel.}
\label{fig:wgg_shear}}
\end{figure*}
%%%%%%%%%%%%%%%%%%%%%%%%%%%%%%%%%%%%%%%%%%%%%%%%%%%%%%%
\subsection{Comparison of the measured and theoretical projected galaxy correlation function}
\label{SV.4}
%%%%%%%%%%%%%%%%%%%%%%%%%%%%%%%%%%%%%%%%%%%%%%%%%%%%%%%
%%%%%%%%%%%%%%%%%%%%%%%%%%%%%%%%%%%%%%%%%%%%%%%%%%%%%%%
In Figure~\ref{fig:wgg_shear} we show again the projected galaxy
correlation functions for the four magnitude bins, but this time we
have averaged the data over the 216 MXXL subcubes. The error bars are
plotted on the mean. The solid lines present the theoretical
predictions. We evaluate the theory as follows: instead of a direct
numerical evaluation of \Eqn{eq:wgg_ave_Limber}, which would require a
model for $P_{\rm gg}(k)$, we determine the projected nonlinear matter
correlation function under the Limber approximation by replacing
$P_{\rm gg}\rightarrow P_{\rm mm}(k)$. We then simply multiply this
quantity by the measured bias, e.g. \Eqn{eq:bgg}:
%%%%%%%%%%%%%%%%%%%%%%%%%%%%%%%%%%%%%%%%%%%%%%%%%%%%%%%
\be 
w^L_{\rm gg}(R) = b_{\rm gg}(R)  w^L_{\rm mm}(R) \ . 
\label{eq:wgg_th}
\ee
%%%%%%%%%%%%%%%%%%%%%%%%%%%%%%%%%%%%%%%%%%%%%%%%%%%%%%%
In computing $w^L_{\rm mm}$ we use the nonlinear matter power spectrum
fitting formula {\tt halofit} \citep{Smithetal2003}.

In \Fig{fig:wgg_shear} we see that on small scales $R<0.1\,\Mpc$, the
predictions underestimate the measured $w_{\rm gg}$ by roughly
$\sim20\%$. These discrepancies are attributed to the fact that {\tt
  halofit} underpredicts the true nonlinear matter power spectrum on
small-scales \citep{Takahashietal2012}. The more recent work of
\citep{Fosalbaetal2015a, Fosalbaetal2015b} shows however that the
revised {\tt halofit} of \citep{Takahashietal2012} overpredicts the
nonlinear power spectrum and consequently also the convergence power
spectrum on small scales. Since the agreement between {\tt halofit}
models and simulation measurements depends on both the mass resolution
and the volume of the simulations -- the closer the simulations'
specifications to those used for the calibration of the
semi-analytical prescriptions, the better the agreement -- we defer
more sophisticated theory predictions to a future work and are
satisfied for now that the original {\tt halofit} of
\citep{Smithetal2003} gives reasonable enough answers for our purpose
here.

It is also interesting to note that the faintest galaxies yield high
projected correlation functions around $1\,\Mpc$, most likely due to
contributions from the satellite galaxies, which inhabit higher-mass
haloes and are therefore more biased.
%%%%%%%%%%%%%%%%%%%%%%%%%%%%%%%%%%%%%%%%%%%%%%%%%%%%%%%
\subsection{The stacked tangential shear of galaxies}
\label{SV.5}
%%%%%%%%%%%%%%%%%%%%%%%%%%%%%%%%%%%%%%%%%%%%%%%%%%%%%%%
%%%%%%%%%%%%%%%%%%%%%%%%%%%%%%%%%%%%%%%%%%%%%%%%%%%%%%%
As mentioned earlier, since the full particle data was not available
for a large set of redshifts, we were not able to perform ray-tracing
simulations. Instead, we use the Born approximation to make lensing
observables from the MXXL data. In real terms, this means that the
convergence is obtained as a weighted line-of-sight integration of the
matter density fluctuations \citep{BartelmannSchneider2001}:
%%%%%%%%%%%%%%%%%%%%%%%%%%%%%%%%%%%%%%%%%%%%%%%%%%%%%%%
\be 
\kappa(\btheta) = \frac{3H_0^2\Omega_m}{2c^2}
\int_{0}^{\chi} d\chi' \frac{\chi'(\chi-\chi')}{\chi}
\frac{\delta(\chi' \btheta,\chi')}{a(\chi')} \ ,
\ee
%%%%%%%%%%%%%%%%%%%%%%%%%%%%%%%%%%%%%%%%%%%%%%%%%%%%%%%
where we have assumed a flat space-time geometry and $H_0$ is the
Hubble constant, $c$ is the speed of light, $\delta$ is the linear
matter density perturbation and $a$ is the expansion factor.

If we now reexamine \Eqn{eq:shear_est_ave} it can be proven that the
azimuthally-averaged tangential shear about a randomly-selected point
$\btheta_0$ which we take to be ${\bf 0}$, may be written in real
space as \citep{Schneider2005}:
%%%%%%%%%%%%%%%%%%%%%%%%%%%%%%%%%%%%%%%%%%%%%%%%%%%%%%%
\be
\gamma_{t,a}(\theta)  =  \overline{\kappa}(\theta) -\kappa_a(\theta) \ .
\label{eq:difd}
\ee
%%%%%%%%%%%%%%%%%%%%%%%%%%%%%%%%%%%%%%%%%%%%%%%%%%%%%%%
If instead of random points we consider the centre of a lens galaxy as
the reference point, then on averaging over all galaxies the above
expression becomes \citep{Guziketal2001}:
%%%%%%%%%%%%%%%%%%%%%%%%%%%%%%%%%%%%%%%%%%%%%%%%%%%%%%%
\be
\langle\widehat{\gamma}_{t,a}^g(R, z_l, z_s)\rangle=\frac{\langle\widehat{\Delta \Sigma}
(R, z_l)\rangle}{\Sigma_{\rm crit}(z_l, z_s)} \ ,
\label{eq:shear_proxy}
\ee
%%%%%%%%%%%%%%%%%%%%%%%%%%%%%%%%%%%%%%%%%%%%%%%%%%%%%%%
where we used the relation $\theta=R/\chi(z_l)$ and the differential
surface mass density is given by
%%%%%%%%%%%%%%%%%%%%%%%%%%%%%%%%%%%%%%%%%%%%%%%%%%%%%%%
\be 
\widehat{\Delta \Sigma}(R, z_l)  = \rho_m^0 \left[\widehat{\widebar{w}}_{\rm gm}(<R,
  z_l)-\widehat{w}_{\rm gm}(R, z_l)\right],
\label{eq:delta_sigma}
\ee
%%%%%%%%%%%%%%%%%%%%%%%%%%%%%%%%%%%%%%%%%%%%%%%%%%%%%%%
In the above $\rho_m^0$ is the comoving matter density of the
Universe, and we have assumed that the circularly-averaged tangential
shear is sourced only by the matter associated with a single lens
galaxy. The critical surface-density for lensing is given by:
%%%%%%%%%%%%%%%%%%%%%%%%%%%%%%%%%%%%%%%%%%%%%%%%%%%%%%%
\ba
\Sigma_{\rm crit}(z_l, z_s) \equiv \frac{c^2}{4\pi
  G}\frac{D_A(z_s)}{D_A(z_l;z_s) D_A(z_l)}\nn
\ea
%%%%%%%%%%%%%%%%%%%%%%%%%%%%%%%%%%%%%%%%%%%%%%%%%%%%%%%
where $D_A(z_s)$, $D_A(z_l)$, and $D_A(z_l; z_s)$ represent the
observer-source, observer-lens, and lens-source angular diameter
distances, respectively. To keep the notation compact, we shall omit
the dependence on $z_l,\,z_s$ and write the critical density simply as
$\Sigma_{\rm crit}$.

From our earlier discussion in \S\ref{SII.2} we see that an
alternative way to compute the stacked tangential shear is through
\Eqn{eq:shear_est_ave}. In the Limber approximation and assuming once
again that only matter associated with the lens galaxy creates the
shear signal we have:
%%%%%%%%%%%%%%%%%%%%%%%%%%%%%%%%%%%%%%%%%%%%%%%%%%%%%%%
\be
\langle\widehat{\gamma}_{t, a}^g(R, z_l, z_s)\rangle=\frac{\rho_m^0}
{\Sigma_{\rm crit}} \int \frac{d^2 k_\perp}{(2\pi)^2}\mathcal{P}_{\rm gm}
(k_\perp, z_l) J_2(k_\perp R).
\label{eq:shear_est_L}
\ee
%%%%%%%%%%%%%%%%%%%%%%%%%%%%%%%%%%%%%%%%%%%%%%%%%%%%%%%
We shall use \Eqn{eq:delta_sigma} to compute the excess surface
density both analytically and from the simulation data. For the
theoretical predictions, we choose to compute $w_{\rm gm}$ in the same
way as we computed $w_{\rm gg}$ in \Eqn{eq:wgg_th}, only this time
using $b_{\rm gm}$ from \Eqn{eq:bgm}. We prefer this approach to that
offered by \Eqn{eq:shear_est_L}, because it allows us to obtain
$\Delta\Sigma$ in the same way for both simulations and the
theory. At larger $R$ this choice is unimportant, however at smaller
radii, owing to the fact that the radial binning does not start at
$R=0$, it plays a more important role and the results from
Eqs.~(\ref{eq:shear_proxy}) and~(\ref{eq:shear_est_L}) differ
systematically.

The right panel of Figure~\ref{fig:wgg_shear} presents a comparison
between the theoretical predictions and measurements of $\Delta\Sigma$
as a function of the transverse spatial scale $R$, for the four
magnitude bins considered. The symbols correspond to the simulations
and the lines to the theory predictions. On large scales, $R>10\Mpc$,
we see that similar to $w_{\rm gg}$, the shear amplitudes in the three
faintest bins are comparable, whereas the brightest galaxies have a
higher amplitude. On smaller scales, $R<0.5\,\Mpc$, we find a
systematic trend: the brighter the galaxies the larger the amplitude
of the tangential shear profile. This finding is in accord with the
GGL measurements from the SDSS by \citet{Mandelbaumetal2006a}.

On closer inspection of the faintest luminosity bin, we see that it
appears to have a somewhat broad, flattish shear profile, with a very
slight second peak at $\sim 1\,\Mpc$, and higher amplitude than the
brighter bins (excepting the brightest bin) on intermediate scales
$2<R<10\,[\Mpc]$. This might be due to the increased relative
abundance of satellite galaxies compared to centrals. The satellite
galaxies are mainly hosted by high-mass haloes and have an average
distance from the centre of the main halo that is roughly on the order
of $\sim0.5\Mpc$, and so we expect that their tangential shear
profiles receive two significant contributions: the first from the
dark matter associated with their own subhalo; the second comes as the
shear profile radius encompasses the central cusp of the main
halo. This qualitatively explains the broadening of the profile of the
faintest bin. We also note that the theory underpredicts the
measurements more significantly than for $w_{\rm gg}$. This can be
attributed to the small-scale inaccuracies of {\tt halofit}
contributing to $\widebar{w}_{\rm gm}(<R)$ at all radii.
%%%%%%%%%%%%%%%%%%%%%%%%%%%%%%%%%%%%%%%%%%%%%%%%%%%%%%%
%%%%%%%%%%%%%%%%%%%%%%%%%%%%%%%%%%%%%%%%%%%%%%%%%%%%%%%
\section{Results II: covariance matrices from the MXXL}
\label{SVI}
%%%%%%%%%%%%%%%%%%%%%%%%%%%%%%%%%%%%%%%%%%%%%%%%%%%%%%%
%%%%%%%%%%%%%%%%%%%%%%%%%%%%%%%%%%%%%%%%%%%%%%%%%%%%%%%
\begin{figure*}
\vspace{0.4cm}
\centering {
\includegraphics[scale=0.56]{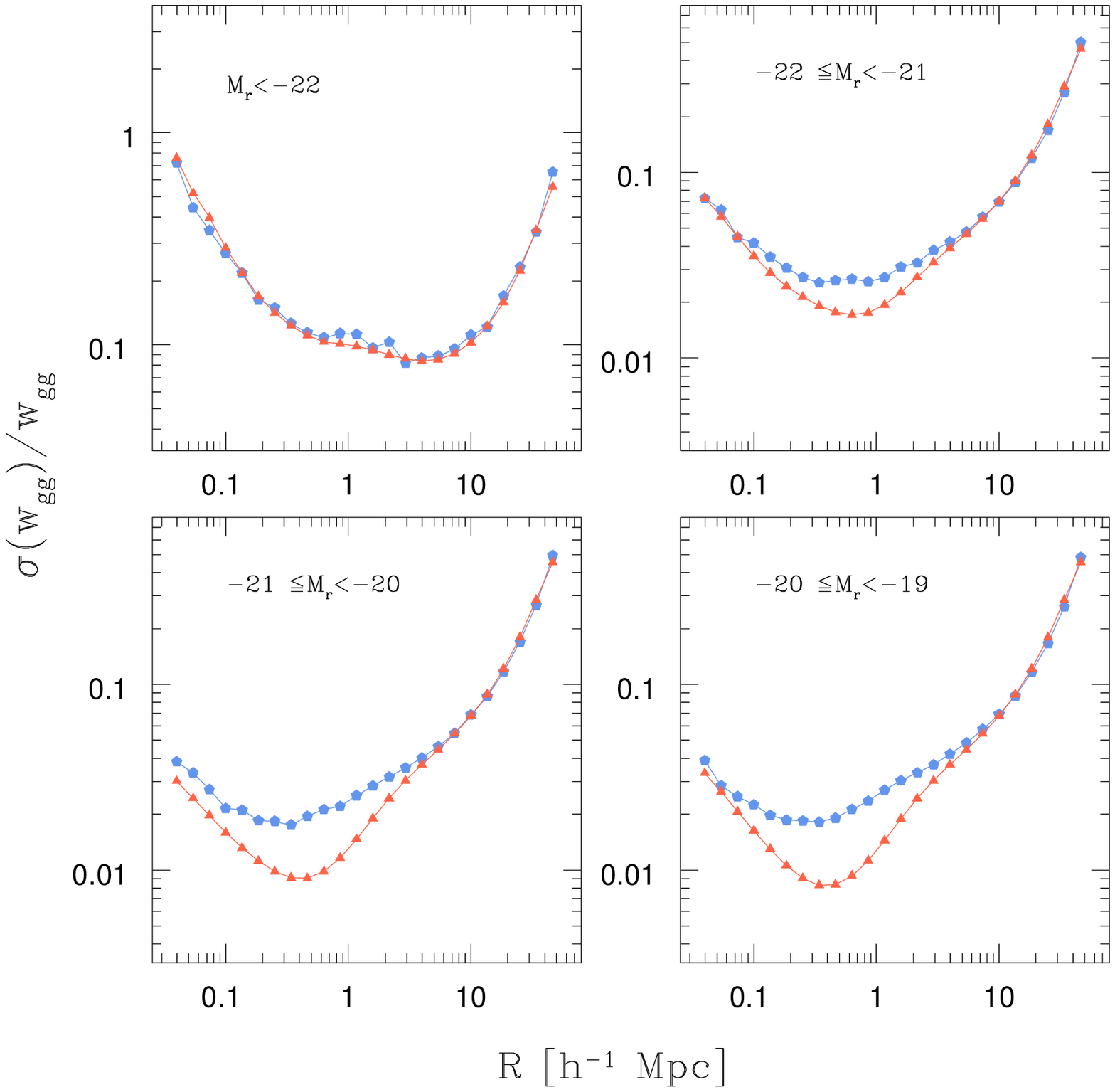}
\caption{The noise-to-signal for the projected galaxy correlation
  function. The red triangles represent the theory prediction, and the
  blue pentagons the simulation measurements.}
\label{fig:NS_GG}}
%%%%%%%%%%%%%%%%%%%%%%%%%%%%%%%%%%%%%%%%%%%%%%%%%%%%%%%
\vspace{0.5cm}
%%%%%%%%%%%%%%%%%%%%%%%%%%%%%%%%%%%%%%%%%%%%%%%%%%%%%%%
\centering {
\includegraphics[scale=0.65]{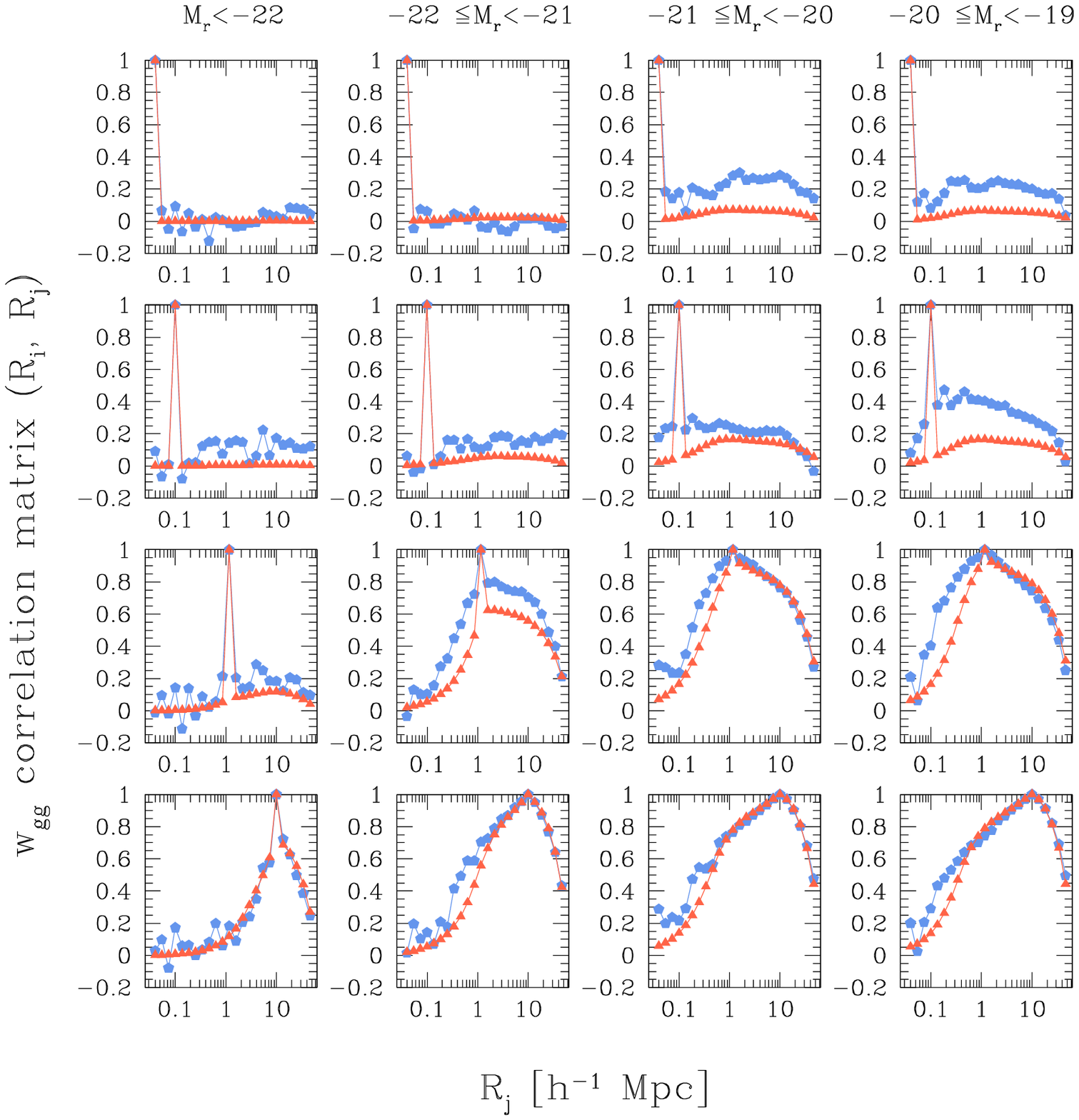}
\caption{The correlation matrix from Eq.~(\ref{eq:def_R}) for the
  projected galaxy correlation function. From top to bottom, the rows
  present four scales: $R_i=0.04, 0.1, 1, 10 \,\Mpc$. From left to
  right, the columns depict the four magnitude bins, as indicated in
  each panel. The symbols are the same as in the figure above.}
\label{fig:Row_GG}}
\end{figure*}
%%%%%%%%%%%%%%%%%%%%%%%%%%%%%%%%%%%%%%%%%%%%%%%%%%%%%%%
%%%%%%%%%%%%%%%%%%%%%%%%%%%%%%%%%%%%%%%%%%%%%%%%%%%%%%%
\subsection{Covariance of the projected correlation function}
\label{SVI.1}
%%%%%%%%%%%%%%%%%%%%%%%%%%%%%%%%%%%%%%%%%%%%%%%%%%%%%%%
%%%%%%%%%%%%%%%%%%%%%%%%%%%%%%%%%%%%%%%%%%%%%%%%%%%%%%%
In Figure~\ref{fig:NS_GG} we present the errors on the mean projected
galaxy correlation function, divided through by the signal, as a
function of the transverse scale $R$, and for the magnitude bins
discussed previously.  The blue pentagons in the figure represent the
noise-to-signal ratio estimated directly from the $N=216$
subcubes. The unbiased estimator of the mean and covariance is:
%%%%%%%%%%%%%%%%%%%%%%%%%%%%%%%%%%%%%%%%%%%%%%%%%%%%%%%
\be
\left<\widehat{\widebar{w}}_{\rm gg}(R_i)\right> \equiv 
\frac{1}{N} \sum_{k=1}^{N}\widehat{\widebar{w}}_{\rm gg,k}(R_i),
\ee
\vspace{-0.5cm}
%%%%%%%%%%%%%%%%%%%%%%%%%%%%%%%%%%%%%%%%%%%%%%%%%%%%%%%
\ba
{\rm Cov}[\widehat{\widebar{w}}_{\rm gg}](R_i, R_j)  & & \hspace{-0.6cm} \equiv \frac{1}{N-1} \sum_{k=1}^{N} 
\left[\widehat{\widebar{w}}_{\rm gg,k}(R_i)
-\left<\widehat{\widebar{w}}_{\rm gg}(R_i)\right>\right]\nn \\
  & & \hspace{-0.6cm}\times  \left[\widehat{\widebar{w}}_{\rm gg,k}(R_j)
-\left<\widehat{\widebar{w}}_{\rm gg}(R_j)\right>\right],
\label{eq:cov_wgg_est}
\ea
%%%%%%%%%%%%%%%%%%%%%%%%%%%%%%%%%%%%%%%%%%%%%%%%%%%%%%%
where $\widehat{\widebar{w}}_{\rm gg,k}$ is the bin-averaged estimate
of the projected correlation function from the $k$th subcube. The
covariance and error on the mean are then simply obtained by further
dividing the right-hand side of Eq.~(\ref{eq:cov_wgg_est}) by the
number of subcubes $N$. The red triangles denote the predictions
obtained from direct evaluation of Eqs.~(\ref{eq:wgg_ave_Limber})
and~(\ref{eq:cov_wgg_L_BA}), where we scaled the variance to the
entire MXXL volume.

The theoretical predictions for the brightest galaxies agree extremely
well with the measured errors. This owes to the fact that on large
scales, $R>10\Mpc$, the errors are determined by the Gaussian part of
the variance. On smaller scales these objects are relatively sparse,
and so the shot-noise contribution to the variance quickly
dominates. Both of these two limits are well characterized by our
formula.

For the fainter magnitude bins there is reasonably good agreement
between our model and the data on both large ($R>10\,\Mpc$) and small
($R<0.1\,\Mpc$) scales. The former is due to the prominence of
Gaussian contributions on large scales, whereas the latter is due to
the shot noise dominance on small scales. However, on intermediate
scales ($0.1\,\Mpc < R <1\,\Mpc$), the agreement is not so good, and
we see that the data have errors roughly a factor of $\sim$2 larger
than the predictions. This is to be expected, since for these scales
the non-Gaussian corrections (e.g. the connected part of the
trispectrum and also the bispectrum), which we neglected in deriving
\Eqn{eq:cov_wgg_L_BA} are significant and have the effect of
increasing the errors.

To compare the off-diagonal elements of the predicted and measured
covariance from Eqs.~(\ref{eq:cov_wgg_L_BA})
and~(\ref{eq:cov_wgg_est}), we choose to examine the correlation
matrix. For any covariance matrix $\textbfss{C}[X]$, the correlation
matrix $\textbfss{r}[X]$, can be defined as:
%%%%%%%%%%%%%%%%%%%%%%%%%%%%%%%%%%%%%%%%%%%%%%%%%%%%%%%
\be
r[X]_{i j}=\frac{C[X]_{i j}}{\sqrt{C[X]_{i i}\,C[X]_{j j}}}\ ,
\label{eq:def_R}
\ee
%%%%%%%%%%%%%%%%%%%%%%%%%%%%%%%%%%%%%%%%%%%%%%%%%%%%%%%
where the subscript ${\rm X}$ is a place-holder for the statistic.
The correlation matrix obeys the constraint $\left|r[X]_{i j}
  \right|\le 1,\,\forall i,j$.

Figure~\ref{fig:Row_GG} shows four rows of the correlation matrix of
$\textbfss{r}[\widehat{\widebar{w}}_{\rm gg}]$ at radii $(R_i, R_j)$
as a function of $R_j$ and at fixed radius $R_i$. The fixed scales
correspond roughly to $R_i = 0.04,\,0.1,\,1,\,10\,\Mpc$, and in each
panel can be quickly determined by noting that
$r[\widehat{\widebar{w}}_{\rm gg}](R_i=R_j)=1$. From left to right,
each column represents a magnitude bin.  

\begin{figure*}
 \centering 
\begin{tabular}{cc}
  \includegraphics[scale=0.52]{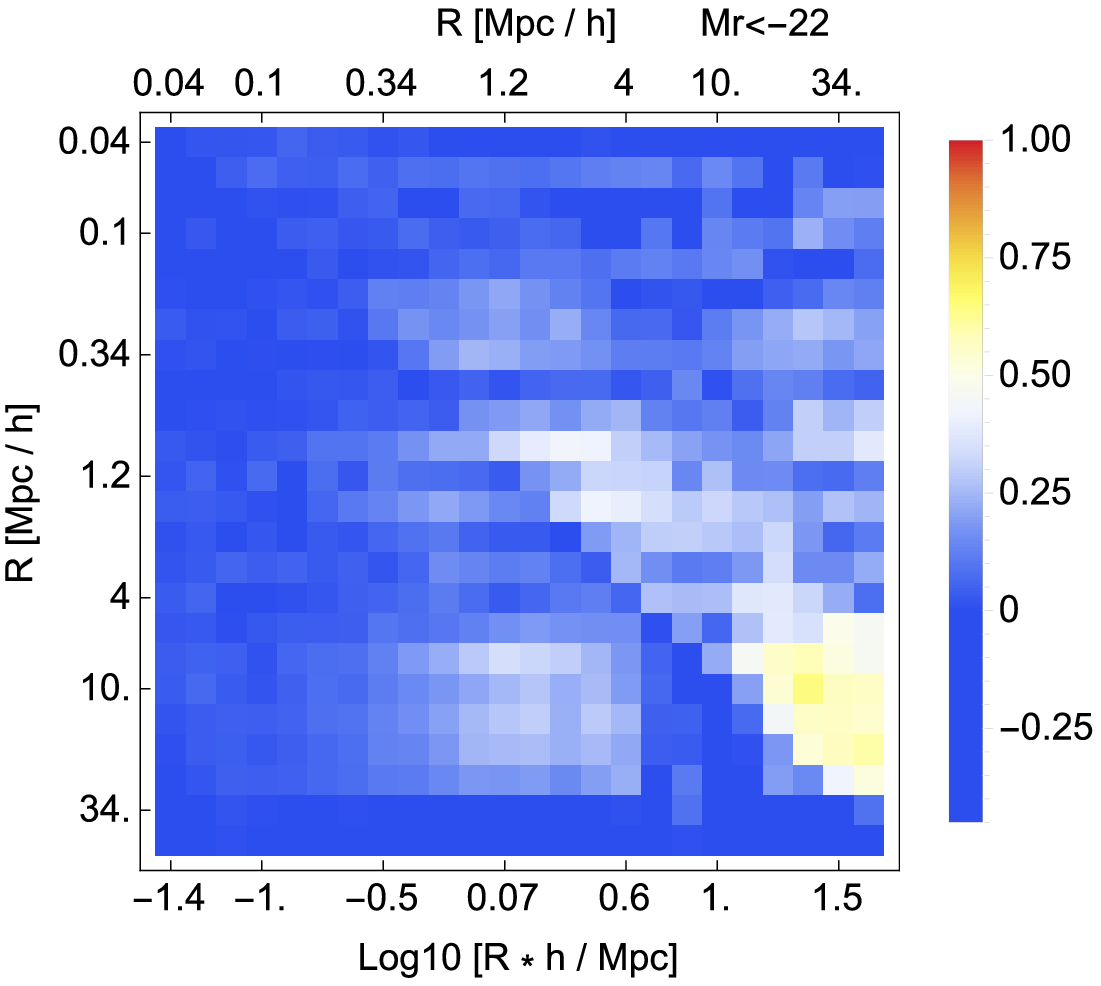}
  \includegraphics[scale=0.52]{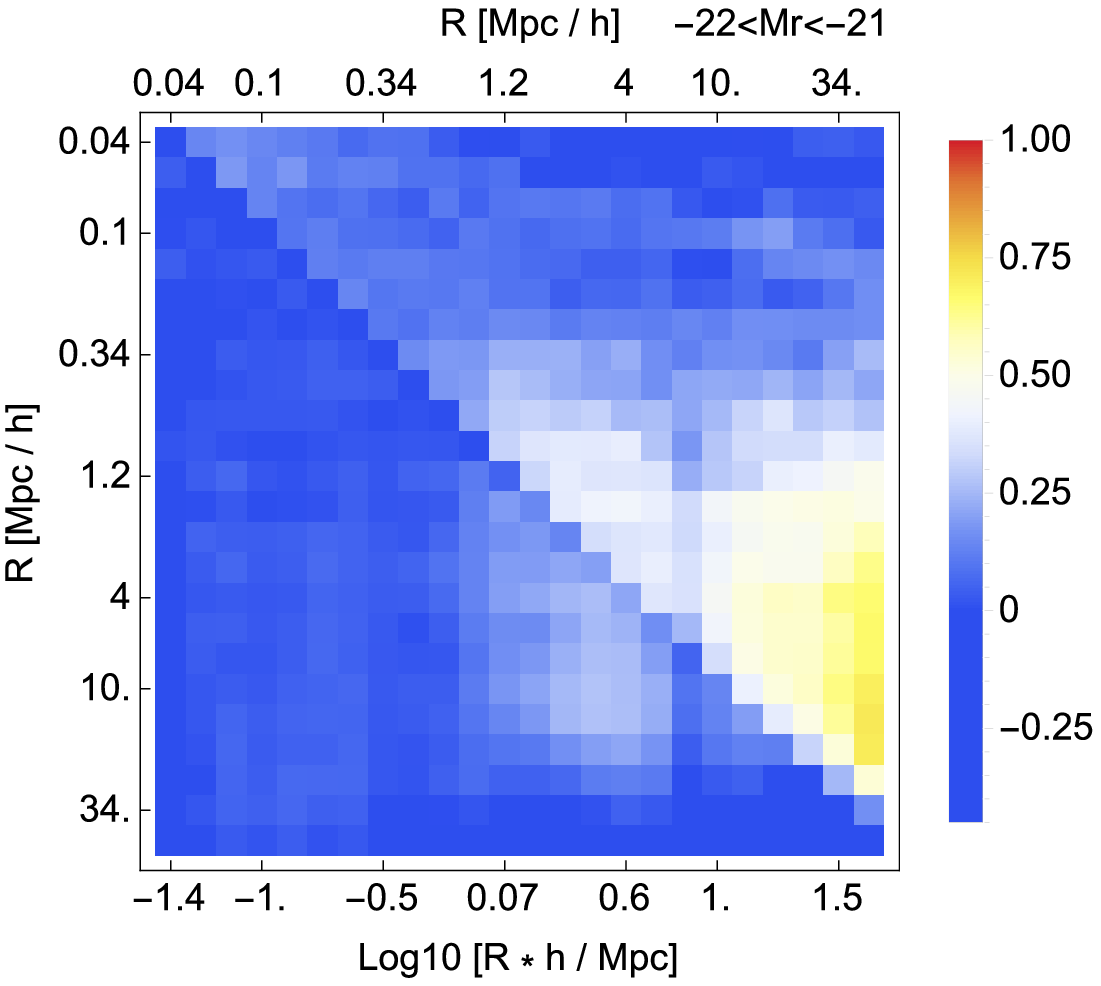}
\end{tabular}
%%%%%%%%%%%%%%%%%%%%%%%%%%%%%%%%%%%%%%%%%%%%%%%%%%%%%%%
\begin{tabular}{cc}
\includegraphics[scale=0.52]{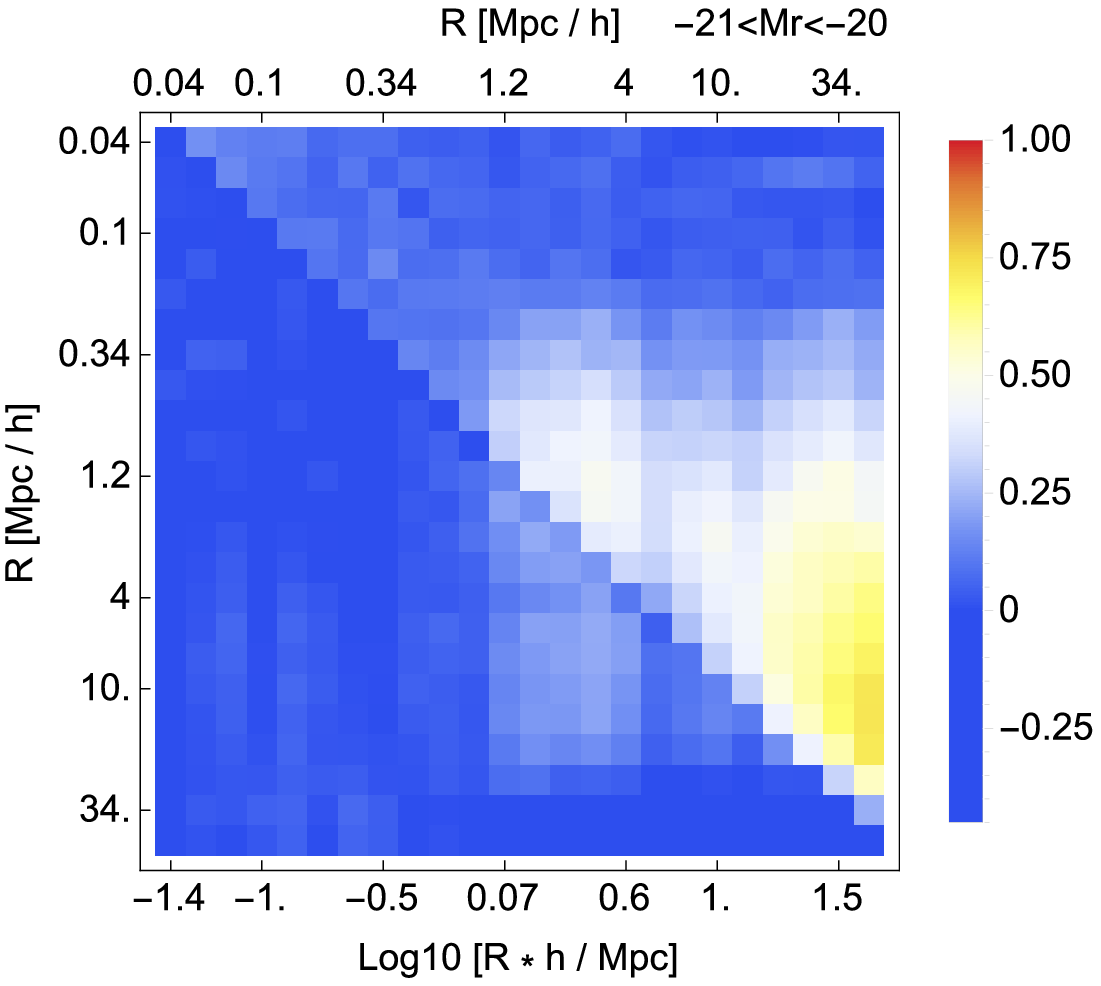}
\includegraphics[scale=0.52]{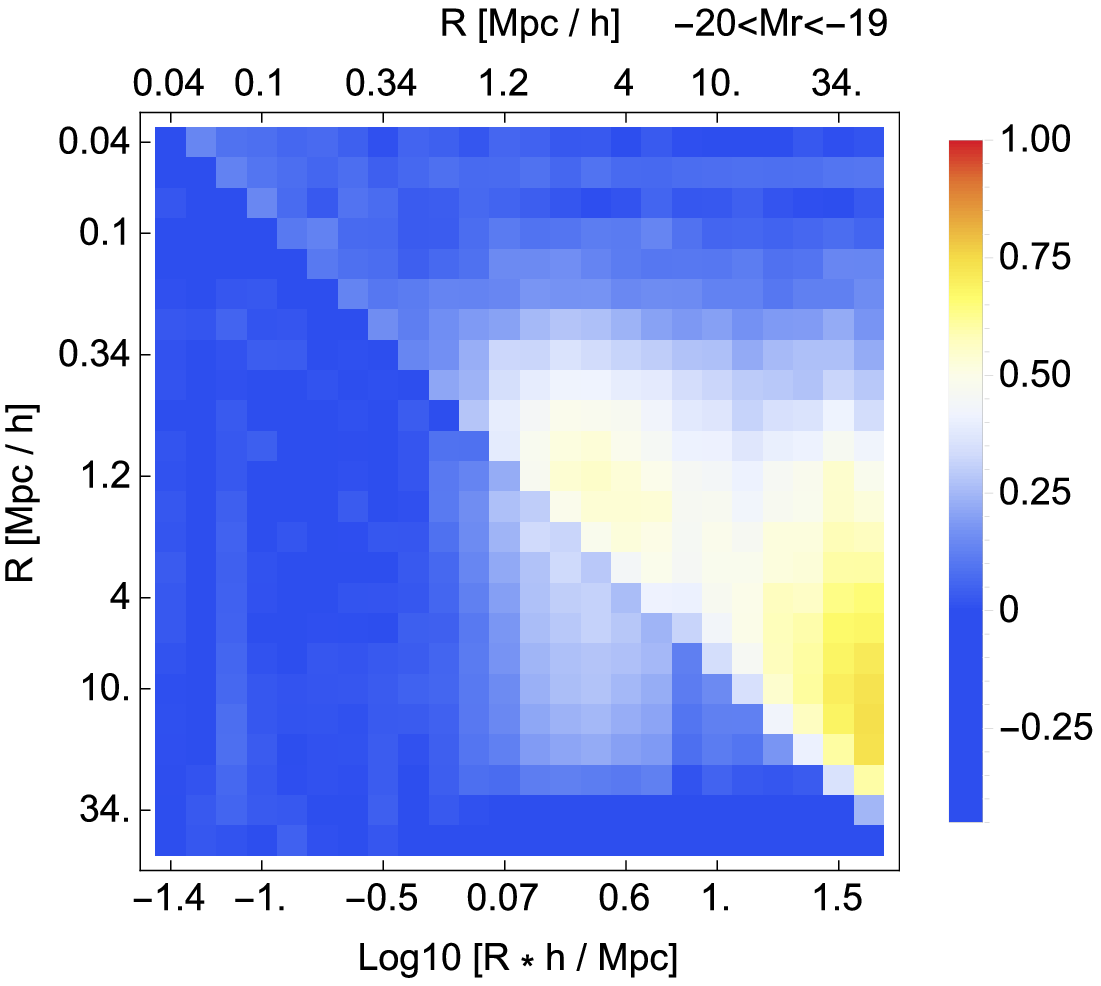} 
\end{tabular}
\caption{The jackknife cross-correlation matrix of $w_{\rm gg}$ and
  $\Delta\Sigma$, estimated from 64 samplings of one of the MXXL
  subcubes.}
\label{fig:GC_GGL_JK}
\end{figure*}

There are some notable trends: when the fixed scale $R_i$ is large
(bottom row of the figure), the neighbouring bins of the matrix are
significantly correlated, and the strength of the correlations is
$r[\widehat{\widebar{w}}_{\rm gg}]>0.5$. There is a decrease in the
correlation coefficient as one considers brighter magnitude
bins. These findings are in good agreement with our Gaussian
model. When the fixed scale is small $(R_i=0.04)$, for the brighter
galaxy bins there is virtually no evidence for bin-to-bin
correlations. Again this is consistent with the predictions of our
model, and is attributed to the fact that for so few objects the
shot-noise errors simply dominate. However, when we consider the
intermediate scales (second and third rows of panels in the figure),
the model predictions underestimate the bin-to-bin correlations that
are exhibited by the data. We interpret this as a sign that the
non-Gaussian contributions to the covariance matrix, which we have
neglected in our model, are significant. For a more intuitive but less
quantitative depiction of the GC correlation matrix, as well as the
GGL one and their cross-correlation, we refer the reader to
Figures~\ref{fig:GC_full_JK}-\ref{fig:GC_GGL_JK}.
%%%%%%%%%%%%%%%%%%%%%%%%%%%%%%%%%%%%%%%%%%%%%%%%%%%%%%%
\begin{figure*}
\vspace{0.4cm}
\centering {
\includegraphics[scale=0.56]{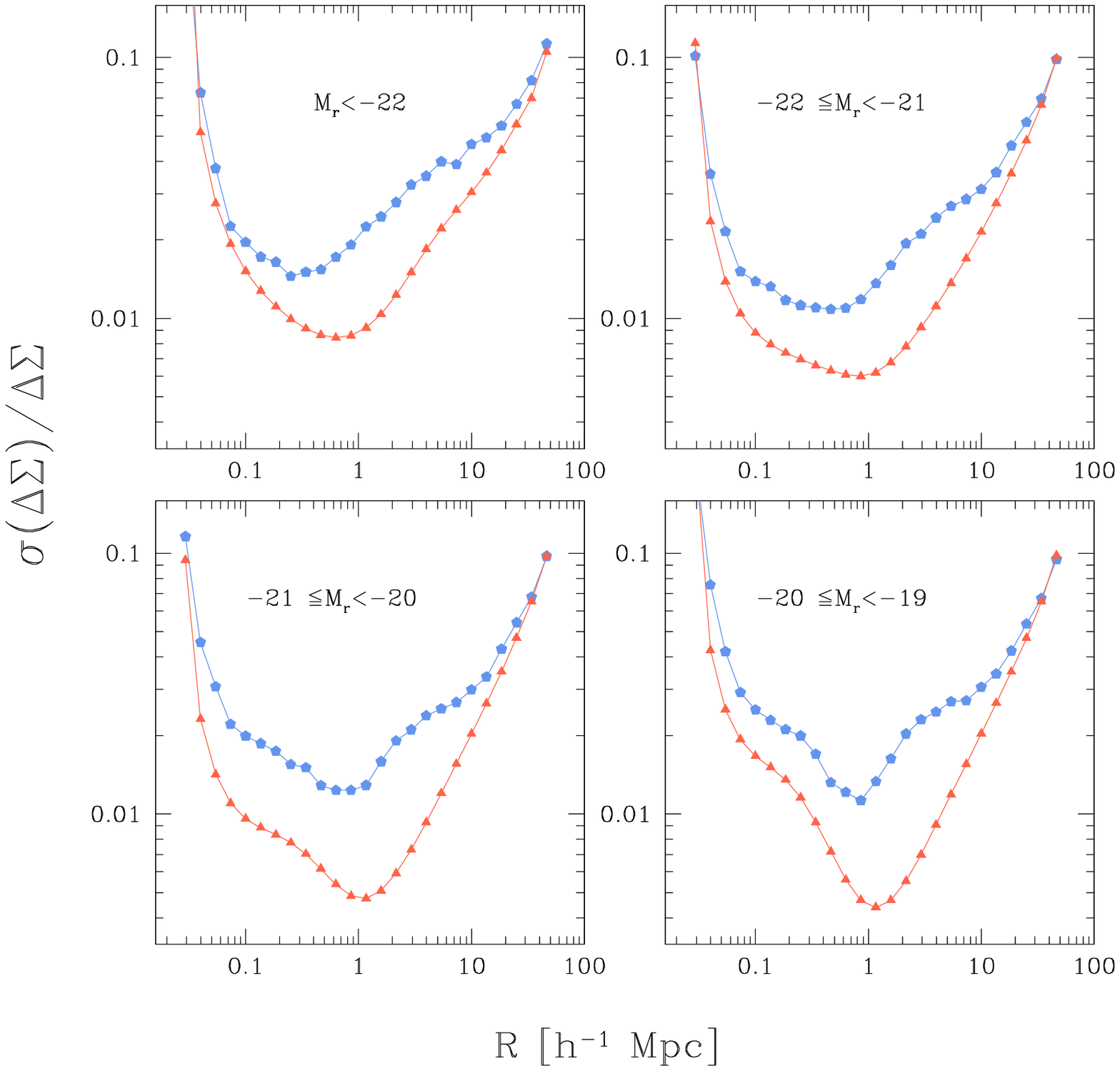}
\caption{The noise-to-signal for the stacked tangential shear. The red
  triangles represent the theory predictions, while the blue pentagons
  correspond to the simulation measurements.}
\label{fig:NS_GGL}}
%%%%%%%%%%%%%%%%%%%%%%%%%%%%%%%%%%%%%%%%%%%%%%%%%%%%%%%
\vspace{0.5cm}
%%%%%%%%%%%%%%%%%%%%%%%%%%%%%%%%%%%%%%%%%%%%%%%%%%%%%%%
\centering {
\includegraphics[scale=0.65]{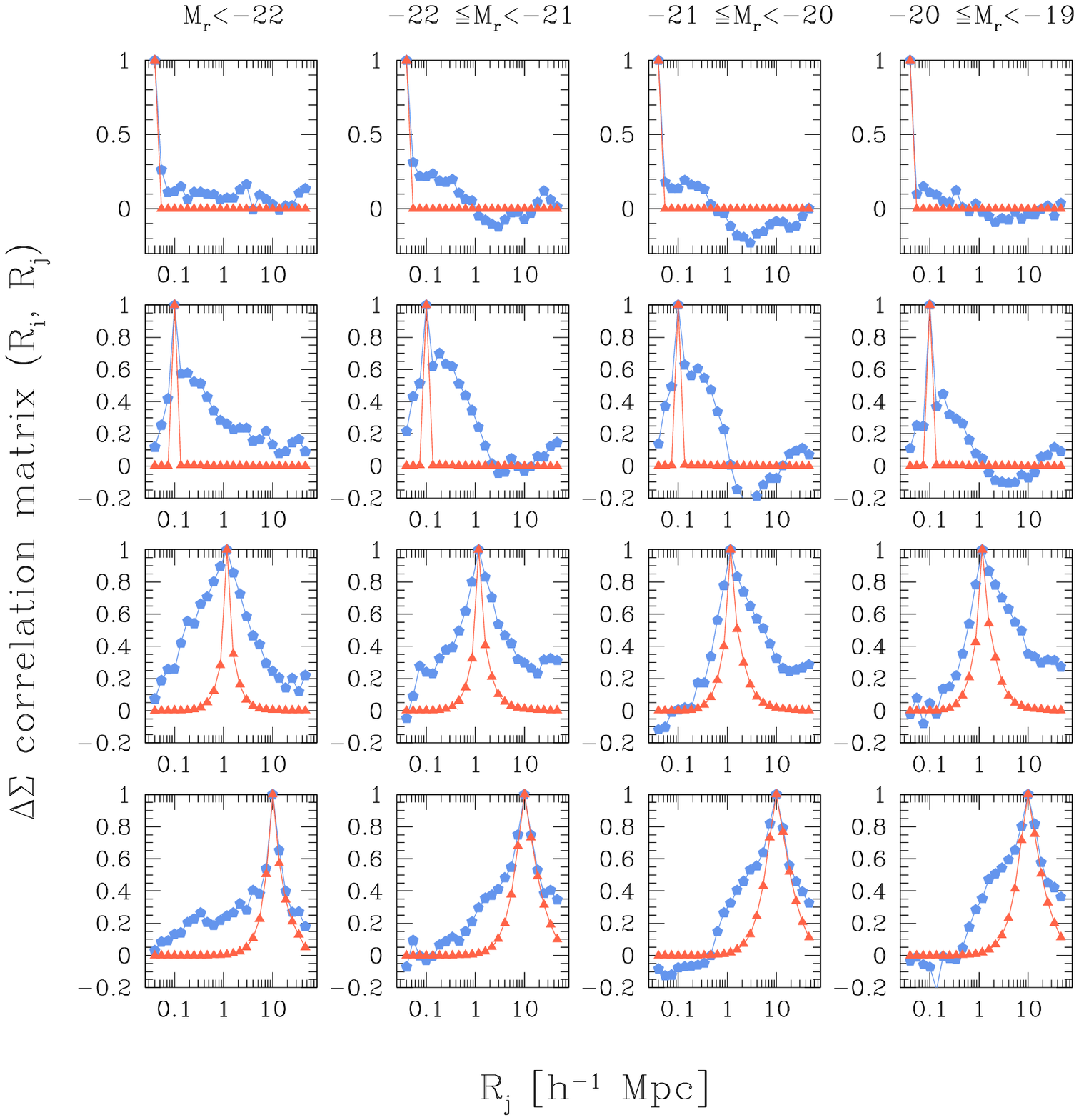}
\caption{The correlation matrix from Eq.~(\ref{eq:def_R}) for the
  stacked tangential shear. Just like in Figure~\ref{fig:Row_GG}, the
  rows present four scales: $R_i=0.04, 0.1, 1, 10 \,\Mpc$. From left
  to right, the columns depict results for increasingly fainter
  galaxies, as indicated in each panel. }
\label{fig:Row_GGL}}
\end{figure*}
%%%%%%%%%%%%%%%%%%%%%%%%%%%%%%%%%%%%%%%%%%%%%%%%%%%%%%%
%%%%%%%%%%%%%%%%%%%%%%%%%%%%%%%%%%%%%%%%%%%%%%%%%%%%%%%
\subsection{Covariance of the stacked tangential shear}
\label{SVI.2}
%%%%%%%%%%%%%%%%%%%%%%%%%%%%%%%%%%%%%%%%%%%%%%%%%%%%%%%
%%%%%%%%%%%%%%%%%%%%%%%%%%%%%%%%%%%%%%%%%%%%%%%%%%%%%%%
Figure~\ref{fig:NS_GGL} presents the errors on the mean of the
stacked tangential shear as a function of the transverse scale
$R$. The blue pentagons in the figure represent the noise-to-signal
ratio estimated directly from the 216 MXXL-subcubes estimated through:
%%%%%%%%%%%%%%%%%%%%%%%%%%%%%%%%%%%%%%%%%%%%%%%%%%%%%%%
\be
\langle\widehat{\widebar{\gamma}}_t^g(R_i)\rangle \equiv 
\frac{1}{\Sigma_{\rm crit}}\frac{1}{N} \sum_{k=1}^{N}
\widehat{\Delta\Sigma}_{k}(R_i), 
\ee
\vspace{-0.5cm}
%%%%%%%%%%%%%%%%%%%%%%%%%%%%%%%%%%%%%%%%%%%%%%%%%%%%%%%
\ba 
{\rm Cov}[\widehat{\widebar{\gamma}}_{t}^g](R_i, R_j)
 & & \hspace{-0.6cm}\equiv \frac{1}{\Sigma_{\rm crit}^2}\frac{1}{N-1}
\hspace{-0.1cm} \sum_{k=1}^{N}
\left[\widehat{\Delta\Sigma}_{k}(R_i)-\langle \widehat{\Delta\Sigma}_{k}
(R_i)\rangle\right] \nn \\
 & & \hspace{-0.6cm}\times \left[\widehat{\Delta\Sigma}_{k}(R_j)-\langle 
\widehat{\Delta\Sigma}_{k}(R_j)\rangle\right],
\label{eq:cov_shear_est}
\ea
%%%%%%%%%%%%%%%%%%%%%%%%%%%%%%%%%%%%%%%%%%%%%%%%%%%%%%%
where $\widehat{\Delta\Sigma}_{k}(R_i)$ is the estimate of the excess
surface density from the $k$th subcube.

\noindent Let us now turn to the theoretical predictions. We note that
\Eqn{eq:cov_shear4} can be rewritten for $\Delta\Sigma$ at the lens
redshift $z_l$ and using the Limber approximation as done in section
\S\ref{SIII.2} and Eq.~(\ref{eq:shear_est_L}). In the following, we
shall ignore the shape-noise contribution from \Eqn{eq:cov_shear4},
since the latter is not an issue for our measurements. What is an
issue however, given that we determine the matter distribution from an
$N$-body simulation of finite resolution, is the particle shot noise
which accompanies $\mathcal{P}_{\rm mm}$. We therefore write the final
expression for the measured bin-averaged covariance of the excess
surface density as:
%%%%%%%%%%%%%%%%%%%%%%%%%%%%%%%%%%%%%%%%%%%%%%%%%%%%%%%
\ba
{\rm Cov}[\widehat{\widebar{\Delta \Sigma}}](R_i, R_j) & & \hspace{-0.5cm} 
= \frac{1}{A_s}(\rho_m^0)^2
\int \frac{d^2k_\perp}{(2\pi)^2}\widebar{J}_2(k_\perp R_i) \widebar{J}_2(k_\perp R_j)\nn \\
& & \hspace{-3.cm}\times \left\{ \left[\mathcal{P}_{\rm mm}(k_\perp)
+\frac{1}{\bar{n}_p}\right] \,\left[\mathcal{P}_{\rm gg}(k_\perp)+
\frac{1}{\bar{n}_g}\right] + \mathcal{P}_{\rm gm}^2(k_\perp)\right\},
\label{eq:cov_delta_sigma}
\ea
%%%%%%%%%%%%%%%%%%%%%%%%%%%%%%%%%%%%%%%%%%%%%%%%%%%%%%%
where all the power spectra are taken at redshift $z_l$; the
transverse area of the survey $A_s$ was introduced in \S\ref{SIII.2};
$\bar{n}_p$ and $\bar{n}_g$ are the mean particle and galaxy densities
in the simulation. The above covariance has dimension of $L^{-4}$, as
expected.

The theory predictions of \Eqn{eq:cov_delta_sigma} are presented as
the red triangles in Figure~\ref{fig:NS_GGL} . As in the case of
$w_{\rm gg}$, we see that the agreement between theory and
measurements is good on large ($R>10\,\Mpc$) and small ($R<0.1\,\Mpc$)
scales. However, on intermediate scales the theory underestimates the
true errors. We also notice the predictions are increasingly poor for
the fainter galaxies. For the brightest galaxies the discrepancy at
$R\sim 1 \Mpc$ is roughly a factor of $\sim 2$, whereas for the
faintest galaxies it is roughly $\sim 3$. Thus, compared to $w_{\rm
  gg}$, the predictions for the stacked shear seem worse, even for the
shot-noise-dominated brightest galaxies. This suggests that the
non-Gaussian contributions to the variance are more important for the
stacked tangential shear than for the projected correlation
function. However, in a real shear survey this discrepancy may not be
so crucial, since the addition of the shape noise term will certainly
be a strong source of noise on small-scales. In Appendix~\S\ref{App3} we
present a theoretical estimation of the impact of shape noise on the
GGL variance and correlation matrix.
%%%%%%%%%%%%%%%%%%%%%%%%%%%%%%%%%%%%%%%%%%%%%%%%%%%%%%%
\begin{figure*}
\centering {
\includegraphics[scale=0.6]{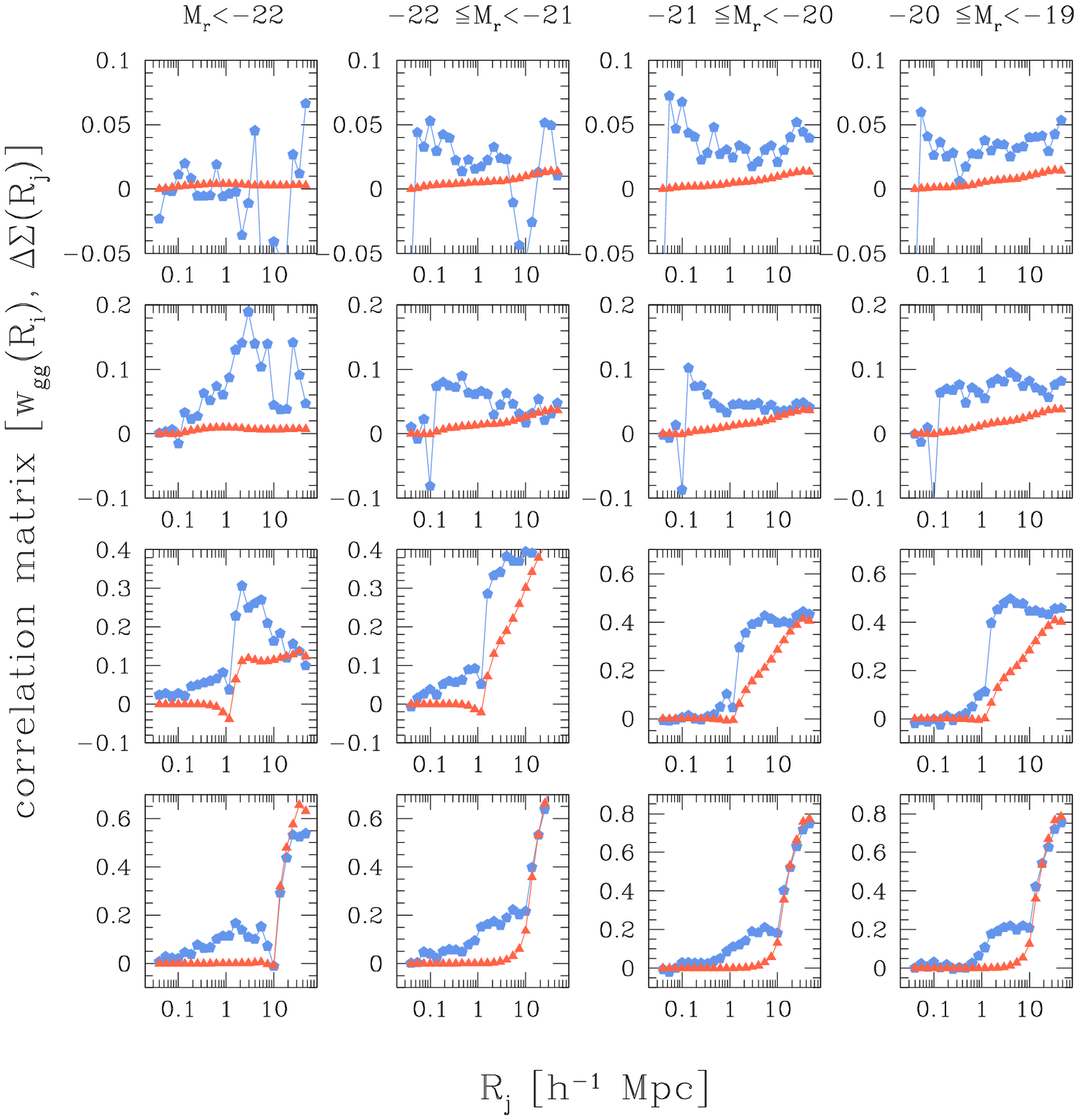}
\caption{The cross-correlation matrix of the projected galaxy
  correlation function and the stacked tangential shear. As in the
  previous Figures~\ref{fig:Row_GG} and~\ref{fig:Row_GGL}, the rows
  present four scales: $R_i=0.04, 0.1, 1, 10 \,\Mpc$. From left to
  right, the columns depict the galaxy bins with decreasing
  brightness. }
\label{fig:Row_GG_GGL_fixedGG}}
\vspace{0.5cm}
\centering {
\includegraphics[scale=0.6]{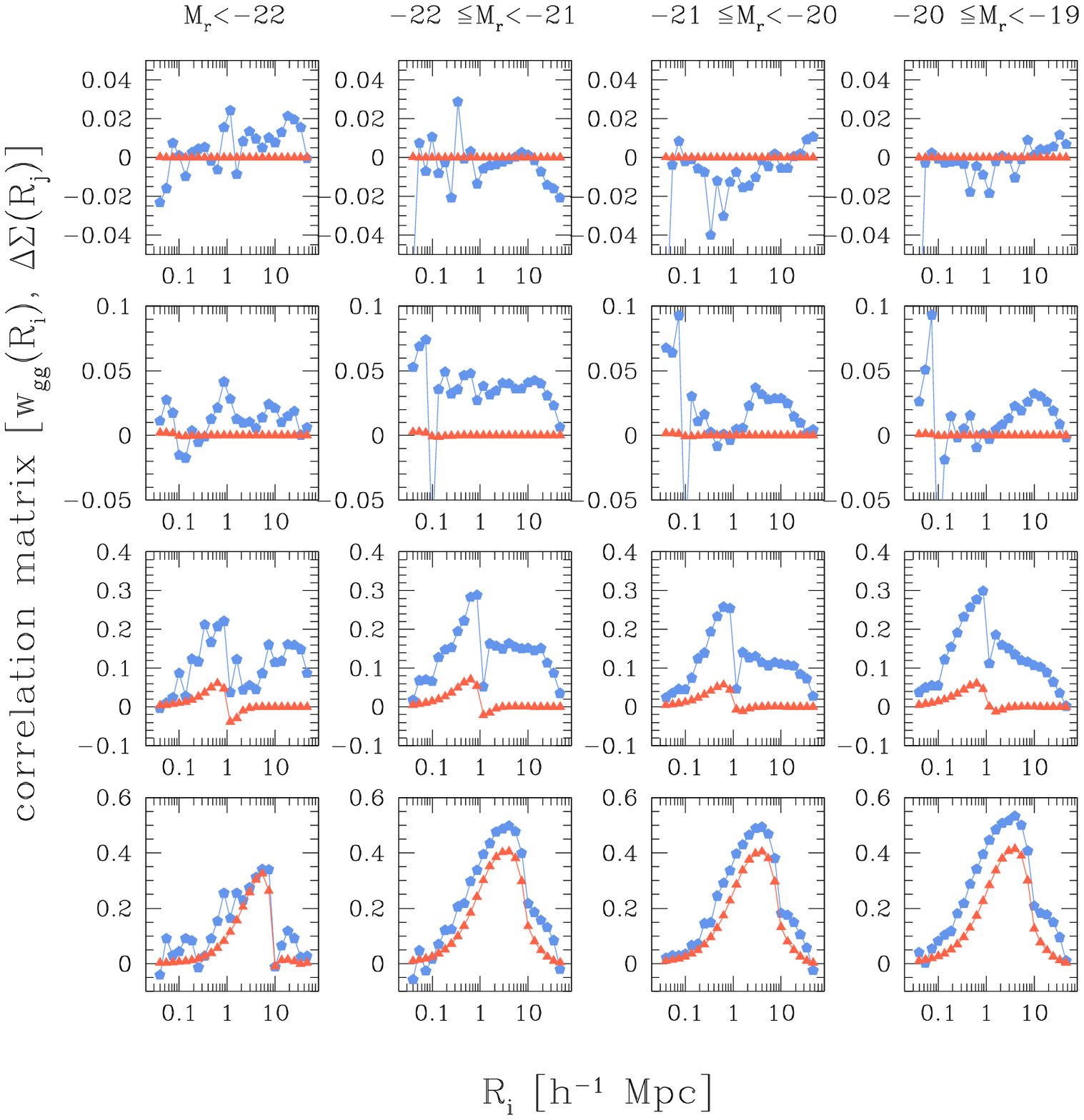}
\caption{The cross-correlation matrix of the projected galaxy
  correlation function and the stacked tangential shear, this time
  fixing $R_j=0.04, 0.1, 1, 10 \,\Mpc$. }
\label{fig:Row_GG_GGL_fixedGGL}}
\end{figure*}
%%%%%%%%%%%%%%%%%%%%%%%%%%%%%%%%%%%%%%%%%%%%%%%%%%%%%%%

Figure~\ref{fig:Row_GGL} presents four rows of the correlation matrix
of $\widehat{\widebar{\gamma}}^g_t$ at $\{R_i, R_j\}$ as a function of
$R_j$ and for fixed $R_i$, with the covariance estimated according to
Eq.~(\ref{eq:cov_shear_est}). The measurements are represented in the
figure as the blue pentagons.  The theoretical predictions are
obtained by evaluating \Eqn{eq:cov_delta_sigma}, and are denoted by
the red triangles. The general trends are similar to those found in
Figure~\ref{fig:Row_GG}: on small scales ($R_i=0.04\,\Mpc$, i.e. the
top row of the figure), the theory and measurements show weak
bin-to-bin correlations and the measurements seem somewhat noisy.
Qualitatively they are consistent with uncorrelated noise. On large
scales ($R_i=10.0\,\Mpc$, i.e. the bottom row of the figure), the
measurements show significant bin-to-bin correlations. However, the
correlations appear to be somewhat weaker than was found for the
projected correlation function for more distant bins. In addition, the
Gaussian predictions of our model do not describe these correlations
as well as in the case of $w_{\rm gg}$. On intermediate scales,
$1<R_i<10 [\Mpc]$, the correlations are significantly stronger for all
magnitude bins than predicted by our theoretical model. Once again,
this suggests that the tangential shear signal is significantly more
non-Gaussian on these scales than the projected galaxy correlation
function.
%%%%%%%%%%%%%%%%%%%%%%%%%%%%%%%%%%%%%%%%%%%%%%%%%%%%%%%
%%%%%%%%%%%%%%%%%%%%%%%%%%%%%%%%%%%%%%%%%%%%%%%%%%%%%%%
\subsection{Cross-covariance of the projected correlation 
function and stacked tangential shear}
\label{SVI.3}
%%%%%%%%%%%%%%%%%%%%%%%%%%%%%%%%%%%%%%%%%%%%%%%%%%%%%%%
%%%%%%%%%%%%%%%%%%%%%%%%%%%%%%%%%%%%%%%%%%%%%%%%%%%%%%%
\begin{figure*}
\centering 
\begin{tabular}{cc}
  \includegraphics[scale=0.52]{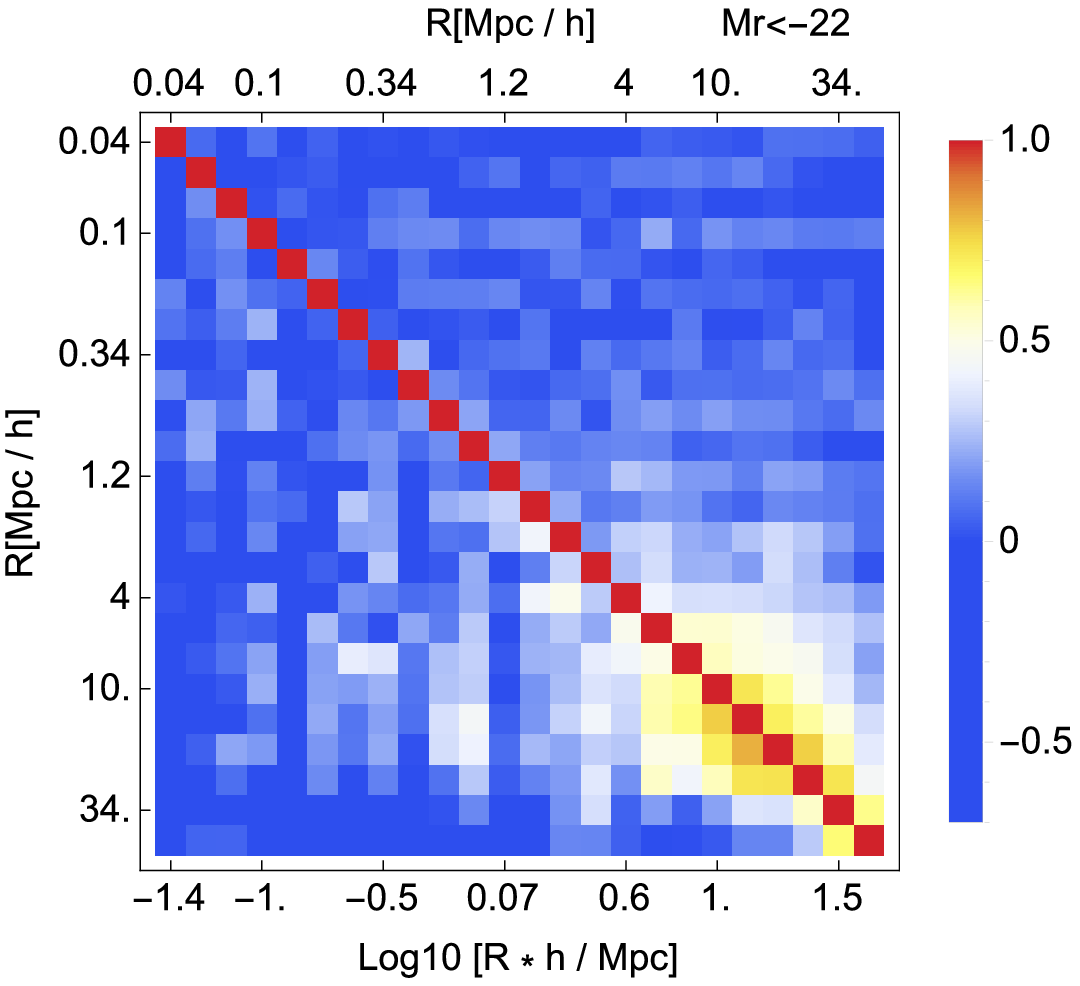}
  \includegraphics[scale=0.52]{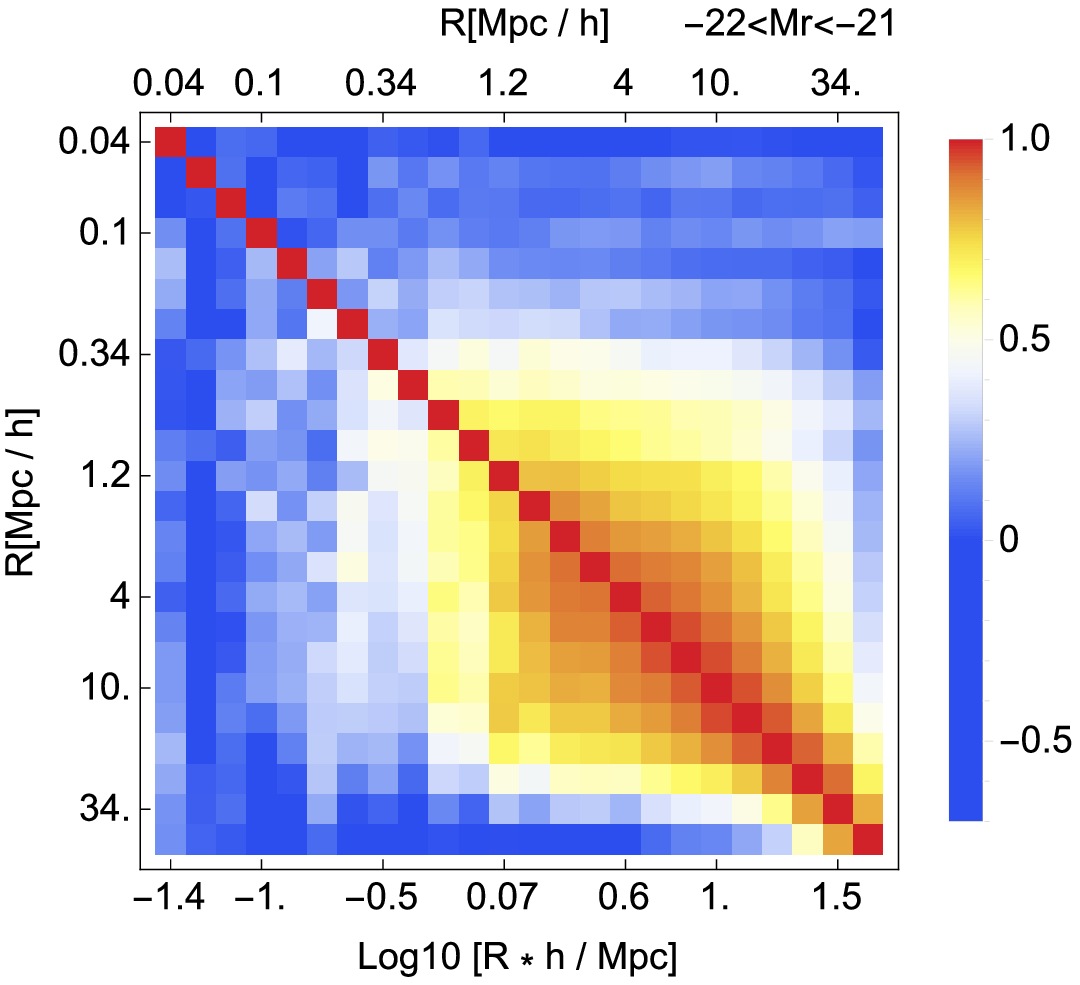}
\end{tabular}
%%%%%%%%%%%%%%%%%%%%%%%%%%%%%%%%%%%%%%%%%%%%%%%%%%%%%%%
\begin{tabular}{cc}
\includegraphics[scale=0.52]{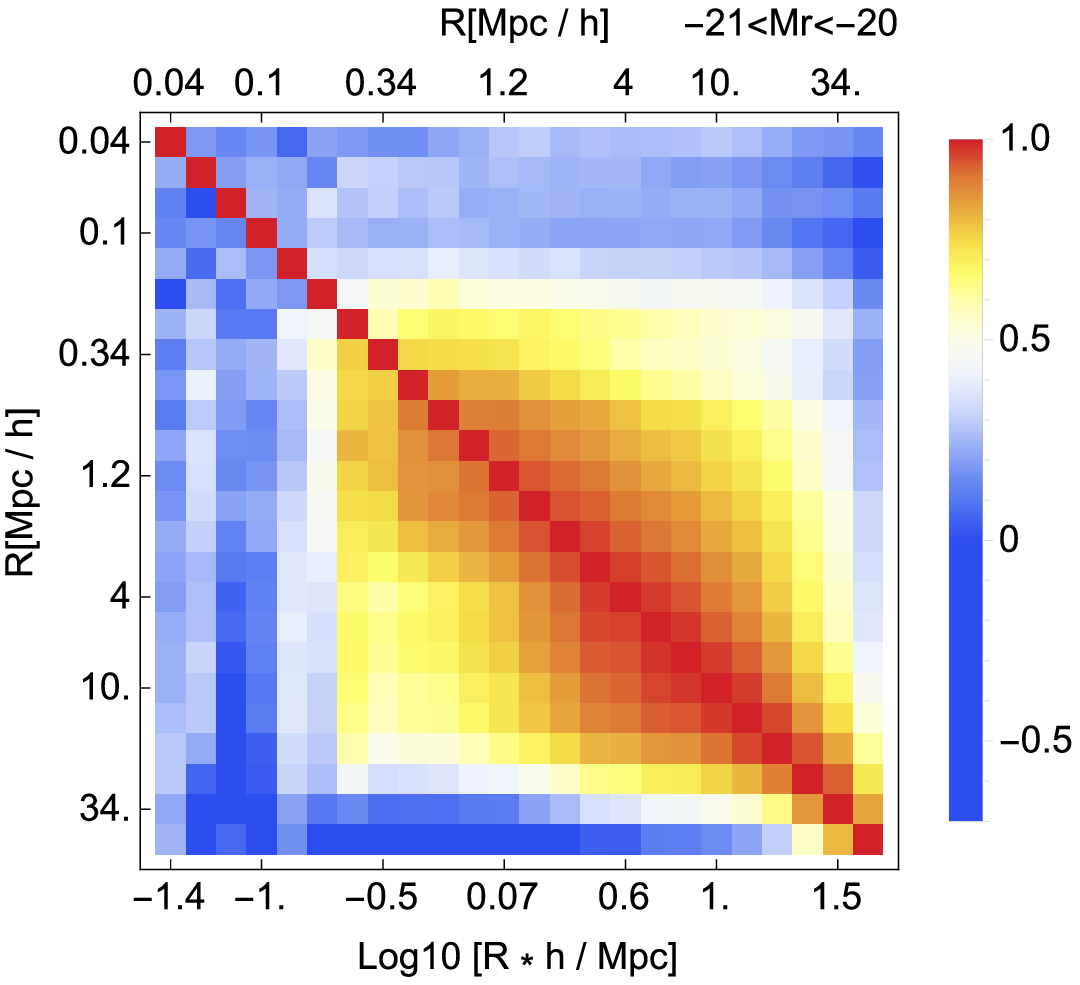}
\includegraphics[scale=0.52]{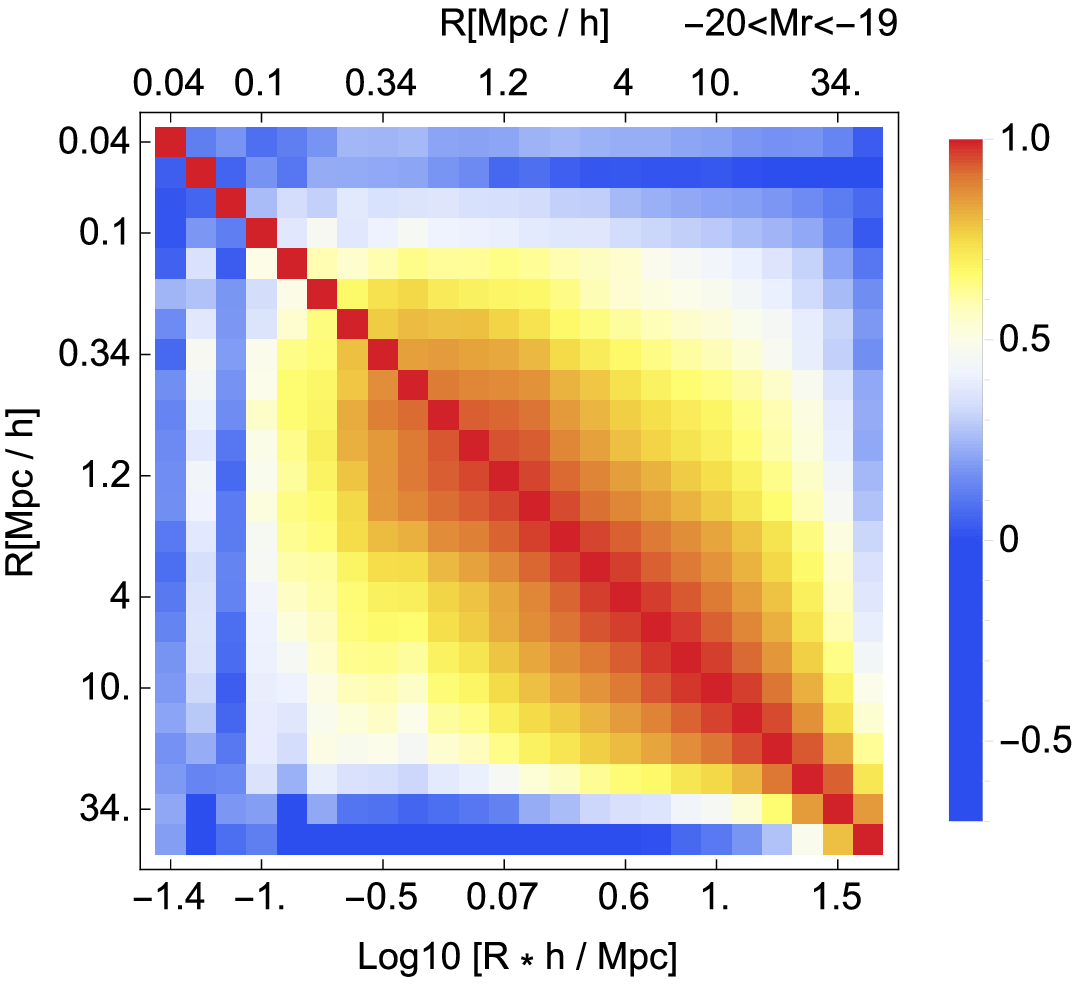}
\end{tabular}
\caption{The correlation matrix of $w_{\rm gg}$ for the four magnitude
  bins considered. The upper right triangle of the matrix represents
  the correlation matrix measured from the whole simulation, i.e. the
  same as in Figure~\ref{fig:Row_GG}. The lower left triangle is the
  correlation matrix obtained from jackknife samplings of one of the
  subcubes of the simulation.}
\label{fig:GC_full_JK}
\end{figure*}
%%%%%%%%%%%%%%%%%%%%%%%%%%%%%%%%%%%%%%%%%%%%%%%%%%%%%%%
%%%%%%%%%%%%%%%%%%%%%%%%%%%%%%%%%%%%%%%%%%%%%%%%%%%%%%%
\begin{figure*}
\centering 
\begin{tabular}{cc}
  \includegraphics[scale=0.52]{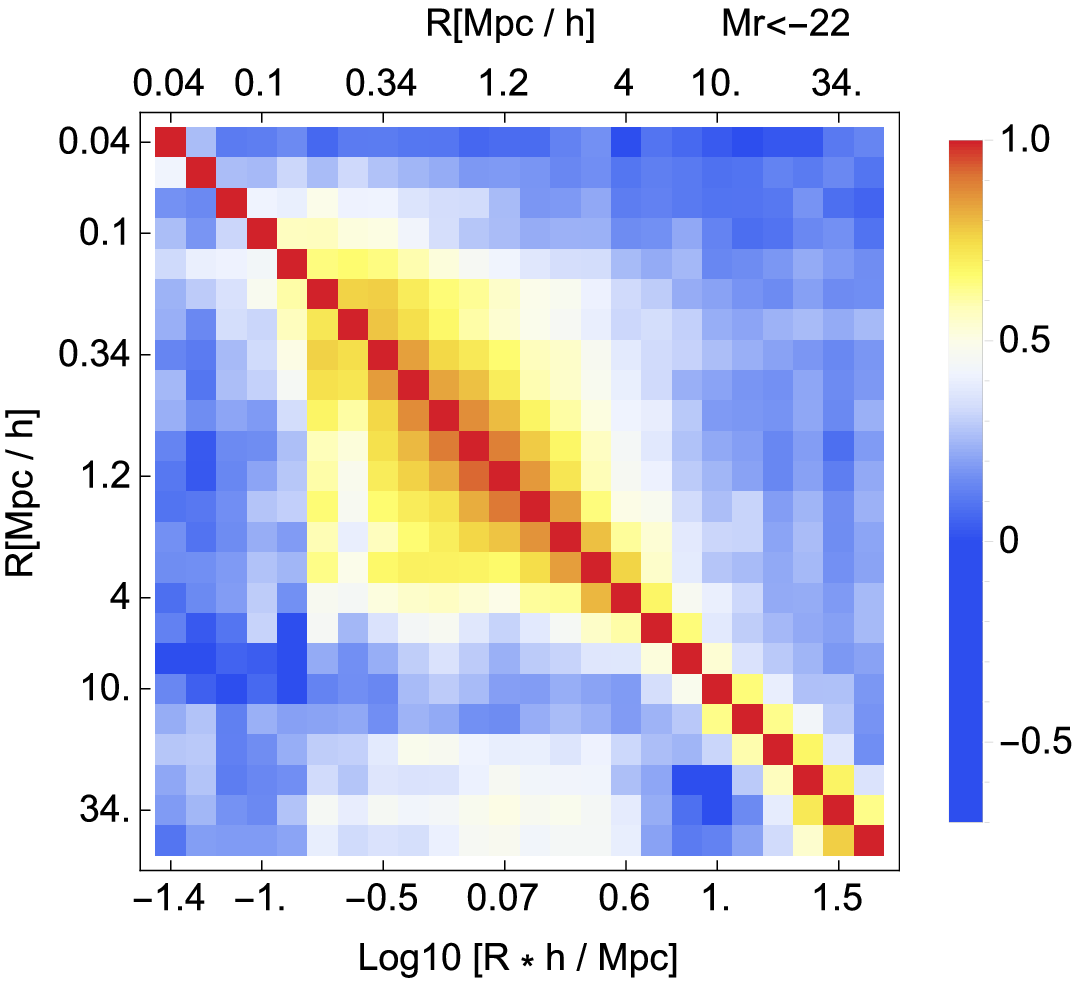}
  \includegraphics[scale=0.52]{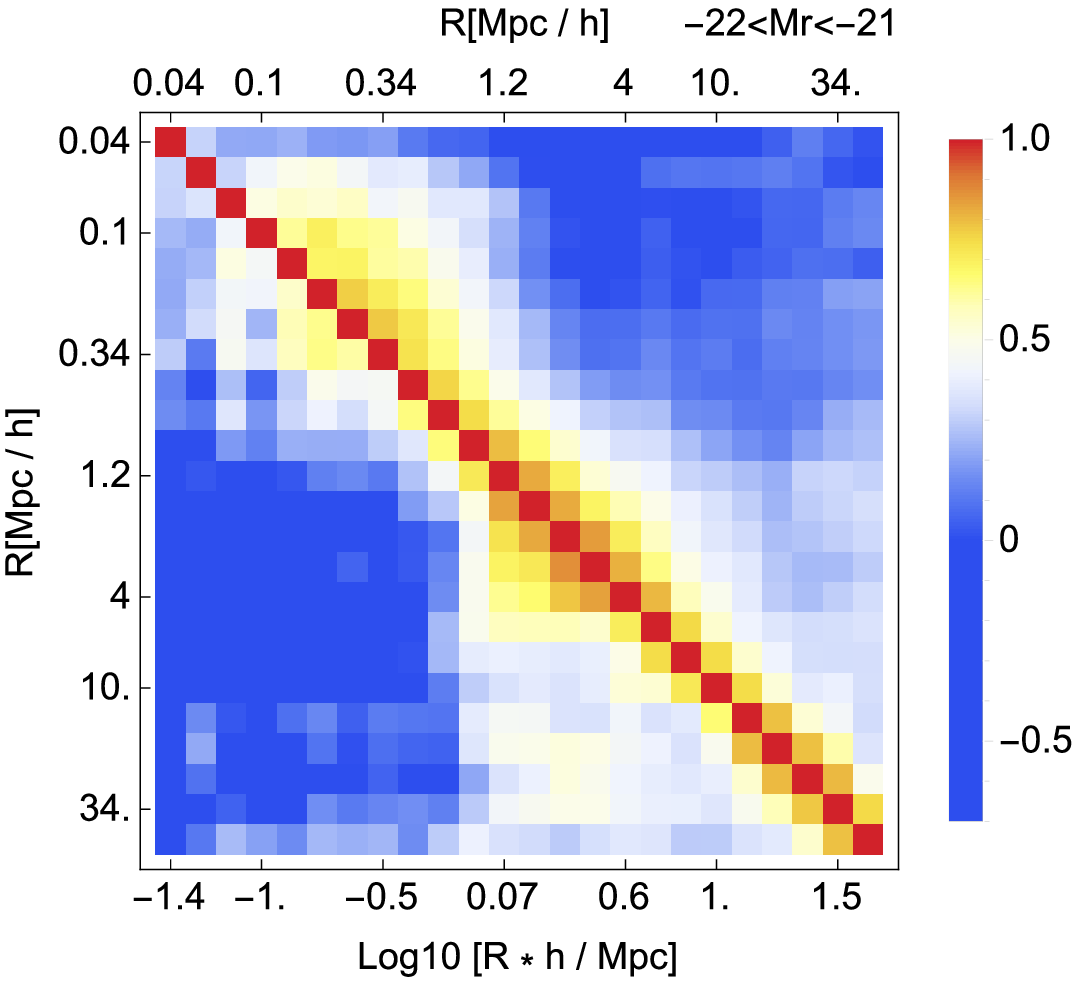}
\end{tabular}
%%%%%%%%%%%%%%%%%%%%%%%%%%%%%%%%%%%%%%%%%%%%%%%%%%%%%%%
\begin{tabular}{cc}
\includegraphics[scale=0.52]{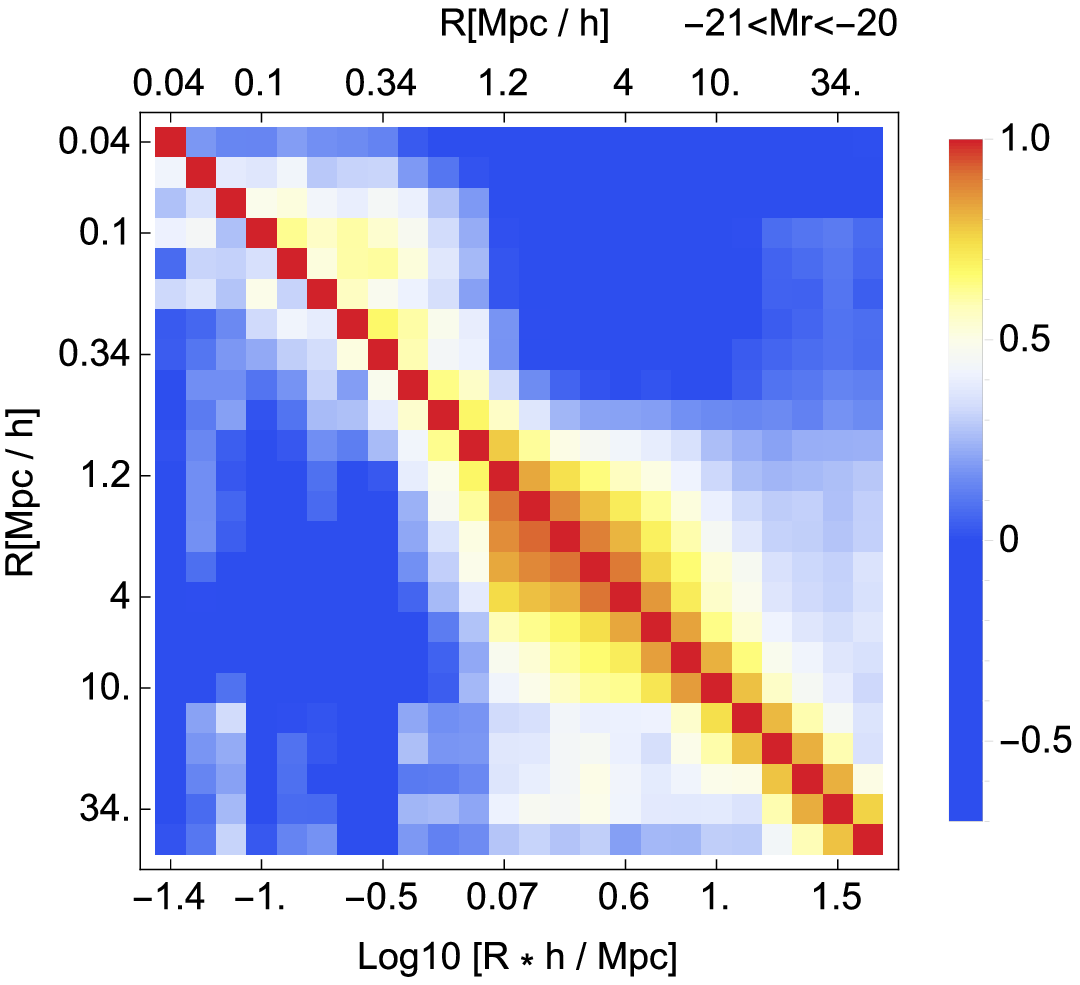}
\includegraphics[scale=0.52]{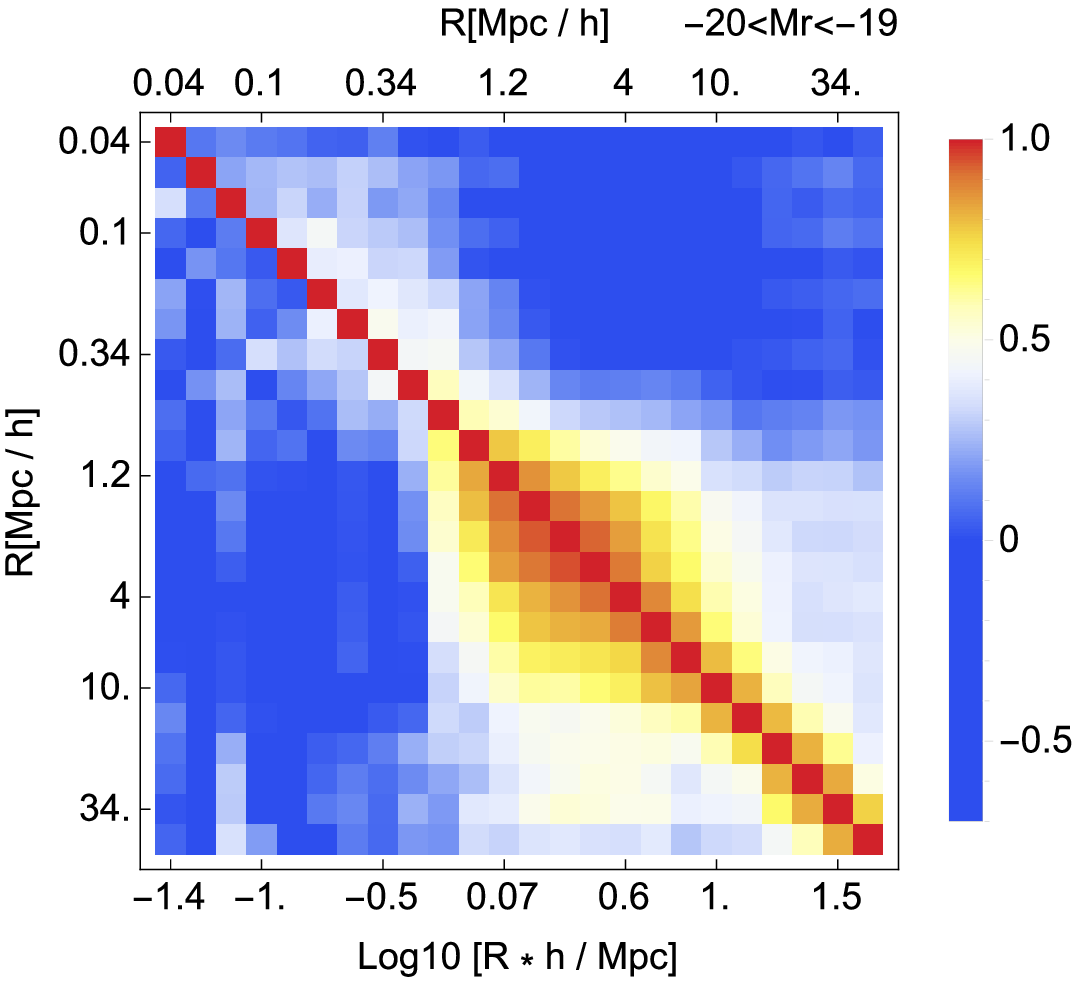}
\end{tabular}
\caption{The same as in the figure above, only for the case of $\Delta\Sigma$. }
\label{fig:GGL_full_JK}
\end{figure*}
%%%%%%%%%%%%%%%%%%%%%%%%%%%%%%%%%%%%%%%%%%%%%%%%%%%%%%%
%%%%%%%%%%%%%%%%%%%%%%%%%%%%%%%%%%%%%%%%%%%%%%%%%%%%%%%
\begin{figure*}
\centering 
\begin{tabular}{cc}
  \includegraphics[scale=0.52]{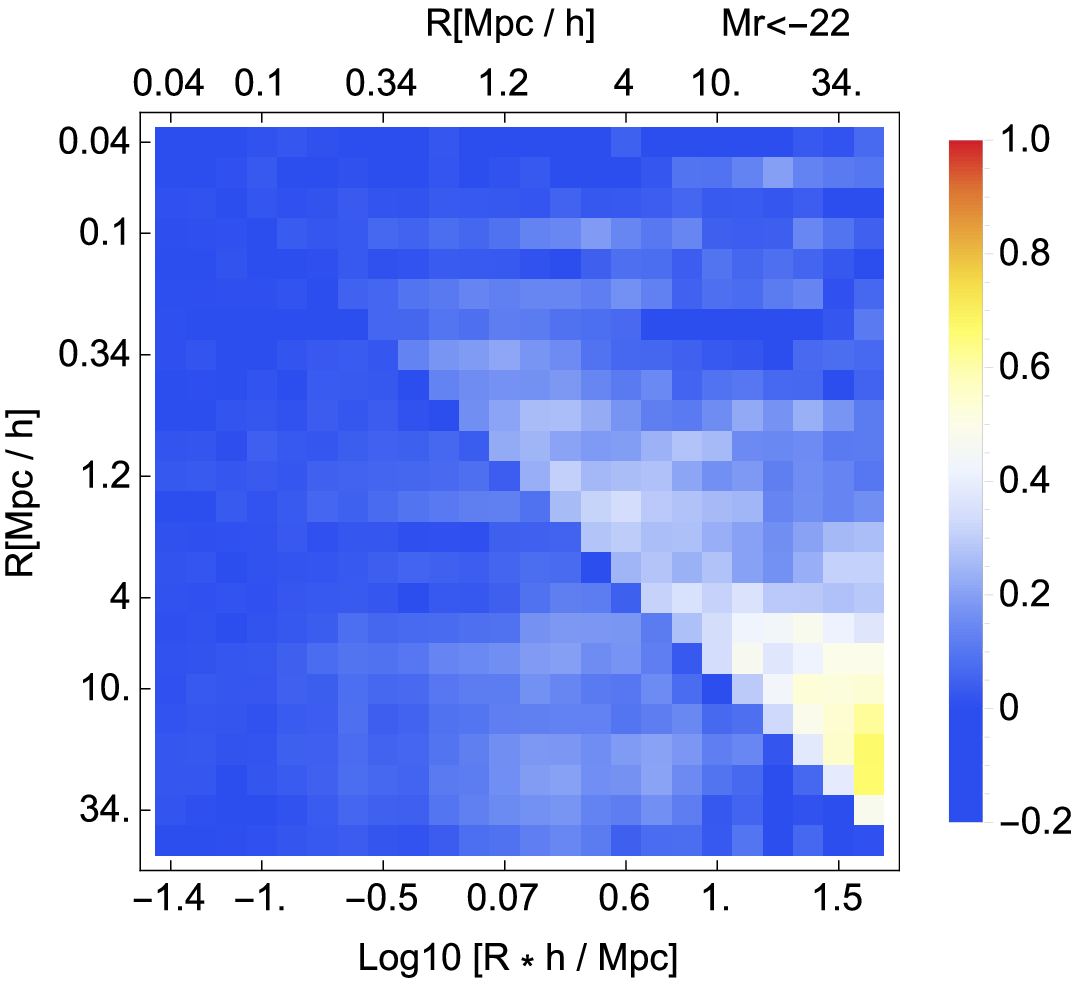}
  \includegraphics[scale=0.52]{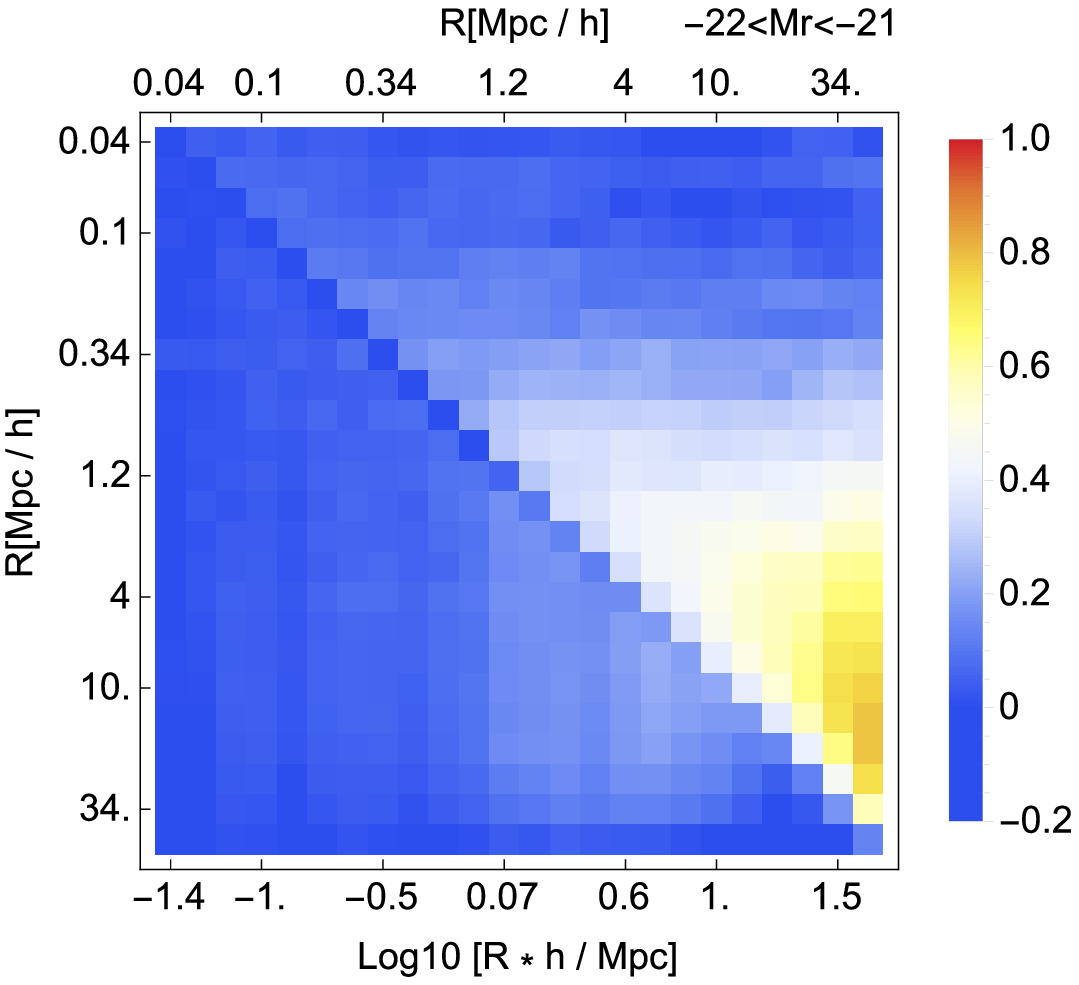}
\end{tabular}
%%%%%%%%%%%%%%%%%%%%%%%%%%%%%%%%%%%%%%%%%%%%%%%%%%%%%%%
\begin{tabular}{cc}
\includegraphics[scale=0.52]{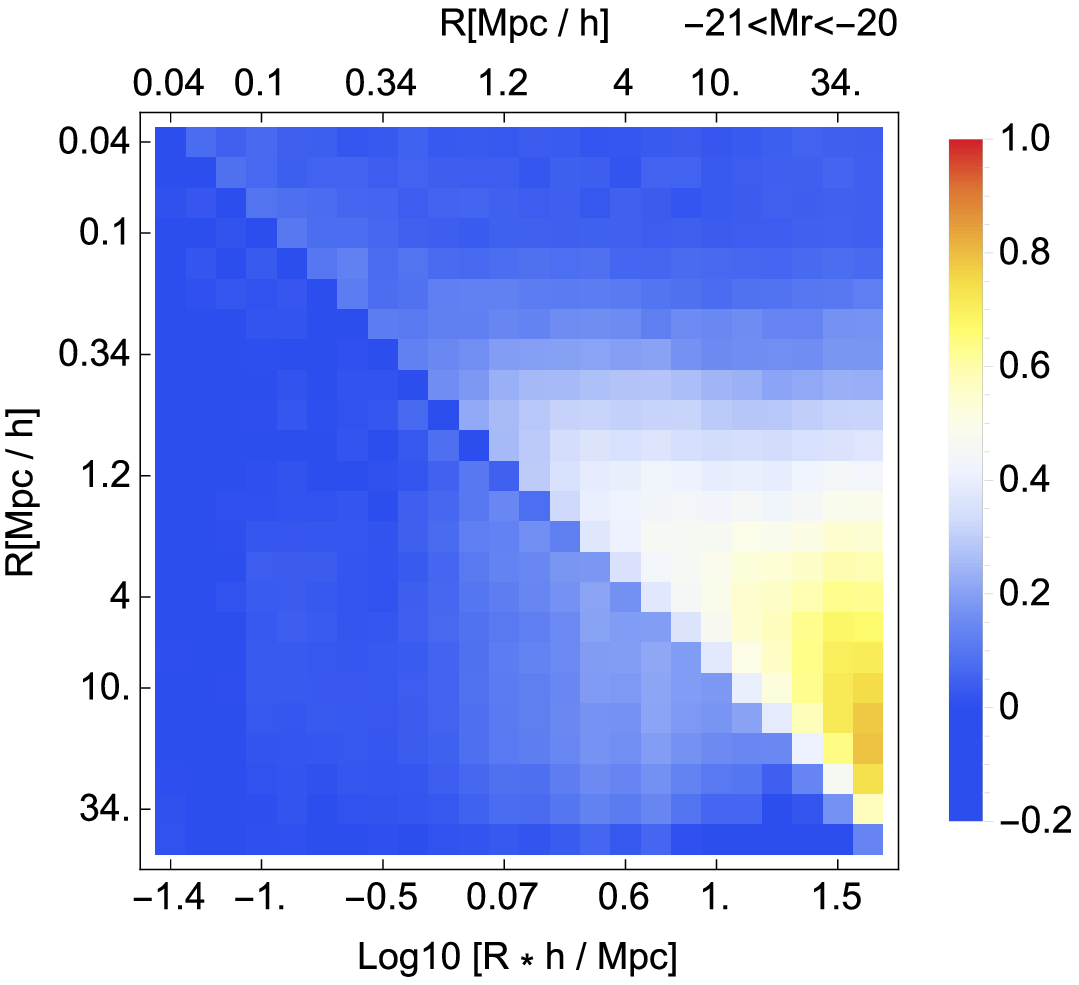}
\includegraphics[scale=0.52]{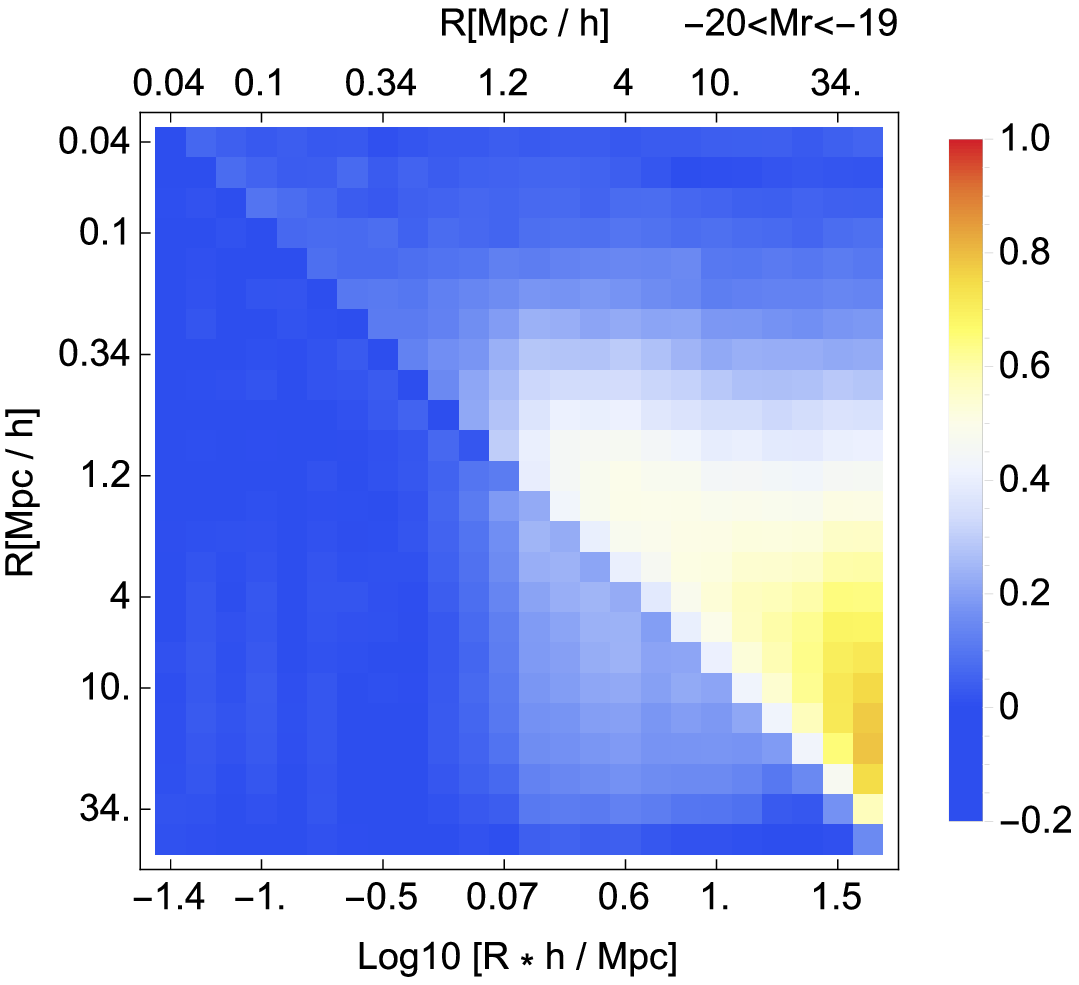} 
\end{tabular}
\caption{The cross-correlation matrix of $w_{\rm gg}$ and
  $\Delta\Sigma$, as computed from the whole MXXL data and presented
  also in Figures~\ref{fig:Row_GG_GGL_fixedGG}
  and~\ref{fig:Row_GG_GGL_fixedGGL}.}
\label{fig:GC_GGL_full}
\end{figure*}
%%%%%%%%%%%%%%%%%%%%%%%%%%%%%%%%%%%%%%%%%%%%%%%%%%%%%%%

%%%%%%%%%%%%%%%%%%%%%%%%%%%%%%%%%%%%%%%%%%%%%%%%%%%%%%%
%%%%%%%%%%%%%%%%%%%%%%%%%%%%%%%%%%%%%%%%%%%%%%%%%%%%%%%
%%%%%%%%%%%%%%%%%%%%%%%%%%%%%%%%%%%%%%%%%%%%%%%%%%%%%%%
In order to perform a joint likelihood analysis of the combination of
the projected galaxy correlation function and the stacked tangential
shear profile, one would start by writing the joint vector of
measurements:
%%%%%%%%%%%%%%%%%%%%%%%%%%%%%%%%%%%%%%%%%%%%%%%%%%%%%%%
\be 
{\bf X}^T = \left( 
\frac{\widehat{\widebar{w}}_{\rm gg}(R_1)}{\sigma[\widehat{\widebar{w}}_{\rm gg}](R_1)},
\dots,
\frac{\widehat{\widebar{w}}_{\rm gg}(R_{\mathcal{N}})}
{\sigma[\widehat{\widebar{w}}_{\rm gg}](R_{\mathcal{N}})},
\dots
\frac{\widehat{\widebar{\gamma}}^g_t(R_{\mathcal{N}})}
{\sigma[\widehat{\widebar{\gamma}}^g_t](R_{\mathcal{N}})} 
\right) \ ,
\ee
%%%%%%%%%%%%%%%%%%%%%%%%%%%%%%%%%%%%%%%%%%%%%%%%%%%%%%%
where we have normalized the measurements by their diagonal errors, so
as to obtain dimensionless numbers. $\mathcal{N}$ is the number of
transverse radial bins. Assuming a Gaussian likelihood, the normalized
data would then be written:
%%%%%%%%%%%%%%%%%%%%%%%%%%%%%%%%%%%%%%%%%%%%%%%%%%%%%%%
\be 
\log {\mathcal L}({\bf X}|{\bphi}) \ \propto 
\sum_{i,j=1}^{\mathcal{N}} [X_i-\bar{X}_i({\bphi})] \ r[X]^{-1}_{ij}
\left[X_j-\bar{X}_j({\bphi})\right] \ ,
\ee
%%%%%%%%%%%%%%%%%%%%%%%%%%%%%%%%%%%%%%%%%%%%%%%%%%%%%%%
where $\bar{X}_i(\bphi)$ is the expectation of the theoretical
model, and $\textbfss{r}[X]^{-1}$ is the inverse correlation matrix.
The correlation matrix is built from four blocks:
%%%%%%%%%%%%%%%%%%%%%%%%%%%%%%%%%%%%%%%%%%%%%%%%%%%%%%%
\be 
\textbfss{r}[X] = \left( 
\begin{array}{cc}
\textbfss{r}[\widehat{\widebar{w}}_{\rm gg}]  
& \textbfss{r}[\widehat{\widebar{w}}_{\rm gg},\widehat{\widebar{\gamma}}^g_t]  \\
\textbfss{r}[\widehat{\widebar{\gamma}}^g_t, \widehat{\widebar{w}}_{\rm gg}] 
& \textbfss{r}[\widehat{\widebar{\gamma}}^g_t]  
\end{array}\right)
\ee
%%%%%%%%%%%%%%%%%%%%%%%%%%%%%%%%%%%%%%%%%%%%%%%%%%%%%%%
where the off-diagonal blocks are simply the transpose of one
another. Again we see that $\left|\textbfss{r}_{X}\right|\le {1}$,
i.e. the absolute value of the determinant must be less than unity,
through the Cauchy-Schwarz inequality. 

Figures~\ref{fig:Row_GG_GGL_fixedGG} and \ref{fig:Row_GG_GGL_fixedGGL}
present four rows and columns through the off-diagonal block
$\textbfss{r}[\widehat{\widebar{w}}_{\rm
    gg},\widehat{\widebar{\gamma}}^g_t]$ of the full correlation
matrix, respectively. Note that unlike the other block matrices, the
off-diagonal block is not symmetric, hence the need to examine the
columns as well as the rows.  Just like before, the blue pentagons
denote the measurements from the 216 MXXL subcubes. The
cross-covariance was estimated similarly to
Eqs.~(\ref{eq:cov_wgg_est}) and~(\ref{eq:cov_shear_est}):
%%%%%%%%%%%%%%%%%%%%%%%%%%%%%%%%%%%%%%%%%%%%%%%%%%%%%%%
\ba
{\rm Cov}[\widehat{\widebar{w}}_{\rm gg}, \widehat{\widebar{\gamma}}_{t}^g](R_i, R_j)
 & & \hspace{-0.6cm}\equiv \frac{1}{\Sigma_{\rm crit}}\frac{1}{N-1}
\hspace{-0.1cm} \sum_{k=1}^{N} \hspace{-0.1cm}\left[\widehat{\widebar{w}}_{\rm gg,k}(R_i)
-\left<\widehat{\widebar{w}}_{\rm gg}(R_i)\right>\right]
 \nn \\
 & & \hspace{-0.6cm}\times \left[\widehat{\Delta\Sigma}_{k}(R_j)-\langle 
\widehat{\Delta\Sigma}_{k}(R_j)\rangle\right].
\label{eq:cross_wgg_shear_est}
\ea
%%%%%%%%%%%%%%%%%%%%%%%%%%%%%%%%%%%%%%%%%%%%%%%%%%%%%%%
Owing to the fact that we use the excess surface mass density as a
proxy for tangential shear, we may rewrite \Eqn{eq:cross_cov_BA} as:
%%%%%%%%%%%%%%%%%%%%%%%%%%%%%%%%%%%%%%%%%%%%%%%%%%%%%%%
\ba
{\rm Cov}[\widehat{\widebar{w}}_{\rm gg}(R_i), \widehat{\widebar{\Delta\Sigma}}(R_j)] 
& & \hspace{-0.6cm}= \frac{2 \rho_m^0}{A_s} 
\int \frac{d^2k_{\perp}}{(2\pi)^2} \widebar{J}_0(k_\perp R_i) \widebar{J}_2(k_\perp R_j) \nn \\
& & \hspace{-0.58cm} \times \hspace{0.1cm}
\mathcal{P}_{\rm gm}(k_\perp)\left[\mathcal{P}_{\rm gg}(k_\perp) +\frac{1}{\bar{n}_g}\right] . 
\label{eq:cross_cov_L_BA2}
\ea
%%%%%%%%%%%%%%%%%%%%%%%%%%%%%%%%%%%%%%%%%%%%%%%%%%%%%%%
The theoretical predictions from this expression are presented in
Figures~\ref{fig:Row_GG_GGL_fixedGG} and~\ref{fig:Row_GG_GGL_fixedGGL} as
the red triangles.

On small scales ($R<0.1\Mpc$, i.e. the top row of the figures), we see
that for all of the magnitude bins considered, the cross-correlation
coefficient is small $(r<0.2)$, and reasonably consistent with the
theoretical predictions from \Eqn{eq:cross_cov_L_BA2}. On large scales
($R\sim10\Mpc$, i.e. the bottom row in the figures), the
cross-correlation coefficient is larger, with $r\sim0.8$ for some of
the bins. Also, the theoretical predictions are qualitatively in
agreement, although the measurements do show a stronger degree of
correlation. On intermediate scales (the middle two rows), the
measurements show stronger bin-to-bin correlations than predicted by
the theory. These correlations on intermediate to large scales are
important to include in any joint likelihood analysis of galaxy
clustering and galaxy-galaxy lensing. Otherwise, the likelihood search
may lead to biased and over-optimistic parameter constraints.
%%%%%%%%%%%%%%%%%%%%%%%%%%%%%%%%%%%%%%%%%%%%%%%%%%%%%%%
\subsection{Testing the jackknife method to estimate covariances}
\label{SVI.4}
%%%%%%%%%%%%%%%%%%%%%%%%%%%%%%%%%%%%%%%%%%%%%%%%%%%%%%%
We now want to see how the correlation matrices estimated from the
entire MXXL data compare to those evaluated through the jackknife
method.

We pick one of the subcubes of the simulation, and generate 64
jackknife samples by removing a region of the subcube and estimating
the projected correlation functions from the remaining data. The
regions that are removed are in the shape of a rectangular
parallelepiped, with the long side aligned with the ${\bf\hat z}$
direction of the subcube, i.e. the projection direction, and the
shorter sides describing a square in the ${\bf\hat x}-{\bf\hat y}$
plane. The latter is obtained by dividing the subcube side in 8 equal
segments along both ${\bf\hat x}$ and ${\bf\hat y}$ directions,
resulting in 64 adjacent squares. Thus the dimensions of each
parallelepiped are $62.5 \times 62.5 \times 500\,[\Mpc]^3$.

The rest of the calculations follow the same path as detailed in
section~\ref{SV.1}. Each jackknife sample requires its own random
catalogue. Just as before, the galaxies and dark matter particles are
subsampled 32 times, in order to reduce the shot noise and the
$\alpha$ parameter describing the relative density of the simulation
and random data. The resulting correlation functions are the average
of the 32 subsample measurements. Once the projected correlation
functions and excess surface density are computed for each jackknife
sample, we combine them to estimate the covariance matrices for
$\Delta\Sigma$ and $w_{\rm gg}$:
%%%%%%%%%%%%%%%%%%%%%%%%%%%%%%%%%%%%%%%%%%%%%%%%%%%%%%%
\ba 
{\rm Cov}[\widehat{X}, \widehat{Y}](R_i, R_j) =\frac{N_{\rm samp}-1}{N_{\rm samp}} 
\sum_{k=1}^{N_{\rm samp}} \left[\widehat{X}_k(R_i) - \langle\widehat{X}_k(R_i)\rangle\right] \nn\\
\times  \left[\widehat{Y}_k(R_j) - \langle\widehat{Y}_k(R_j)\rangle\right] \ , 
\hspace{2.3cm}
\ea
\label{eq:cov_jk}
%%%%%%%%%%%%%%%%%%%%%%%%%%%%%%%%%%%%%%%%%%%%%%%%%%%%%%%
where $N_{\rm samp}=64$; $X$ and $Y$ stand in for $w_{\rm gg}$ or
$\Delta\Sigma$ and the mean is the average of the jackknife samples:
%%%%%%%%%%%%%%%%%%%%%%%%%%%%%%%%%%%%%%%%%%%%%%%%%%%%%%%
\be
\langle \widehat{X}(R)\rangle = \frac{1}{N_{\rm samp}} 
\sum_{k=1}^{N_{\rm samp}} \widehat{X}_k(R) \ .
\nn \ee
%%%%%%%%%%%%%%%%%%%%%%%%%%%%%%%%%%%%%%%%%%%%%%%%%%%%%%%
In \fig{fig:GC_full_JK} we present a comparison between our original
measurement of the correlation matrix of $w_{\rm gg}$ and the one
obtained through jackknife resamplings, for the four magnitude bins
that we have considered. In each panel, the upper right half depicts
the original correlation matrix, while the lower left shows the
jackknife result. The agreement is remarkable, with the jackknife
result being slightly noisier than the full one. Similarly,
\fig{fig:GGL_full_JK} shows the same comparison for the excess surface
density. \figs{fig:GC_GGL_full}{fig:GC_GGL_JK} present the original
and jackknife cross-correlation matrices respectively. In all figures
the jackknife result appears a little more correlated than the
original measurement, but in general the agreement is
surprising. Although not shown, we have also compared the variance for
$w_{\rm gg}$ and $\Delta \Sigma$: the findings were very similar, with
the jackknife variance being a little noisier, but essentially
matching the original measurement rather well.
%%%%%%%%%%%%%%%%%%%%%%%%%%%%%%%%%%%%%%%%%%%%%%%%%%%%%%%
%%%%%%%%%%%%%%%%%%%%%%%%%%%%%%%%%%%%%%%%%%%%%%%%%%%%%%%
%%%%%%%%%%%%%%%%%%%%%%%%%%%%%%%%%%%%%%%%%%%%%%%%%%%%%%%
%%%%%%%%%%%%%%%%%%%%%%%%%%%%%%%%%%%%%%%%%%%%%%%%%%%%%%%
\section{Conclusions}
\label{SVII}
%%%%%%%%%%%%%%%%%%%%%%%%%%%%%%%%%%%%%%%%%%%%%%%%%%%%%%%
%%%%%%%%%%%%%%%%%%%%%%%%%%%%%%%%%%%%%%%%%%%%%%%%%%%%%%%

In this paper we have explored the required ingredients for performing
a joint likelihood analysis of galaxy clustering and galaxy-galaxy
lensing. The combination of these probes has the potential to enable
simultaneous constraints of the cosmological and galaxy formation
model parameters.

In \S\ref{SII}, we developed an analytic framework to predict the
projected correlation function and the stacked tangential shears of a
population of galaxies. We provided both exact and Limber-approximated
expressions for the observables in the flat-sky limit. In
appendix~\ref{app:cov} we presented the full derivations of the auto-
and cross-covariance matrices. In \S\ref{SIII} we summarized the
results for the case where the underlying density fluctuations are
Gaussian and modulated by a shot-noise contribution. The two necessary
ingredients for the results of sections \S\ref{SII} and \S\ref{SIII}
are the galaxy-galaxy and galaxy-matter power spectra or alternatively
a model for the bias defined by \Eqns{eq:bgg}{eq:bgm}.

The theoretical predictions of the models were then tested against
measurements from numerical simulations of structure formation. In
\S\ref{SIV} we provided a brief overview of the MXXL simulation. The
galaxy catalogues were generated using the Garching semi-analytic
model (SAM) of galaxy formation, which enabled us to compute the $u$,
$g$, $r$, $i$, $z$ SDSS absolute magnitudes. As a diagnostic of the
SAM galaxies we presented the evolution of the galaxy luminosity
function for the five bands. We also explored certain properties of
the galaxies as a function of absolute $r$ band magnitude ($M_r$),
allowing us to understand the impact of the finite mass resolution of
the MXXL simulation on the derived galaxy properties. Through detailed
comparison with the Millennium simulation, we determined that
resolution effects were qualitatively important only for magnitudes
fainter than $M_r > -19$.

In \S\ref{SV} we described our methods for estimating the projected
correlations from the MXXL. This required the implementation of fast
and efficient algorithms, since the dark matter distribution was
represented by over 300 billion dark matter particles and the galaxy
catalogue contained more than a billion objects. We thus developed a
parallel k-D tree-algorithm for computing the correlation functions of
the subsampled matter and galaxy distributions.

We split the galaxy catalogues into four sub-samples based on their
$M_r$, with the constraint that $M_r<-19$. We also sliced the dark
matter and galaxy catalogues into a series of 216 subcubes of volume
$500\Mpccube$. As a step towards modelling the GC and GGL signals, we
examined the projected galaxy-galaxy and galaxy-matter correlation
functions using the estimator proposed by \citet{LandySzalay1993}. We
also computed the bias coefficients as a function of scale from these
statistics. We found that on large scales, $R>10\Mpc$, the bias was
consistent with being linear and deterministic.  On smaller scales,
the bias possessed a complex scale dependence, which varied with the
magnitude bin through the larger number of satellites and their
spatial distribution, as well as the decreasing host mass for the
centrals in the fainter bins. We also found that the cross-correlation
coefficient was significantly greater than unity on small scales. This
could be explained as a consequence of an exclusion effect -- no two
galaxies may be closer than the sum of their respective virial radii.

We next computed the excess surface-mass density $\Delta\Sigma$, which
is directly proportional to the tangential shear, and the projected
galaxy correlation function for each of the magnitude bins and for
each of the 216 subcubes. These were compared to the theoretical
predictions in the Limber approximation. The evaluation of the theory
required a model for the nonlinear matter power spectrum and the
scale-dependence of the bias. For the former we used the {\tt halofit}
code \citep{Smithetal2003}, and for the latter we took directly the
measured bias. We found that on small scales the theory underpredicted
the measurements by $\lesssim 20\%$ for $w_{\rm gg}$ and by $\lesssim
30 \%$ for $\Delta\Sigma$. This was attributed to the diminished
accuracy of {\tt halofit} on small scales. Both functions displayed
complicated features for galaxies in the faintest magnitude bin
$M_r>-20$. This was again attributed to the increased abundance of
satellite galaxies.

In \S\ref{SVI} we used the measured estimates of $w_{\rm gg}$ and
$\Delta\Sigma$ from the subcubes to construct the auto- and cross-
covariance matrices. We found that on large and small scales the
errors were reasonably well described by the Gaussian plus shot-noise
model. However, on intermediate scales, $0.1<R<10\Mpc$, the measured
errors were significantly larger by a factor of $\sim2$--3.  This
suggests that in order to accurately model the covariance matrix one
needs to take account of the non-Gaussian contributions from the bi-
and trispectrum. Note that in the case of GGL, the presence of shape
noise does alleviate some of the non-Gaussianities on intermediate
scales, as shown in Appendix \S\ref{App3}.

Importantly, we found that the cross-covariance between $w_{\rm gg}$
and $\Delta\Sigma$ was not zero. The elements of the cross-covariance
matrix showed significant bin-to-bin correlations with $r\sim0.8$ for
some elements on large-scales. This result is important, since in a
number of previous analyses neglected the cross-correlation of GC and
GGL. Our results suggest that if these correlations are ignored then a
standard likelihood analysis would lead to biased and over-optimistic
constraints on the parameters of the models. 

Finally, we compared the auto- and cross-covariances obtained from the
jackknife samplings of one of the subcubes of MXXL with the full
simulation covariances. We concluded that the jackknife method is
remarkably effective, for both auto- and cross-covariances.

In the future, more work will be required to establish an accurate
model of the GC and GGL signals on scales $R<5\Mpc$. We note that,
whilst the SAM model of galaxy formation is not to be taken as an
exact replica of reality, it nevertheless captures many of the effects
that we expect to have to model when constraining model parameters
with real data. It will also be important to determine how well the
combination of GC and GGL can actually break the degeneracies between
the galaxy-formation and cosmological parameters. We shall aim to
explore this in a future paper.  In the meantime, we note that
\citet{Eifleretal2014} have explored clustering and lensing probe
combinations and have included additional non-Gaussian terms in the
modelling of the covariances. However, they have assumed a rather
simplistic modelling of the galaxy bias, which we showed here it is
not the case. This leads us to suspect that their results will be
somewhat over-optimistic.

Another possibility is the exploration of the $\Upsilon$ statistic
\citep{Baldaufetal2010,Mandelbaumetal2013}, which was developed to
remove the complicated scale-dependences of the bias.  However, what
is not clear, is whether the $\Upsilon$ statistic in combination with
galaxy clustering can provide constraints on both the cosmological and
galaxy formation model that are competitive with the standard
statistics. We leave this for future work.

%%%%%%%%%%%%%%%%%%%%%%%%%%%%%%%%%%%%%%%%%%%%%%%%%%%%%%%

\section*{Acknowledgements}

The authors would like thank Simon White, Eiichiro Komatsu, Peter
Schneider, and Gary Bernstein for useful discussions, Bruno Henriques
for providing the Millennium simulation galaxy catalogue used in
\Fig{fig:gal_prop_Bruno}. We are also grateful to Volker Springel,
Adrian Jenkins, Carlos Frenk and Carlton Baugh for their contribution
to the MXXL simulation used in this study.The MXXL was carried out on
Juropa at the J\"ulich Supercomputer Centre in Germany. LM thanks MPA
for its kind hospitality while this work was being performed. LM was
supported by the DFG through the grant MA 4967/1-2. RES acknowledges
support from ERC Advanced Grant 246797 GALFORMOD.

%%%%%%%%%%%%%%%%%%%%%%%%%%%%%%%%%%%%%%%%%%%%%%%%%%%%%%%
%\bibliographystyle{mn2e.NEW} 
%\bibliography{XXL}

\begin{thebibliography}{65}
\expandafter\ifx\csname natexlab\endcsname\relax\def\natexlab#1{#1}\fi

\bibitem[{{Angulo} {et~al}\mbox{.}(2012){Angulo}, {Springel}, {White},
  {Jenkins}, {Baugh}, \& {Frenk}}]{Anguloetal2012}
{Angulo} R.~E., {Springel} V., {White} S.~D.~M., {Jenkins} A., {Baugh} C.~M.,
  {Frenk} C.~S., 2012, \mnras, 426, 2046

\bibitem[{{Angulo} {et~al}\mbox{.}(2014){Angulo}, {White}, {Springel}, \&
  {Henriques}}]{Anguloetal2014}
{Angulo} R.~E., {White} S.~D.~M., {Springel} V., {Henriques} B., 2014, \mnras,
  442, 2131

\bibitem[{{Baldauf} {et~al}\mbox{.}(2010){Baldauf}, {Smith}, {Seljak}, \&
  {Mandelbaum}}]{Baldaufetal2010}
{Baldauf} T., {Smith} R.~E., {Seljak} U., {Mandelbaum} R., 2010, \prd, 81,
  063531

\bibitem[{{Bartelmann} \& {Schneider}(2001)}]{BartelmannSchneider2001}
{Bartelmann} M., {Schneider} P., 2001, \physrep, 340, 291

\bibitem[{{Blanton} {et~al}\mbox{.}(2003){Blanton}, {Hogg}, {Bahcall},
  {Brinkmann}, {Britton}, {Connolly}, {Csabai}, {Fukugita}, {Loveday},
  {Meiksin}, {Munn}, {Nichol}, {Okamura}, {Quinn}, {Schneider}, {Shimasaku},
  {Strauss}, {Tegmark}, {Vogeley}, \& {Weinberg}}]{Blantonetal2003}
{Blanton} M.~R. {et~al.}, 2003, \apj, 592, 819

\bibitem[{{Brainerd}, {Blandford} \& {Smail}(1996){Brainerd}, {Blandford}, \&
  {Smail}}]{Brainerdetal1996}
{Brainerd} T.~G., {Blandford} R.~D., {Smail} I., 1996, \apj, 466, 623

\bibitem[{{Cacciato} {et~al}\mbox{.}(2009){Cacciato}, {van den Bosch}, {More},
  {Li}, {Mo}, \& {Yang}}]{Cacciatoetal2009}
{Cacciato} M., {van den Bosch} F.~C., {More} S., {Li} R., {Mo} H.~J., {Yang}
  X., 2009, \mnras, 394, 929

\bibitem[{{Cacciato} {et~al}\mbox{.}(2013){Cacciato}, {van den Bosch}, {More},
  {Mo}, \& {Yang}}]{Cacciatoetal2013}
{Cacciato} M., {van den Bosch} F.~C., {More} S., {Mo} H., {Yang} X., 2013,
  \mnras, 430, 767

\bibitem[{{Dekel} \& {Lahav}(1999)}]{DekelLahav1999}
{Dekel} A., {Lahav} O., 1999, \apj, 520, 24

\bibitem[{{dell'Antonio} \& {Tyson}(1996)}]{Dell'Antonioetal1996}
{dell'Antonio} I.~P., {Tyson} J.~A., 1996, \apjl, 473, L17

\bibitem[{{Eifler} {et~al}\mbox{.}(2014){Eifler}, {Krause}, {Schneider}, \&
  {Honscheid}}]{Eifleretal2014}
{Eifler} T., {Krause} E., {Schneider} P., {Honscheid} K., 2014, \mnras, 440,
  1379

\bibitem[{{Fosalba} {et~al}\mbox{.}(2015{\natexlab{a}}){Fosalba}, {Crocce},
  {Gazta{\~n}aga}, \& {Castander}}]{Fosalbaetal2015a}
{Fosalba} P., {Crocce} M., {Gazta{\~n}aga} E., {Castander} F.~J.,
  2015{\natexlab{a}}, \mnras, 448, 2987

\bibitem[{{Fosalba} {et~al}\mbox{.}(2015{\natexlab{b}}){Fosalba},
  {Gazta{\~n}aga}, {Castander}, \& {Crocce}}]{Fosalbaetal2015b}
{Fosalba} P., {Gazta{\~n}aga} E., {Castander} F.~J., {Crocce} M.,
  2015{\natexlab{b}}, \mnras, 447, 1319

\bibitem[{{Griffiths} {et~al}\mbox{.}(1996){Griffiths}, {Casertano}, {Im}, \&
  {Ratnatunga}}]{Griffithsetal1996}
{Griffiths} R.~E., {Casertano} S., {Im} M., {Ratnatunga} K.~U., 1996, \mnras,
  282, 1159

\bibitem[{{Guo} {et~al}\mbox{.}(2011){Guo}, {White}, {Boylan-Kolchin}, {De
  Lucia}, {Kauffmann}, {Lemson}, {Li}, {Springel}, \& {Weinmann}}]{Guoetal2011}
{Guo} Q. {et~al.}, 2011, \mnras, 413, 101

\bibitem[{{Guzik} \& {Seljak}(2001)}]{Guziketal2001}
{Guzik} J., {Seljak} U., 2001, \mnras, 321, 439

\bibitem[{{Guzik} \& {Seljak}(2002)}]{Guziketal2002}
{Guzik} J., {Seljak} U., 2002, \mnras, 335, 311

\bibitem[{{Hayashi} \& {White}(2008)}]{HayashiWhite2008}
{Hayashi} E., {White} S.~D.~M., 2008, \mnras, 388, 2

\bibitem[{{Henriques} {et~al}\mbox{.}(2012){Henriques}, {White}, {Lemson},
  {Thomas}, {Guo}, {Marleau}, \& {Overzier}}]{Henriquesetal2012}
{Henriques} B.~M.~B., {White} S.~D.~M., {Lemson} G., {Thomas} P.~A., {Guo} Q.,
  {Marleau} G.-D., {Overzier} R.~A., 2012, \mnras, 421, 2904

\bibitem[{{Hirata} {et~al}\mbox{.}(2004){Hirata}, {Mandelbaum}, {Seljak},
  {Guzik}, {Padmanabhan}, {Blake}, {Brinkmann}, {Bud{\'a}vari}, {Connolly},
  {Csabai}, {Scranton}, \& {Szalay}}]{Hirataetal2004}
{Hirata} C.~M. {et~al.}, 2004, \mnras, 353, 529

\bibitem[{{Hoekstra} {et~al}\mbox{.}(2002){Hoekstra}, {van Waerbeke},
  {Gladders}, {Mellier}, \& {Yee}}]{Hoekstraetal2002}
{Hoekstra} H., {van Waerbeke} L., {Gladders} M.~D., {Mellier} Y., {Yee}
  H.~K.~C., 2002, \apj, 577, 604

\bibitem[{{Hoekstra}, {Yee} \& {Gladders}(2004){Hoekstra}, {Yee}, \&
  {Gladders}}]{Hoekstraetal2004}
{Hoekstra} H., {Yee} H.~K.~C., {Gladders} M.~D., 2004, \apj, 606, 67

\bibitem[{{Hudson} {et~al}\mbox{.}(2015){Hudson}, {Gillis}, {Coupon},
  {Hildebrandt}, {Erben}, {Heymans}, {Hoekstra}, {Kitching}, {Mellier},
  {Miller}, {Van Waerbeke}, {Bonnett}, {Fu}, {Kuijken}, {Rowe}, {Schrabback},
  {Semboloni}, {van Uitert}, \& {Velander}}]{Hudsonetal2015}
{Hudson} M.~J. {et~al.}, 2015, \mnras, 447, 298

\bibitem[{{Hudson} {et~al}\mbox{.}(1998){Hudson}, {Gwyn}, {Dahle}, \&
  {Kaiser}}]{Hudsonetal1998}
{Hudson} M.~J., {Gwyn} S.~D.~J., {Dahle} H., {Kaiser} N., 1998, \apj, 503, 531

\bibitem[{{Jarvis}, {Bernstein} \& {Jain}(2004){Jarvis}, {Bernstein}, \&
  {Jain}}]{Jarvisetal2004}
{Jarvis} M., {Bernstein} G., {Jain} B., 2004, \mnras, 352, 338

\bibitem[{{Jeong}, {Komatsu} \& {Jain}(2009){Jeong}, {Komatsu}, \&
  {Jain}}]{Jeongetal2009}
{Jeong} D., {Komatsu} E., {Jain} B., 2009, \prd, 80, 123527

\bibitem[{{Johnston} {et~al}\mbox{.}(2007){Johnston}, {Sheldon}, {Wechsler},
  {Rozo}, {Koester}, {Frieman}, {McKay}, {Evrard}, {Becker}, \&
  {Annis}}]{Johnstonetal2007}
{Johnston} D.~E. {et~al.}, 2007, ArXiv e-prints astro-ph/0709.1159

\bibitem[{{Landy} \& {Szalay}(1993)}]{LandySzalay1993}
{Landy} S.~D., {Szalay} A.~S., 1993, \apj, 412, 64

\bibitem[{{Leauthaud et al.}(2012)}]{Leauthaudetal2012}
{Leauthaud et al.}, 2012, \apj, 744, 159

\bibitem[{{Mandelbaum} {et~al}\mbox{.}(2006{\natexlab{a}}){Mandelbaum},
  {Hirata}, {Broderick}, {Seljak}, \& {Brinkmann}}]{Mandelbaumetal2006b}
{Mandelbaum} R., {Hirata} C.~M., {Broderick} T., {Seljak} U., {Brinkmann} J.,
  2006{\natexlab{a}}, \mnras, 370, 1008

\bibitem[{{Mandelbaum} {et~al}\mbox{.}(2005){Mandelbaum}, {Hirata}, {Seljak},
  {Guzik}, {Padmanabhan}, {Blake}, {Blanton}, {Lupton}, \&
  {Brinkmann}}]{Mandelbaumetal2005}
{Mandelbaum} R. {et~al.}, 2005, \mnras, 361, 1287

\bibitem[{{Mandelbaum} {et~al}\mbox{.}(2006{\natexlab{b}}){Mandelbaum},
  {Seljak}, {Cool}, {Blanton}, {Hirata}, \& {Brinkmann}}]{Mandelbaumetal2006c}
{Mandelbaum} R., {Seljak} U., {Cool} R.~J., {Blanton} M., {Hirata} C.~M.,
  {Brinkmann} J., 2006{\natexlab{b}}, \mnras, 372, 758

\bibitem[{{Mandelbaum} {et~al}\mbox{.}(2006{\natexlab{c}}){Mandelbaum},
  {Seljak}, {Kauffmann}, {Hirata}, \& {Brinkmann}}]{Mandelbaumetal2006a}
{Mandelbaum} R., {Seljak} U., {Kauffmann} G., {Hirata} C.~M., {Brinkmann} J.,
  2006{\natexlab{c}}, \mnras, 368, 715

\bibitem[{{Mandelbaum} {et~al}\mbox{.}(2013){Mandelbaum}, {Slosar}, {Baldauf},
  {Seljak}, {Hirata}, {Nakajima}, {Reyes}, \& {Smith}}]{Mandelbaumetal2013}
{Mandelbaum} R., {Slosar} A., {Baldauf} T., {Seljak} U., {Hirata} C.~M.,
  {Nakajima} R., {Reyes} R., {Smith} R.~E., 2013, \mnras, 432, 1544

\bibitem[{{McKay et al.}(2001)}]{McKayetal2001}
{McKay et al.}, 2001, ArXiv e-prints astro-ph/0108013

\bibitem[{{Miyatake} {et~al}\mbox{.}(2013){Miyatake}, {More}, {Mandelbaum},
  {Takada}, {Spergel}, {Kneib}, {Schneider}, {Brinkmann}, \&
  {Brownstein}}]{Miyatakeetal2013}
{Miyatake} H. {et~al.}, 2013, ArXiv e-prints astro-ph/1311.1480

\bibitem[{{Moore} {et~al}\mbox{.}(2001){Moore}, {Connolly}, {Genovese}, {Gray},
  {Grone}, {Kanidoris}, {Nichol}, {Schneider}, {Szalay}, {Szapudi}, \&
  {Wasserman}}]{Mooreetal2001}
{Moore} A.~W. {et~al.}, 2001, in Mining the Sky, {Banday} A.~J., {Zaroubi} S.,
  {Bartelmann} M., eds., p.~71

\bibitem[{{More} {et~al}\mbox{.}(2014){More}, {Miyatake}, {Mandelbaum},
  {Takada}, {Spergel}, {Brownstein}, \& {Schneider}}]{Moreetal2014}
{More} S., {Miyatake} H., {Mandelbaum} R., {Takada} M., {Spergel} D.,
  {Brownstein} J., {Schneider} D.~P., 2014, ArXiv e-prints astro-ph/1407.1856

\bibitem[{{More} {et~al}\mbox{.}(2013){More}, {van den Bosch}, {Cacciato},
  {More}, {Mo}, \& {Yang}}]{Moreetal2013}
{More} S., {van den Bosch} F.~C., {Cacciato} M., {More} A., {Mo} H., {Yang} X.,
  2013, \mnras, 430, 747

\bibitem[{{Nakajima} {et~al}\mbox{.}(2012){Nakajima}, {Mandelbaum}, {Seljak},
  {Cohn}, {Reyes}, \& {Cool}}]{Nakajimaetal2012}
{Nakajima} R., {Mandelbaum} R., {Seljak} U., {Cohn} J.~D., {Reyes} R., {Cool}
  R., 2012, \mnras, 420, 3240

\bibitem[{{Parker} {et~al}\mbox{.}(2007){Parker}, {Hoekstra}, {Hudson}, {van
  Waerbeke}, \& {Mellier}}]{Parkeretal2007}
{Parker} L.~C., {Hoekstra} H., {Hudson} M.~J., {van Waerbeke} L., {Mellier} Y.,
  2007, \apj, 669, 21

\bibitem[{{Saghiha} {et~al}\mbox{.}(2012){Saghiha}, {Hilbert}, {Schneider}, \&
  {Simon}}]{Saghihaetal2012}
{Saghiha} H., {Hilbert} S., {Schneider} P., {Simon} P., 2012, \aap, 547, A77

\bibitem[{{Schneider}(1998)}]{Schneider1998}
{Schneider} P., 1998, \apj, 498, 43

\bibitem[{{Schneider}(2005)}]{Schneider2005}
{Schneider} P., 2005, ArXiv e-prints astro-ph/0509252

\bibitem[{{Seljak} {et~al}\mbox{.}(2005){Seljak}, {Makarov}, {Mandelbaum},
  {Hirata}, {Padmanabhan}, {McDonald}, {Blanton}, {Tegmark}, {Bahcall}, \&
  {Brinkmann}}]{Seljaketal2005}
{Seljak} U. {et~al.}, 2005, \prd, 71, 043511

\bibitem[{{Seljak} \& {Warren}(2004)}]{SeljakWarren2004}
{Seljak} U., {Warren} M.~S., 2004, \mnras, 355, 129

\bibitem[{{Sheldon} {et~al}\mbox{.}(2004){Sheldon}, {Johnston}, {Frieman},
  {Scranton}, {McKay}, {Connolly}, {Budav{\'a}ri}, {Zehavi}, {Bahcall},
  {Brinkmann}, \& {Fukugita}}]{Sheldonetal2004}
{Sheldon} E.~S. {et~al.}, 2004, \aj, 127, 2544

\bibitem[{{Sheldon} {et~al}\mbox{.}(2009{\natexlab{a}}){Sheldon}, {Johnston},
  {Masjedi}, {McKay}, {Blanton}, {Scranton}, {Wechsler}, {Koester}, {Hansen},
  {Frieman}, \& {Annis}}]{Sheldonetal2009b}
{Sheldon} E.~S. {et~al.}, 2009{\natexlab{a}}, \apj, 703, 2232

\bibitem[{{Sheldon} {et~al}\mbox{.}(2009{\natexlab{b}}){Sheldon}, {Johnston},
  {Scranton}, {Koester}, {McKay}, {Oyaizu}, {Cunha}, {Lima}, {Lin}, {Frieman},
  {Wechsler}, {Annis}, {Mandelbaum}, {Bahcall}, \&
  {Fukugita}}]{Sheldonetal2009a}
{Sheldon} E.~S. {et~al.}, 2009{\natexlab{b}}, \apj, 703, 2217

\bibitem[{{Smith}(2009)}]{Smith2009}
{Smith} R.~E., 2009, \mnras, 400, 851

\bibitem[{{Smith}(2012)}]{Smith2012}
{Smith} R.~E., 2012, \mnras, 426, 531

\bibitem[{{Smith} \& {Marian}(2014)}]{SmithMarian2014}
{Smith} R.~E., {Marian} L., 2014, ArXiv e-prints astro-ph/1406.1800

\bibitem[{{Smith} \& {Marian}(2015)}]{SmithMarian2015}
{Smith} R.~E., {Marian} L., 2015, ArXiv e-prints astro-ph/1503.06830

\bibitem[{{Smith} {et~al}\mbox{.}(2003){Smith}, {Peacock}, {Jenkins}, {White},
  {Frenk}, {Pearce}, {Thomas}, {Efstathiou}, \& {Couchman}}]{Smithetal2003}
{Smith} R.~E. {et~al.}, 2003, \mnras, 341, 1311

\bibitem[{{Spergel et al.}(2003)}]{Spergeletal2003}
{Spergel et al.}, 2003, \apjs, 148, 175

\bibitem[{{Spergel et al.}(2007)}]{Spergeletal2007}
{Spergel et al.}, 2007, \apjs, 170, 377

\bibitem[{{Springel} {et~al}\mbox{.}(2005){Springel}, {White}, {Jenkins},
  {Frenk}, {Yoshida}, {Gao}, {Navarro}, {Thacker}, {Croton}, {Helly},
  {Peacock}, {Cole}, {Thomas}, {Couchman}, {Evrard}, {Colberg}, \&
  {Pearce}}]{Springeletal2005}
{Springel} V. {et~al.}, 2005, \nat, 435, 629

\bibitem[{{Takahashi} {et~al}\mbox{.}(2012){Takahashi}, {Sato}, {Nishimichi},
  {Taruya}, \& {Oguri}}]{Takahashietal2012}
{Takahashi} R., {Sato} M., {Nishimichi} T., {Taruya} A., {Oguri} M., 2012,
  \apj, 761, 152

\bibitem[{{Tyson} {et~al}\mbox{.}(1984){Tyson}, {Valdes}, {Jarvis}, \&
  {Mills}}]{Tysonetal1984}
{Tyson} J.~A., {Valdes} F., {Jarvis} J.~F., {Mills}, Jr. A.~P., 1984, \apjl,
  281, L59

\bibitem[{{van den Bosch} {et~al}\mbox{.}(2013){van den Bosch}, {More},
  {Cacciato}, {Mo}, \& {Yang}}]{vandenBoschetal2013}
{van den Bosch} F.~C., {More} S., {Cacciato} M., {Mo} H., {Yang} X., 2013,
  \mnras, 430, 725

\bibitem[{{van Uitert} {et~al}\mbox{.}(2011){van Uitert}, {Hoekstra},
  {Velander}, {Gilbank}, {Gladders}, \& {Yee}}]{vanUitertetal2011}
{van Uitert} E., {Hoekstra} H., {Velander} M., {Gilbank} D.~G., {Gladders}
  M.~D., {Yee} H.~K.~C., 2011, \aap, 534, A14

\bibitem[{{Velander et al.}(2014)}]{Velanderetal2014}
{Velander et al.}, 2014, \mnras, 437, 2111

\bibitem[{{Wilson} {et~al}\mbox{.}(2001){Wilson}, {Kaiser}, {Luppino}, \&
  {Cowie}}]{Wilsonetal2001}
{Wilson} G., {Kaiser} N., {Luppino} G.~A., {Cowie} L.~L., 2001, \apj, 555, 572

\bibitem[{{Yang}, {Mo} \& {van den Bosch}(2003){Yang}, {Mo}, \& {van den
  Bosch}}]{Yangetal2003}
{Yang} X., {Mo} H.~J., {van den Bosch} F.~C., 2003, \mnras, 339, 1057

\bibitem[{{Yoo} {et~al}\mbox{.}(2006){Yoo}, {Tinker}, {Weinberg}, {Zheng},
  {Katz}, \& {Dav{\'e}}}]{Yooetal2006}
{Yoo} J., {Tinker} J.~L., {Weinberg} D.~H., {Zheng} Z., {Katz} N., {Dav{\'e}}
  R., 2006, \apj, 652, 26

\end{thebibliography}
%\setlength{\itemindent}{-5mm}

%%%%%%%%%%%%%%%%%%%%%%%%%%%%%%%%%%%%%%%%%%%%%%%%%%%%%%%
%%%%%%%%%%%%%%%%%%%%%%%%%%%%%%%%%%%%%%%%%%%%%%%%%%%%%%%
%%%%%%%%%%%%%%%%%%%%%%%%%%%%%%%%%%%%%%%%%%%%%%%%%%%%%%%
\appendix
\onecolumn
%%%%%%%%%%%%%%%%%%%%%%%%%%%%%%%%%%%%%%%%%%%%%%%%%%%%%%%
\section{Derivation of the auto- and cross-covariance matrices}
\label{app:cov}
%%%%%%%%%%%%%%%%%%%%%%%%%%%%%%%%%%%%%%%%%%%%%%%%%%%%%%%
%%%%%%%%%%%%%%%%%%%%%%%%%%%%%%%%%%%%%%%%%%%%%%%%%%%%%%%
\subsection{Derivation of the covariance matrix of stacked tangential shear}
\label{app:ggl-cov}
%%%%%%%%%%%%%%%%%%%%%%%%%%%%%%%%%%%%%%%%%%%%%%%%%%%%%%%
%%%%%%%%%%%%%%%%%%%%%%%%%%%%%%%%%%%%%%%%%%%%%%%%%%%%%%%
To simplify the notation, from now on we shall consider that all 2D
integrals are implicitly done on the survey area $\Omega_s$ without
explicitly stating so in every case.

\noindent Using Eq.~(\ref{eq:gal_den}) we write:
%%%%%%%%%%%%%%%%%%%%%%%%%%%%%%%%%%%%%%%%%%%%%%%%%%%%%%%
\ba
\langle\widehat{\gamma}_t^g(\btheta_1)\widehat{\gamma}_t^g(\btheta_2)\rangle & = & 
\frac{1}{N_{g}^{2}}\int d^2 x \,d^2 y\sum_{i,j=1}^{N_g}\langle \delta_D(\bx-\bx_i)
 \delta_D(\by-\bx_j) \gamma_t(\bx+\btheta_1 ; \bx)\gamma_t(\by+\btheta_2 ; \by)\rangle \nonumber \\
& = &\frac{1}{N_{g}^{2}}\left[ \int d^2 x \langle n_g(\bx)\gamma_t(\bx+\btheta_1 ; \bx)
\gamma_t(\bx+\btheta_2 ; \bx)\rangle + \int d^2x \,d^2y \,\langle n_g(\bx) n_g(\by) 
\gamma_t(\bx+\btheta_1 ; \bx) \gamma_t(\by+\btheta_2 ; \by) \rangle \right]. 
\label{eq:cov_shear1}
\ea
%%%%%%%%%%%%%%%%%%%%%%%%%%%%%%%%%%%%%%%%%%%%%%%%%%%%%%%
In the above double sum, the special case where $i=j$ gives rise to
the 3-point function $\langle
n_g(\bx)\gamma_t(\bx+\btheta_1 ; \bx)\gamma_t(\bx+\btheta_2 ; \bx)\rangle =
\bar{n}_g [
  \langle\gamma_t(\bx+\btheta_1 ; \bx)\gamma_t(\bx+\btheta_2 ; \bx)\rangle +
  \langle \delta_g(\bx)\gamma_t(\bx+\btheta_1 ; \bx)\gamma_t(\bx+\btheta_2 ; \bx)\rangle]$.  
Similarly, the case where $i \neq j$ yields a 4-point function
which can be expressed as
%%%%%%%%%%%%%%%%%%%%%%%%%%%%%%%%%%%%%%%%%%%%%%%%%%%%%%%
$ \langle n_g(\bx) n_g(\by)
\gamma_t(\bx+\btheta_1 ; \bx)\gamma_t(\by+\btheta_2 ; \by)\rangle =
\bar{n}_g^2\,\{\langle
\gamma_t(\bx+\btheta_1 ; \bx)\gamma_t(\by+\btheta_2 ; \by)\rangle
+\langle\delta_g(\bx)\delta_g(\by)\gamma_t(\bx+\btheta_1 ; \bx)\gamma_t(\by+\btheta_2 ; \by)\rangle
+ \langle \delta_g(\bx)
\gamma_t(\bx+\btheta_1 ; \bx)\gamma_t(\by+\btheta_2 ; \by)\rangle + \langle
\delta_g(\by) \gamma_t(\bx+\btheta_1 ; \bx)\gamma_t(\by+\btheta_2 ; \by) \rangle
\}.$ Since this work assumes the Gaussianity of all cosmological
fields, we ignore their odd-point functions, as well as the connected
functions. After decomposing the 4-point function in the above
relation in products of 2-point functions, in accordance to Wick's
theorem, we write Eq.~(\ref{eq:cov_shear1}) as
%%%%%%%%%%%%%%%%%%%%%%%%%%%%%%%%%%%%%%%%%%%%%%%%%%%%%%%
\ba
\langle\widehat{\gamma}_t^g(\btheta_1)\widehat{\gamma}_t^g(\btheta_2)\rangle & = & 
\frac{1}{\Omega_s^2}\left\{\frac{1}{\bar{n}_g}\int d^2 x \,\langle \gamma_t(\bx+\btheta_1 ; \bx) 
\gamma_t(\bx+\btheta_2 ; \bx)\rangle + \int d^2x\, d^2y \left[\langle \gamma_t(\bx+\btheta_1 ; \bx)
\gamma_t(\by+\btheta_2 ; \by)\rangle   \right.\right. \nonumber \\
& + &  \left.\left. \langle\delta_g(\bx) \delta_g(\by)\rangle \langle 
\gamma_t(\bx+\btheta_1 ; \bx)\gamma_t(\by + \btheta_2 ; \by)\rangle 
+ \langle\delta_g(\bx)\gamma_t(\bx+\btheta_1 ; \bx)\rangle \langle \delta_g(\by)
\gamma_t(\by+\btheta_2 ; \by)\rangle \right.\right. \nonumber\\
& + & \left.\left. \langle\delta_g(\bx)\gamma_t(\by+\btheta_2 ; \by)\rangle
\langle\delta_g(\by)\gamma_t(\bx+\btheta_1 ; \bx)\rangle \right]\frac{ }{ }\right\}. 
\label{eq:cov_shear2}
\ea 
%%%%%%%%%%%%%%%%%%%%%%%%%%%%%%%%%%%%%%%%%%%%%%%%%%%%%%%
We now proceed to the computation of each of the five terms in
Eq.~(\ref{eq:cov_shear2}), omitting for now the inverse square of the
survey area that multiplies the whole covariance. Our final goal is in
fact the azimuthal average of the covariance in
Eq.~(\ref{eq:cov_shear_def}).
%%%%%%%%%%%%%%%%%%%%%%%%%%%%%%%%%%%%%%%%%%%%%%%%%%%%%%%
We first introduce the shape-noise contribution to the covariance of
the tangential shear at $\btheta_1, \btheta_2$ with respect to the
origins $\btheta_{01}, \btheta_{02}$:
%%%%%%%%%%%%%%%%%%%%%%%%%%%%%%%%%%%%%%%%%%%%%%%%%%%%%%%
\be 
\langle \gamma_t(\btheta_1; \btheta_{01})\gamma_t(\btheta_2;
\btheta_{02})\rangle=\frac{\sigma_{\gamma}^2}{2 \bar{n}_s}
\cos[2(\phi_{\btheta_1-\btheta_{01}}-\phi_{\btheta_2-\btheta_{02}})]\,\delta_D(\btheta_1-\btheta_2),
\label{eq:gamma_t_shape_noise}
\ee
%%%%%%%%%%%%%%%%%%%%%%%%%%%%%%%%%%%%%%%%%%%%%%%%%%%%%%%
where we defined
$\sigma_{\gamma}^2/2\equiv\sigma_{\gamma_1}^2=\sigma_{\gamma_2}^2$ as
the variance per shear component in the measurement of one source
galaxy. We assume the source galaxies to have a mean angular density
of $\bar{n}_s$. In the above we also assumed that the shape noise does
not correlate different positions in the source plane, i.e. we ignore
for example the effect of intrinsic alignments which might introduce
precisely such correlations.
\vspace{0.3cm}
% DO T1 HERE
%%%%%%%%%%%%%%%%%%%%%%%%%%%%%%%%%%%%%%%%%%%%%%%%%%%%%%%

\noindent Starting with the first term, we employ Eq.~(\ref{eq:tan_shear}) to
write:
%%%%%%%%%%%%%%%%%%%%%%%%%%%%%%%%%%%%%%%%%%%%%%%%%%%%%%%
\ba
T_1(\btheta_1, \btheta_2) & \equiv & \frac{1}{\bar{n}_g}\int d^2 x \,\langle \gamma_t(\bx+\btheta_1 ; \bx) 
\gamma_t(\bx+\btheta_2 ; \bx)\rangle  =  \frac{\Omega_s}{\bar{n}_g} \langle 
\gamma_t(\btheta_1 ; {\bf 0})\gamma_t(\btheta_2 ; {\bf 0})\rangle \nonumber \\
& = & \frac{\Omega_s}{\bar{n}_g}\left[\int \frac{d^2l}{(2\pi)^2} \cos[2(\phi_{\btheta_1}-\phi_{\bl})]
\cos[2(\phi_{\btheta_2}-\phi_{\bl})]\mathcal{C}_{\kappa \kappa}(l)+ \frac{\sigma_{\gamma}^2}
    {2 \bar{n}_s}\cos[2(\phi_{\btheta_1}-\phi_{\btheta_2})]\delta_D(\btheta_1-\btheta_2)\right], \nonumber 
\ea
%%%%%%%%%%%%%%%%%%%%%%%%%%%%%%%%%%%%%%%%%%%%%%%%%%%%%%%
Writing that
$\delta_D(\btheta_1-\btheta_2)=\delta_D(\theta_1-\theta_2)\delta_D(\phi_{\btheta_1}-\phi_{\btheta_2})/\theta_1$,
we compute the azimuthal average of the above equation using
Eqs.~(\ref{eq:aa}) and (\ref{eq:Bessel2}). Thus the first term of the
covariance of the stacked shear estimator is:
%%%%%%%%%%%%%%%%%%%%%%%%%%%%%%%%%%%%%%%%%%%%%%%%%%%%%%%
\ba
T_{1, a}(\theta_1, \theta_2) \equiv \int_0^{2\pi}\frac{d\phi_{\btheta_1}}{2\pi}\frac{d\phi_{\btheta_2}}{2\pi}
T_1(\btheta_1, \btheta_2) & = & \frac{\Omega_s}{\bar{n}_g}\left[\int \frac{d^2l}{(2\pi)^2} 
\mathcal{C}_{\kappa \kappa}(l) J_2(l\theta_1) J_2(l\theta_2) + \frac{\sigma_{\gamma}^2}{2 \bar{n}_s} 
\frac{\delta_D(\theta_1-\theta_2)} {2\pi\theta_1} \right] \nonumber \\
&= &\frac{\Omega_s}{\bar{n}_g}\int \frac{d^2l}{(2\pi)^2} \left[\mathcal{C}_{\kappa \kappa}(l) 
+ \frac{\sigma_{\gamma}^2}{2 \bar{n}_s}\right] J_2(l\theta_1) J_2(l\theta_2),
\label{eq:T1aa}
\ea
%%%%%%%%%%%%%%%%%%%%%%%%%%%%%%%%%%%%%%%%%%%%%%%%%%%%%%%
where we have put the shape noise term in a similar form to the
rest of the first term by using the Bessel identity:
%%%%%%%%%%%%%%%%%%%%%%%%%%%%%%%%%%%%%%%%%%%%%%%%%%%%%%%
\[
\frac{\delta_D(\theta_1-\theta_2)}{2\pi\theta_1}=\int \frac{d^2l}{(2\pi)^2}
J_2(l\theta_1)J_2(l\theta_2).
\]
%%%%%%%%%%%%%%%%%%%%%%%%%%%%%%%%%%%%%%%%%%%%%%%%%%%%%%%
% NOW DO T2
%%%%%%%%%%%%%%%%%%%%%%%%%%%%%%%%%%%%%%%%%%%%%%%%%%%%%%%
\vspace{0.2cm} \par\noindent The second term of the covariance matrix
is a constant, equal in fact to $0$.  This can be seen through a
simple change of variables: $\bx'=\bx+\btheta_1$ and
$\by'=\by+\btheta_2$ which leaves the Jacobian unchanged, $d^2x'=d^2x,
\, d^2y'=d^2y$. Relabelling $\bx'=\bx$ and $\by'=\by$, we write:
%%%%%%%%%%%%%%%%%%%%%%%%%%%%%%%%%%%%%%%%%%%%%%%%%%%%%%%
\ba
T_2(\btheta_1, \btheta_2) & \equiv &\int d^2x \,d^2y\,\langle\gamma_t(\bx+\btheta_1 ; \bx)
\gamma_t(\by+\btheta_2 ; \by)\rangle \nonumber \\
& = &\int d^2x \int d^2y \,\langle\gamma_t(\bx ; \bx-\btheta_1)
\gamma_t(\by ; \by-\btheta_2)\rangle=\langle\int d^2x\,\gamma_t(\bx ; \bx-\btheta_1)
\int d^2y\,\gamma_t(\by ; \by-\btheta_2)\rangle=0.
\label{eq:T2}
\ea
%%%%%%%%%%%%%%%%%%%%%%%%%%%%%%%%%%%%%%%%%%%%%%%%%%%%%%%
In the above we have used the ergodic theorem and implicitly assumed
that the survey area is sufficiently large so that the tangential
shear integrated over it behaves like the ensemble average of the
tangential shear, and is therefore equal to 0.
%%%%%%%%%%%%%%%%%%%%%%%%%%%%%%%%%%%%%%%%%%%%%%%%%%%%%%%
%NOW DO T3
%%%%%%%%%%%%%%%%%%%%%%%%%%%%%%%%%%%%%%%%%%%%%%%%%%%%%%%
\vspace{0.2cm} \par\noindent We now turn our attention to the
calculation of the third term of Eq.~(\ref{eq:cov_shear2}), henceforth
denoted as $T_3$.
%%%%%%%%%%%%%%%%%%%%%%%%%%%%%%%%%%%%%%%%%%%%%%%%%%%%%%%
\be
T_3(\btheta_1, \btheta_2)\equiv\int d^2x \,d^2y\, \langle\delta_g(\bx) \delta_g(\by)\rangle 
\langle \gamma_t(\bx+\btheta_1 ; \bx)\gamma_t(\by + \btheta_2 ; \by)\rangle. 
\label{eq:T3def}
\ee
%%%%%%%%%%%%%%%%%%%%%%%%%%%%%%%%%%%%%%%%%%%%%%%%%%%%%%%
Using the invariance under translation due to the homogeneity of the
Universe, we have that $\langle\delta_g(\bx) \delta_g(\by)\rangle =
\langle\delta_g(\bx-\by)\delta_g({\bf 0})\rangle$, and $\langle
\gamma_t(\bx+\btheta_1 ; \bx)\gamma_t(\by + \btheta_2 ; \by)\rangle =
\langle \gamma_t(\bx-\by+\btheta_1 ; \bx-\by) \gamma_t(\btheta_2 ;{\bf
  0})\rangle$. Changing variables $\bx'=\bx-\by$ and relabelling
$\bx'=\bx$, we write
%%%%%%%%%%%%%%%%%%%%%%%%%%%%%%%%%%%%%%%%%%%%%%%%%%%%%%%
\be
T_3(\btheta_1, \btheta_2)=\Omega_s\int d^2 x\,\langle\delta_g(\bx)\delta_g({\bf 0})
\rangle \langle \gamma_t(\bx+\btheta_1 ; \bx)\gamma_t(\btheta_2 ; {\bf 0})\rangle 
\label{eq:T3def2}
\ee
%%%%%%%%%%%%%%%%%%%%%%%%%%%%%%%%%%%%%%%%%%%%%%%%%%%%%%%
Note that $T_3$ also has a contribution from shape noise, which we
address below. We shall evaluate Eq.~(\ref{eq:T3def2}) in Fourier
space. The Fourier transforms of the shear and galaxy density
fluctuations are written as four integrals over the wave vectors which
can be reduced to two integrals, by using the definitions of the
convergence and galaxy power spectra in Eq.~(\ref{eq:power_defs}):
%%%%%%%%%%%%%%%%%%%%%%%%%%%%%%%%%%%%%%%%%%%%%%%%%%%%%%%
\ba
T_3(\btheta_1, \btheta_2) & = & \Omega_s \left\{\int \frac{d^2l_1}{(2\pi)^2}\frac{d^2l_2}{(2\pi)^2} 
\mathcal{C}_{g g}(l_1) \,\mathcal{C}_{\kappa\kappa}(l_2)\,e^{i\,\bl_2\cdot(\btheta_1-\btheta_2)} 
\cos[2(\phi_{\btheta_1}-\phi_{\bl_2})] \cos[2(\phi_{\btheta_2}-\phi_{\bl_2})]\int d^2x\, 
e^{i \,\bx\cdot(\bl_1+\bl_2)} \right.\nonumber \\
& + & \left.\frac{\sigma_{\gamma}^2}{2 \bar{n}_s}\,\int d^2x\,\langle\delta_g(\bx)\delta_g({\bf 0})\rangle\,
\cos[2(\phi_{\btheta_1}-\phi_{\btheta_2})]\,\delta_D(\bx+\btheta_1-\btheta_2) \right\} \nonumber \\
& = & \Omega_s \left\{ \int \frac{d^2l}{(2\pi)^2} \mathcal{C}_{g g}(l) \,\mathcal{C}_{\kappa\kappa}(l)
\,e^{i\,\bl\cdot(\btheta_1-\btheta_2)} \cos[2(\phi_{\btheta_1}-\phi_{\bl})] \cos[2(\phi_{\btheta_2}-\phi_{\bl})] 
\right. \nonumber \\
& + & \left.\frac{\sigma_{\gamma}^2}{2 \bar{n}_s} \int \frac{d^2 l}{(2\pi)^2} \mathcal{C}_{g g}(l)
 e^{i \, \bl\cdot(\btheta_2-\btheta_1)} \cos[2(\phi_{\btheta_1}-\phi_{\btheta_2})] \right\}.
\label{eq:T3}
\ea
%%%%%%%%%%%%%%%%%%%%%%%%%%%%%%%%%%%%%%%%%%%%%%%%%%%%%%%
We now calculate the azimuthal average of this term, using just as
before Eq.~(\ref{eq:Bessel2}). For the shape noise term we also change
variables to $\phi'_{\btheta_{1,2}}=\phi_{\btheta_{1,2}}-\phi_{\bl}$.
%%%%%%%%%%%%%%%%%%%%%%%%%%%%%%%%%%%%%%%%%%%%%%%%%%%%%%%
\ba
T_{3, a}(\theta_1, \theta_2) & \equiv & \int_0^{2\pi}\frac{d\phi_{\btheta_1}}{2\pi}
\frac{d\phi_{\btheta_2}}{2\pi} T_3(\btheta_1, \btheta_2) = \Omega_s \left\{\int \frac{d^2l}{(2\pi)^2} 
\mathcal{C}_{g g}(l) \,\mathcal{C}_{\kappa\kappa}(l) J_2(l\theta_1) J_2(l \theta_2) \right.\nonumber\\
& + & \left.\frac{\sigma_{\gamma}^2}{2 \bar{n}_s} \int \frac{d^2 l}{(2\pi)^2} \mathcal{C}_{g g}(l)
\int \frac{d\phi'_{\btheta_1}}{2\pi} \frac{d\phi'_{\btheta_2}}{2\pi}\,e^{i\,[l\theta_2\cos(\phi'_{\btheta_2})
-l\theta_1 \cos(\phi'_{\btheta_1})]} \cos[2(\phi'_{\btheta_1}-\phi'_{\btheta_2})] \right\} \nonumber \\
& = & \Omega_s\left\{ \int \frac{d^2l}{(2\pi)^2} \mathcal{C}_{g g}(l) \,\left[ \mathcal{C}_{\kappa\kappa}(l) 
+ \frac{\sigma_{\gamma}^2}{2 \bar{n}_s} \right] J_2(l\theta_1) J_2(l \theta_2) \right\}.
\label{eq:T3aa}
\ea
%%%%%%%%%%%%%%%%%%%%%%%%%%%%%%%%%%%%%%%%%%%%%%%%%%%%%%%
% NOW DO T4
%%%%%%%%%%%%%%%%%%%%%%%%%%%%%%%%%%%%%%%%%%%%%%%%%%%%%%%
\vspace{0.2cm} \par\noindent Moving on to the fourth term of
Eq.~(\ref{eq:cov_shear2}), we take advantage of the homogeneity of the
Universe and of Eq.~(\ref{eq:shear_est_ave0}) to write:
%%%%%%%%%%%%%%%%%%%%%%%%%%%%%%%%%%%%%%%%%%%%%%%%%%%%%%%
\be
T_4(\btheta_1, \btheta_2) \equiv \int d^2x \,d^2y\,\langle\delta_g(\bx)\gamma_t(\bx+\btheta_1 ; \bx)
\rangle \langle \delta_g(\by)
\gamma_t(\by+\btheta_2 ; \by)\rangle = \Omega_s^2\,\langle \delta_g({\bf 0})\gamma_t(\btheta_1 ; {\bf 0})
\rangle\langle \delta_g({\bf 0})\gamma_t(\btheta_2 ; {\bf 0})\rangle=\Omega_s^2\,\langle
\widehat{\gamma}_t^g(\btheta_1)\rangle \langle\widehat{\gamma}_t^g(\btheta_2)\rangle
\label{eq:T4}
\ee
%%%%%%%%%%%%%%%%%%%%%%%%%%%%%%%%%%%%%%%%%%%%%%%%%%%%%%%
Note how this fourth term exactly cancels the second one of the
covariance in Eq.~(\ref{eq:cov_shear_def}), once we remember the
factor of $1/\Omega_s^2$ that we deliberately left out. 
%%%%%%%%%%%%%%%%%%%%%%%%%%%%%%%%%%%%%%%%%%%%%%%%%%%%%%%
% NOW DO T5
%%%%%%%%%%%%%%%%%%%%%%%%%%%%%%%%%%%%%%%%%%%%%%%%%%%%%%%
\vspace{0.2cm} \par\noindent Proceeding to the fifth term, we once
again use the invariance under translation due to the inhomogeneity of
the Universe to write: $\langle\delta_g(\bx)\gamma_t(\by+\btheta_2 ;
\by)\rangle = \langle\delta_g(\bx-\by)\gamma_t(\btheta_2; {\bf
  0})\rangle$, and $\langle\delta_g(\by)\gamma_t(\bx+\btheta_1 ;
\bx)\rangle=\langle\delta_g(\by-\bx)\gamma_t(\btheta_1; {\bf
  0})\rangle$. Making the change of variables $\bx'=\bx-\by$ to
eliminate one of the surface-area integrals on the right-hand side,
and relabelling $\bx'=\bx$, we obtain:
%%%%%%%%%%%%%%%%%%%%%%%%%%%%%%%%%%%%%%%%%%%%%%%%%%%%%%%
\be
T_5(\btheta_1, \btheta_2) \equiv \int d^2x\, d^2y\, \langle\delta_g(\bx)
\gamma_t(\by+\btheta_2 ; \by)\rangle\,\langle\delta_g(\by)\gamma_t(\bx+\btheta_1 ; \bx)\rangle 
=  \Omega_s \int d^2x\,\langle\delta_g(\bx)\gamma_t(\btheta_2 ; {\bf 0})\rangle 
\langle\delta_g(-\bx)\gamma_t(\btheta_1 ; {\bf 0})\rangle
\label{eq:T5def}.
\ee
%%%%%%%%%%%%%%%%%%%%%%%%%%%%%%%%%%%%%%%%%%%%%%%%%%%%%%%
Using Eqs.~(\ref{eq:tan_shear}),~(\ref{eq:power_defs}), we can further write
%%%%%%%%%%%%%%%%%%%%%%%%%%%%%%%%%%%%%%%%%%%%%%%%%%%%%%%
\be
T_5(\btheta_1, \btheta_2) = \Omega_s \int \frac{d^2l}{(2\pi)^2}
\mathcal{C}_{g\kappa}^2(l) \cos[2(\phi_{\btheta_1}-\phi_{\bl})]
\cos[2(\phi_{\btheta_2}-\phi_{\bl})]\,e^{i\,\bl \cdot (\btheta_1+\btheta_2)}.
\label{eq:T5}
\ee
%%%%%%%%%%%%%%%%%%%%%%%%%%%%%%%%%%%%%%%%%%%%%%%%%%%%%%%
The azimuthal average of this expression can be obtained with the help
of Eq.~(\ref{eq:Bessel2}):
%%%%%%%%%%%%%%%%%%%%%%%%%%%%%%%%%%%%%%%%%%%%%%%%%%%%%%%
\be
T_{5, a}(\theta_1, \theta_2) \equiv \int_0^{2\pi}\frac{d\phi_{\btheta_1}}{2\pi}\frac{d\phi_{\btheta_2}}{2\pi}
T_5(\btheta_1, \btheta_2) = \Omega_s \int \frac{d^2l}{(2\pi)^2}
\mathcal{C}_{g\kappa}^2(l) J_2(l\theta_1) J_2(l\theta_2).
\label{eq:T5aa}
\ee
%%%%%%%%%%%%%%%%%%%%%%%%%%%%%%%%%%%%%%%%%%%%%%%%%%%%%%
%%%%%%%%%%%%%%%%%%%%%%%%%%%%%%%%%%%%%%%%%%%%%%%%%%%%%%%
\par\noindent Finally, putting together all the terms in
Eqs.~(\ref{eq:T1aa}),~(\ref{eq:T2},~(\ref{eq:T3aa},~(\ref{eq:T4}),
and~(\ref{eq:T5aa}), as well as the factor $\frac{1}{\Omega_s^2}$ that
we left out earlier, we reexpress Eq.~(\ref{eq:cov_shear2}) as
%%%%%%%%%%%%%%%%%%%%%%%%%%%%%%%%%%%%%%%%%%%%%%%%%%%%%%%
\be
{\rm Cov}[\widehat{\gamma}_t^{g}](\theta_1, \theta_2) = \frac{1}{\Omega_s}
\int_0^{\infty} \frac{d^2 l}{(2\pi)^2}\,  J_2(l\theta_1) J_2(l\theta_2) \left\{
\left[\mathcal{C}_{\kappa \kappa}(l) + \frac{\sigma_{\gamma}^2}{2 \bar{n}_s} \right] 
\left[\mathcal{C}_{\rm gg}(l) + \frac{1}{\bar{n}_g}\right] 
 +  \mathcal{C}_{g\kappa}^2(l) \right\}. 
\label{eq:cov_shear3}
\ee
%%%%%%%%%%%%%%%%%%%%%%%%%%%%%%%%%%%%%%%%%%%%%%%%%%%%%%%
This equation concludes the theoretical predictions for the stacked
tangential shear estimator and its covariance.
%%%%%%%%%%%%%%%%%%%%%%%%%%%%%%%%%%%%%%%%%%%%%%%%%%%%%%%
%%%%%%%%%%%%%%%%%%%%%%%%%%%%%%%%%%%%%%%%%%%%%%%%%%%%%%%
%%%%%%%%%%%%%%%%%%%%%%%%%%%%%%%%%%%%%%%%%%%%%%%%%%%%%%%
\subsection{Derivation of the covariance matrix of the 3D-galaxy 
correlation function}
\label{app:gc-cov}
%%%%%%%%%%%%%%%%%%%%%%%%%%%%%%%%%%%%%%%%%%%%%%%%%%%%%%%
%%%%%%%%%%%%%%%%%%%%%%%%%%%%%%%%%%%%%%%%%%%%%%%%%%%%%%%
%%%%%%%%%%%%%%%%%%%%%%%%%%%%%%%%%%%%%%%%%%%%%%%%%%%%%%%
We define a general estimator for the galaxy overdensity field with
respect to a random galaxy catalogue as:
%%%%%%%%%%%%%%%%%%%%%%%%%%%%%%%%%%%%%%%%%%%%%%%%%%%%%%%
\be
\widehat{F_g}(\bx)\equiv\frac{w(\bx)}{\sqrt{A}}[n_g(\bx)-\alpha n_R(\bx)],
\label{eq:def_est_ngal}
\ee
%%%%%%%%%%%%%%%%%%%%%%%%%%%%%%%%%%%%%%%%%%%%%%%%%%%%%%%
where $n_g(\bx)$ is the galaxy number density field; $w(\bx)$ is a
weight function; $n_R(\bx)$ is the number density of a mock galaxy
catalogue, used to compute any excess correlation in our data set;
$\alpha$ quantifies how dense the random catalogue is compared to the
galaxy data, i.e. the smaller $\alpha$, the better the estimator:
%%%%%%%%%%%%%%%%%%%%%%%%%%%%%%%%%%%%%%%%%%%%%%%%%%%%%%% 
\[
\alpha=\frac{\langle n_g(\bx)\rangle}{\langle n_R(\bx)\rangle}
\approx\frac{\int d^3 x\,\bar{n}_g(\bx)}{\int d^3x\,\bar{n}_R(\bx)}.
\]
%%%%%%%%%%%%%%%%%%%%%%%%%%%%%%%%%%%%%%%%%%%%%%%%%%%%%%%
In the above, the mean density of galaxies can vary across the survey,
if the survey is a small volume-limited one or simply flux-limited.
Note that the above definition of $\alpha$ insures that the density
fluctuation estimator is unbiased, i.e. $\langle
\widehat{F_g}(\bx)\rangle=0$.
Finally, the normalization constant is:
%%%%%%%%%%%%%%%%%%%%%%%%%%%%%%%%%%%%%%%%%%%%%%%%%%%%%%%
\be
A\equiv\int d^3x\, \bar{n}_g^2(\bx)w^2(\bx).
\ee
%%%%%%%%%%%%%%%%%%%%%%%%%%%%%%%%%%%%%%%%%%%%%%%%%%%%%%%
This estimator for the galaxy overdensity field has been reviewed in
the recent work of \cite{SmithMarian2014}, who showed that the galaxy
correlation function is related to the covariance of the galaxy
overdensity field estimator:
%%%%%%%%%%%%%%%%%%%%%%%%%%%%%%%%%%%%%%%%%%%%%%%%%%%%%%%
\be
\langle \widehat{F_g}(\bx_1)\widehat{F_g}(\bx_2)\rangle = \frac{w(\bx_1)w(\bx_2)
\bar{n}_g(\bx_1)\bar{n}_g(\bx_2)}{A}\left[\xi_{\rm gg}(\bx_1, \bx_2)+\frac{(1+\alpha)}
{\bar{n}_g(\bx_2)}\delta_D(\bx_1-\bx_2)\right].
\label{eq:est_xigg_v0}
\ee
%%%%%%%%%%%%%%%%%%%%%%%%%%%%%%%%%%%%%%%%%%%%%%%%%%%%%%%
In the following, we shall assume that: i) our survey is
volume-limited and large enough so that the galaxy number density
field has constant mean density; ii) the random catalogue also has
constant mean density; iii) the weights are constant and equal to
$1$. Whilst this means that our estimator is sub-optimal in terms of
maximizing the signal-to-noise for the galaxy correlation function, it
is perfectly adequate for the purpose of this work. Under these
assumptions, we define the estimator for the galaxy correlation
function as:
%%%%%%%%%%%%%%%%%%%%%%%%%%%%%%%%%%%%%%%%%%%%%%%%%%%%%%%
\be 
\widehat{\xi}_{\rm gg}(\bx)=\int d^3 y\, \widehat{F_g}(\by)
\widehat{F_g}(\by+\bx), \hspace{0.5cm}\mbox{with}\hspace{0.1cm} \bx \neq {\bf 0} 
\label{eq:est_xigg_v1}
\ee
%%%%%%%%%%%%%%%%%%%%%%%%%%%%%%%%%%%%%%%%%%%%%%%%%%%%%%%
where we ignore the 0-lag correlation, affected by a shot noise term
as seen in Eq.~(\ref{eq:est_xigg_v0}). The reason for this is that our
measurements from numerical simulations will not contain the 0-lag
contribution. The normalization constant also has the simple
expression $A=\bar{n}_g^2 V_s$, where $V_s$ is the survey volume. The
estimator is unitless, and in the limit that the survey volume is
large, it is also unbiased. 
Its covariance is given by:
%%%%%%%%%%%%%%%%%%%%%%%%%%%%%%%%%%%%%%%%%%%%%%%%%%%%%%%
\be 
{\rm Cov}[\widehat{\xi}_{\rm gg}](\br_1, \br_2)=\int d^3x\, d^3y\,\left[\langle \widehat{F_g}(\bx)
\widehat{F_g}(\bx+\br_1) \widehat{F_g}(\by) \widehat{F_g}(\by+\br_2)\rangle
-\langle\widehat{F_g}(\bx)\widehat{F_g}(\bx+\br_1)\rangle
\langle\widehat{F_g}(\by)\widehat{F_g}(\by+\br_2)\rangle \right]
\label{eq:def_cov_xi}
\ee
%%%%%%%%%%%%%%%%%%%%%%%%%%%%%%%%%%%%%%%%%%%%%%%%%%%%%%%
To shorten the notation, we shall denote any function $ g(\bx_1)\equiv
g_1$, $g(\bx_1, \bx_2)\equiv g_{12}$, $g(\bx_1, \bx_2, \bx_3)\equiv
g_{123}$ and so on. We shall also temporarily drop the subscript $g$
from `galaxy' in the correlation functions. The above covariance
requires us to compute the 4-point function $\langle \widehat{F}_1
\widehat{F}_2 \widehat{F}_3 \widehat{F}_4\rangle$. Since the details
for this calculation are fully explained in \cite{SmithMarian2014}, here we
just write directly the result, retaining the weights and varying mean
densities for the moment in order to preserve the generality of the
calculation. For the same reason, we also retain some terms
originating in the 0-lag correlations, and drop them later.
%%%%%%%%%%%%%%%%%%%%%%%%%%%%%%%%%%%%%%%%%%%%%%%%%%%%%%%
\ba
\langle \widehat{F}_1 \widehat{F}_2 \widehat{F}_3
\widehat{F}_4\rangle & = &\frac{\prod_{i=1}^{4}(w_i\bar{n}_i)}{A^2}\left\{\xi_{1234}
+\frac{\xi_{123}}{\bar{n}_4}(\delta_D^{14}+\delta_D^{24}+\delta_D^{34})+\frac{\xi_{124}}
{\bar{n}_3}(\delta_D^{13}+\delta_D^{23})+\frac{\xi_{134}}{\bar{n}_2}\delta_D^{12}
+\xi_{12}\xi_{34}+\xi_{13}\xi_{24}+\xi_{14}\xi_{23} \right.\nonumber \\
& + & \left. \frac{\xi_{12}}{\bar{n}_3\bar{n}_4}
(\delta_D^{13}\delta_D^{24} + \delta_D^{14}\delta_D^{23} +\delta_D^{24}\delta_D^{23} +
\delta_D^{13}\delta_D^{14}) + \frac{\xi_{13}}{\bar{n}_2\bar{n}_4}(\delta_D^{12}\delta_D^{14} 
+ \delta_D^{12}\delta_D^{34}) + \frac{\xi_{14}}{\bar{n}_2\bar{n}_3}\delta_D^{12}\delta_D^{13}
+ \frac{1+\alpha}{\bar{n}_2}\xi_{34}\delta_D^{12} \right. \nonumber \\ 
& + & \left. \frac{1+\alpha}{\bar{n}_3}(\xi_{14}\delta_D^{23}+\xi_{24}\delta_D^{13}) 
+ \frac{1+\alpha}{\bar{n}_4}(\xi_{12}\delta_D^{34} + \xi_{13}\delta_D^{24} + \xi_{23}\delta_D^{14}) 
+ \frac{(1+\alpha)^2}{\bar{n}_3\bar{n}_4}(\delta_D^{13}\delta_D^{24}+\delta_D^{14}\delta_D^{23})
\right. \nonumber \\ 
& + & \left. \frac{(1+\alpha)^2}{\bar{n}_2\bar{n}_4}\delta_D^{12}\delta_D^{34}
 + \frac{1+\alpha^3}{\bar{n}_2\bar{n}_3\bar{n}_4}\delta_D^{12}\delta_D^{13}\delta_D^{14}
\right\}.
\label{eq:4p}
\ea
%%%%%%%%%%%%%%%%%%%%%%%%%%%%%%%%%%%%%%%%%%%%%%%%%%%%%%%
We also need the ingredient
$\langle\widehat{F}_1\widehat{F}_2\rangle\langle\widehat{F}_3\widehat{F}_4\rangle$
which is shown in \cite{SmithMarian2014} to be
%%%%%%%%%%%%%%%%%%%%%%%%%%%%%%%%%%%%%%%%%%%%%%%%%%%%%%%
\be
\langle\widehat{F}_1\widehat{F}_2\rangle\langle\widehat{F}_3\widehat{F}_4\rangle=
\frac{\prod_{i=1}^{4}(w_i\bar{n}_i)}{A^2}\left[\xi_{12}\xi_{34} + \frac{1+\alpha}{\bar{n}_2}
\xi_{34}\delta_D^{12} + \frac{1+\alpha}{\bar{n}_4}\xi_{12}\delta_D^{34} + \frac{(1+\alpha)^2}
{\bar{n}_2\bar{n}_4}\delta_D^{12}\delta_D^{34}\right].
\label{eq:2p2p}
\ee
%%%%%%%%%%%%%%%%%%%%%%%%%%%%%%%%%%%%%%%%%%%%%%%%%%%%%%%
Putting together Eqs.~(\ref{eq:4p}) and~(\ref{eq:2p2p}) and rearanging the terms, we write
%%%%%%%%%%%%%%%%%%%%%%%%%%%%%%%%%%%%%%%%%%%%%%%%%%%%%%%
\ba
\langle \widehat{F}_1 \widehat{F}_2 \widehat{F}_3
\widehat{F}_4\rangle - \langle\widehat{F}_1\widehat{F}_2\rangle
\langle\widehat{F}_3\widehat{F}_4\rangle &=& \frac{\prod_{i=1}^{4}(w_i\bar{n}_i)}{A^2}
\left\{\left(\xi_{1234}
+\frac{\xi_{123}}{\bar{n}_4}(\delta_D^{14}+\delta_D^{24}+\delta_D^{34})+\frac{\xi_{124}}
{\bar{n}_3}(\delta_D^{13}+\delta_D^{23})+\frac{\xi_{134}}{\bar{n}_2}\delta_D^{12} \right)
+ \frac{\xi_{14}}{\bar{n}_2\bar{n}_3}\delta_D^{12} \delta_D^{13} \right.\nonumber\\
& + &\left. \left[\xi_{13}+\frac{1+\alpha}{\bar{n}_3}\delta_D^{13}\right]
\left[\xi_{24}+\frac{1+\alpha} {\bar{n}_4}\delta_D^{24}\right]
+\left[\xi_{14}+\frac{1+\alpha}{\bar{n}_4}\delta_D^{14}\right]
\left[\xi_{23}+\frac{1+\alpha}{\bar{n}_3}\delta_D^{23}\right]  \right.\nonumber\\
& + & \left. \frac{\xi_{12}}{\bar{n}_3\bar{n}_4} (\delta_D^{13}\delta_D^{24} + 
\delta_D^{14}\delta_D^{23} +\delta_D^{24}\delta_D^{23} + \delta_D^{13}\delta_D^{14}) 
+ \frac{\xi_{13}\delta_D^{12}}{\bar{n}_2\bar{n}_4}(\delta_D^{14}+\delta_D^{34}) 
+ \frac{1+\alpha^3}{\bar{n}_2\bar{n}_3\bar{n}_4}\delta_D^{12}\delta_D^{13}\delta_D^{14}
\right\}
\label{eq:cov_gg_whole}
\ea
%%%%%%%%%%%%%%%%%%%%%%%%%%%%%%%%%%%%%%%%%%%%%%%%%%%%%%%
Just as in the case of the stacked tangential shear of galaxies from
the previous section, here too we work under the assumption of
Gaussianity of the galaxy density field. In the above equation, the
first term in round brackets on the right-hand side vanishes under this
assumption, i.e. we ignore the connected 4-point correlation function,
as well as the 3-point functions of the galaxy density field. We shall
also ignore all terms with contributions from the 0-lag correlations,
e.g. terms containing $\delta_D^{12}$ and $\delta_D^{34}$, and also
contributions from the bin containing the value $\br={\bf 0}$,
e.g. the terms with the products $\delta_D^{24} \delta_D^{23}$ and
$\delta_D^{13} \delta_D^{14}$. The reason for the latter is that the
binning of the simulation data starts from values larger than 0.

To write down the covariance for the estimator defined by
Eq.~(\ref{eq:est_xigg_v1}), we set the ${\bar n}_i={\bar n}_g,\,
w_i=1, \:\: (i=1, 2, 3, 4)$, and we combine
Eqs.~(\ref{eq:def_cov_xi}),~(\ref{eq:cov_gg_whole}) discarding all the
above-mentioned terms:
%%%%%%%%%%%%%%%%%%%%%%%%%%%%%%%%%%%%%%%%%%%%%%%%%%%%%%%
\ba
{\rm Cov}[\widehat{\xi}_{\rm gg}](\br_1, \br_2)&=&\int d^3x\, d^3y\,\left\{\left[\xi_{\rm gg}(\bx, \by)
+\frac{1+\alpha}{\bar{n}_g}\delta_D(\bx-\by)\right]\left[\xi_{\rm gg}(\bx+\br_1, \by+\br_2)+
\frac{1+\alpha}{\bar{n}_g}\delta_D(\bx+\br_1-\by-\br_2)\right]\right.\nonumber \\
& + & \left.\left[\xi_{\rm gg}(\bx, \by+\br_2)+\frac{1+\alpha}{\bar{n}_g}\delta_D(\bx-\by-\br_2)
\right]\left[\xi_{\rm gg}(\bx+\br_1, \by)+\frac{1+\alpha}{\bar{n}_g}\delta_D(\bx+\br_1-\by)\right]
 \right. \nonumber \\
& + & \left. \frac{\xi_{\rm gg}(\br_1)}{\bar{n}_g^2} \left[\delta_D(\bx-\by)
\delta_D(\bx+\br_1-\by-\br_2) +\delta_D(\bx-\by-\br_2)
\delta_D(\bx+\br_1-\by)\right]\frac{}{}\right\},
\label{eq:cov_xi_gg_rough}
\ea
%%%%%%%%%%%%%%%%%%%%%%%%%%%%%%%%%%%%%%%%%%%%%%%%%%%%%%%
where we have used the homogeneity of the Universe to write
$\xi_{\rm gg}(\br_1, \br_2)=\xi_{\rm gg}(\br_1-\br_2)$. A little further
manipulation of the above equation gives us the expression for the
covariance of the galaxy correlation function estimator in
Eq.~(\ref{eq:est_xigg_v1})
%%%%%%%%%%%%%%%%%%%%%%%%%%%%%%%%%%%%%%%%%%%%%%%%%%%%%%%
\ba
{\rm Cov}[\widehat{\xi}_{\rm gg}](\br_1, \br_2)& = & \frac{1}{V_s} \left\{
\int d^3x \, \left[\xi_{\rm gg}(\bx)\xi_{\rm gg}(\bx+\br_1-\br_2) + \xi_{\rm gg}(\bx+\br_1)
\xi_{\rm gg}(\bx-\br_2)\right] + \frac{2(1+\alpha)}{\bar{n}_g}\left[\xi_{\rm gg}
  (\br_1+\br_2)+\xi_{\rm gg}(\br_1-\br_2)\right] \right. \nonumber \\ 
& + & \left. \frac{1}{\bar{n}_g^2}\left[\xi_{\rm gg}(\br_1)\left[\delta_D(\br_1-\br_2)+
    \delta_D(\br_1+\br_2) \right] +(1+\alpha)^2\left[\delta_D(\br_1-\br_2) 
+ \delta_D(\br_1+\br_2)\right]\right]\frac{}{}\right\}.
\label{eq:cov_xi_gg}
\ea
%%%%%%%%%%%%%%%%%%%%%%%%%%%%%%%%%%%%%%%%%%%%%%%%%%%%%%%
Note that this expression is in accord with the work of
\cite{Smith2009}, with the lag-0 and bin-0 contributions excluded.

\par\noindent Just like in the previous section on stacked tangential
shear, we are actually interested in the azimuthally-averaged
covariance. To obtain the azimuthal average, we shall work in
cylindrical coordinates, suitable to the projected correlation
functions discussed in subsection \S\ref{SII.3}. Writing
$\delta_D(\br_1-\br_2)=\frac{\delta_D(R_1-R_2)}{R_1}\delta_D(\chi_1-\chi_2)
\delta_D(\phi_{\bR_1}-\phi_{\bR_2})$ and
$\delta_D(\br_1+\br_2)=\frac{\delta_D(R_1-R_2)}{R_1}
\delta_D(\chi_1-\chi_2)\delta_D(\phi_{\bR_1}-\phi_{\bR_2}+\pi)$, we
compute
%%%%%%%%%%%%%%%%%%%%%%%%%%%%%%%%%%%%%%%%%%%%%%%%%%%%%%%
\ba
{\rm Cov}[\widehat{\xi}_{\rm gg}](r_1, r_2) & = &\int_0^{2\pi} \frac{d\phi_{\bR_1}}{2\pi}
\int_0^{2\pi} \frac{d\phi_{\bR_2}}{2\pi}{\rm Cov}[\widehat{\xi}_{\rm gg}](\br_1, \br_2) \nonumber \\
& = & \frac{1}{V_s}\left\{\int_0^{\infty} \frac{dk_\perp}{(2\pi)^2}\,k_\perp J_0(k_\perp R_1)J_0(k_\perp R_2)
\int_{-\infty}^{\infty} dk_z \mathcal{P}_{\rm gg}(k)\left[\mathcal{P}_{\rm gg}(k)+\frac{2(1+\alpha)}{\bar{n}_g}
\right]\left[e^{i\,k_z(\chi_1-\chi_2)} + e^{i\,k_z(\chi_1+\chi_2)}\right]\right.\nonumber \\
& + & \left.\frac{2}{\bar{n}_g^2}\left[\left[\xi_{\rm gg}(r_1)+ 
(1+\alpha)^2\right]\delta_D(\chi_1-\chi_2)
\frac{\delta_D(R_1-R_2)}{2\pi R_1} \right]\right\}
\label{eq:cov_xigg_AA}
\ea
%%%%%%%%%%%%%%%%%%%%%%%%%%%%%%%%%%%%%%%%%%%%%%%%%%%%%%%
where we have used $k=\sqrt{k^2_\perp + k_z^2}$,
$r_i=\sqrt{R_i^2+\chi_i^2},\:i=1,2$ for more compact notation, and
Eq.~(\ref{eq:Pgg}) to express $\xi_{\rm gg}$ in terms of
$\mathcal{P}_{\rm gg}$. Eq.~(\ref{eq:cov_xigg_AA}) represents the
azimuthally-averaged covariance matrix of the galaxy correlation
function estimator introduced in Eq.~(\ref{eq:est_xigg_v1}).

%%%%%%%%%%%%%%%%%%%%%%%%%%%%%%%%%%%%%%%%%%%%%%%%%%%%%%%
%%%%%%%%%%%%%%%%%%%%%%%%%%%%%%%%%%%%%%%%%%%%%%%%%%%%%%%
%%%%%%%%%%%%%%%%%%%%%%%%%%%%%%%%%%%%%%%%%%%%%%%%%%%%%%%
\subsection{Cross-covariance matrix of stacked tangential 
shear and the projected galaxy correlation function}
\label{app:ggl-gc-cov}
%%%%%%%%%%%%%%%%%%%%%%%%%%%%%%%%%%%%%%%%%%%%%%%%%%%%%%%
%%%%%%%%%%%%%%%%%%%%%%%%%%%%%%%%%%%%%%%%%%%%%%%%%%%%%%%
We shall compute the cross-covariance of the estimators for the
angular correlation function of galaxies and the stacked tangential
shear. We take this approach because the angular correlation function
in a thin redshift shell -- which is our assumption for the lens
population throughout this study -- is in fact equal to the projected
correlation function.

Eq.~(\ref{eq:def_cov_cross}) defines the cross-covariance, which can
be further written as:
%%%%%%%%%%%%%%%%%%%%%%%%%%%%%%%%%%%%%%%%%%%%%%%%%%%%%%%
\be
{\rm Cov}[\widehat{w}_{\rm gg}(\btheta_1), \widehat{\gamma}^g_t(\btheta_2)] = 
\frac{1}{N_g} \int_{\Omega_s} d^2x\, d^2y \left\{\langle \widehat{F}_g(\bx)
\widehat{F}_g(\bx+\btheta_1)n_g(\by) \gamma_t(\btheta_2+\by;\by)\rangle
-\langle \widehat{F}_g(\bx)\widehat{F}_g(\bx+\btheta_1)\rangle
\, \langle n_g(\by)\gamma_t(\btheta_2+\by;\by)\rangle \right\},
\label{eq:cov_cross_v0}
\ee
%%%%%%%%%%%%%%%%%%%%%%%%%%%%%%%%%%%%%%%%%%%%%%%%%%%%%%%
where $\btheta_1$ and $\btheta_2$ are two angular position vectors,
and the estimator for the galaxy overdensity field is defined by
Eq.~(\ref{eq:def_est_ngal}), only in this case for 2D vectors.  The
normalization constant is defined just as before $A_{2D}=\int d^2x\,
\bar{n}_g^2(\bx)w^2(\bx)$. Here we shall assume unitary weights for
which $A_{2D}=\Omega_s\bar{n}_g^2=N_g^2/\Omega_s$.

The first factor in the curly brackets on the right-hand-side of the
above equation can be written:
%%%%%%%%%%%%%%%%%%%%%%%%%%%%%%%%%%%%%%%%%%%%%%%%%%%%%%%
\ba
\langle \widehat{F}_1 \widehat{F}_2 n_3 \gamma_{t 4;3}\rangle
& \equiv & \langle \widehat{F}_g(\bz_1) \widehat{F}_g(\bz_2) 
n_g(\bz_3) \gamma_t(\bz_4+\bz_3; \bz_3) \rangle  \nn \\
& = & \frac{1}{A_{2D}} \left\{\langle n_{g1} n_{g2} n_{g3} \gamma_{t 4;3}\rangle -\alpha 
\left[ \langle n_{g1} n_{R2} n_{g3} \gamma_{t 4;3}\rangle + \langle n_{R1} n_{g2} n_{g3} 
\gamma_{t 4;3}\rangle \right] + \alpha^2 \langle n_{R1} n_{R2} n_{g3} \gamma_{t 4;3}
\rangle \right\}\label{eq:RH1_def}
\ea
%%%%%%%%%%%%%%%%%%%%%%%%%%%%%%%%%%%%%%%%%%%%%%%%%%%%%%%
We compute these terms using the expansion of the galaxy number
density as a sum of Dirac delta functions at the positions of the
galaxies, e.g. Eq.~(\ref{eq:gal_den}). Starting with the first term,
%%%%%%%%%%%%%%%%%%%%%%%%%%%%%%%%%%%%%%%%%%%%%%%%%%%%%%%
\ba
\langle n_{g1} n_{g2} n_{g3} \gamma_{t 4;3}\rangle & = & \sum_{i,j,k} \langle \delta_D(\bz_1-\bx_i)
\delta_D(\bz_2-\bx_j)\delta_D(\bz_3-\bx_k)\gamma_t(\bz_4; \bz_3)\rangle \nn \\
& \equiv & T_{i\neq j \neq k} + T_{i=j\neq k} + T_{i=k \neq j} + T_{j=k \neq i} + T_{i=j=k}
\label{eq:T1_terms}
\ea
%%%%%%%%%%%%%%%%%%%%%%%%%%%%%%%%%%%%%%%%%%%%%%%%%%%%%%%
Considering now the term with $i\neq j \neq k$, we use the fluctuation
in the number density of galaxies
$n_g(\bz)=\bar{n}_g[1+\delta_g(\bz)]$ just like in sections
\S\ref{SII.1}. Dropping the subscript `g' for simplicity, we write:
%%%%%%%%%%%%%%%%%%%%%%%%%%%%%%%%%%%%%%%%%%%%%%%%%%%%%%%
\ba
T_{i\neq j \neq k} & = &\bar{n}_g^3\langle(1+\delta_1)(1+\delta_2)(1+\delta_3)\gamma_{t 4;3}
\rangle \nn \\
& = & \bar{n}_g^3 \left[\langle\delta_1 \delta_2 \delta_3 \gamma_{t 4;3}\rangle + 
\langle \delta_1 \delta_2\gamma_{t 4;3}\rangle + \langle\delta_1 \delta_3 \gamma_{t 4;3}\rangle
+\langle \delta_2\delta_3\gamma_{t 4;3}\rangle + \langle\delta_1\gamma_{t 4;3}\rangle
+ \langle \delta_2\gamma_{t 4;3}\rangle + \langle \delta_3 \gamma_{t 4;3}\rangle 
+ \langle \gamma_{t 4;3} \rangle \right] \nn \\
& = &  \bar{n}_g^3 \left[\langle\delta_1 \delta_2 \delta_3 \gamma_{t 4;3}\rangle_c + 
\langle\delta_1 \delta_2\rangle \langle \delta_3 \gamma_{t 4;3}\rangle + 
\langle\delta_1 \delta_3\rangle \langle \delta_2 \gamma_{t 4;3}\rangle +
\langle\delta_2 \delta_3\rangle \langle \delta_1 \gamma_{t 4;3}\rangle +
\langle \delta_1 \delta_2\gamma_{t 4;3}\rangle + \langle\delta_1 \delta_3 \gamma_{t 4;3}\rangle
+\langle \delta_2\delta_3\gamma_{t 4;3}\rangle \right.\nn \\
& + & \left.\langle\delta_1\gamma_{t 4;3}\rangle
+ \langle \delta_2\gamma_{t 4;3}\rangle + \langle \delta_3 \gamma_{t 4;3}\rangle\right],
\ea
%%%%%%%%%%%%%%%%%%%%%%%%%%%%%%%%%%%%%%%%%%%%%%%%%%%%%%%
where we applied Wick's theorem to the 4-point function in order to
arrive at the last line. The subscript `c' in the remaining 4-point
function indicates that it is connected. We also used
$\langle\gamma_t\rangle=0$ to remove the last term on the second line,
and will repeat this step henceforth without mentioning it explicitly.
We proceed similarly with the other four terms in
Eq.~(\ref{eq:T1_terms}).
%%%%%%%%%%%%%%%%%%%%%%%%%%%%%%%%%%%%%%%%%%%%%%%%%%%%%%%
\ba
T_{i=j\neq k} & = & \bar{n}_g^2 \delta_D^{12}\,\langle(1+\delta_1)(1+\delta_3)\gamma_{t 4;3}\rangle 
= \bar{n}_g^2 \delta_D^{12} \left[\langle\delta_1\gamma_{t 4;3}\rangle + 
\langle\delta_3\gamma_{t 4;3}\rangle + \langle\delta_1 \delta_3 \gamma_{t 4;3}\rangle \right],
\nn\\
 T_{i=k \neq j} & = & \bar{n}_g^2 \delta_D^{13} \left[\langle\delta_1\gamma_{t 4;3}\rangle + 
\langle\delta_2\gamma_{t 4;3}\rangle + \langle\delta_1 \delta_2 \gamma_{t 4;3}\rangle \right],
\nn \\
T_{j=k \neq i} & = &\bar{n}_g^2 \delta_D^{23} \left[\langle\delta_1\gamma_{t 4;3}\rangle + 
\langle\delta_2\gamma_{t 4;3}\rangle + \langle\delta_1 \delta_2 \gamma_{t 4;3}\rangle \right],
\nn \\
T_{i=j=k} & = &\bar{n}_g\,\delta_D^{12} \delta_D^{13}\,\langle \delta_1\gamma_{t 4;3}\rangle.
\ea
%%%%%%%%%%%%%%%%%%%%%%%%%%%%%%%%%%%%%%%%%%%%%%%%%%%%%%%
Putting together all these terms, we rewrite Eq.~(\ref{eq:T1_terms}) as
%%%%%%%%%%%%%%%%%%%%%%%%%%%%%%%%%%%%%%%%%%%%%%%%%%%%%%%
\ba
\langle n_{g1} n_{g2} n_{g3} \gamma_{t 4;3}\rangle & = &\bar{n}_g^3 \left\{
\langle\delta_1 \delta_2 \delta_3 \gamma_{t 4;3}\rangle_c + 
\langle \delta_1 \delta_2\gamma_{t 4;3}\rangle\left(1+\frac{\delta_D^{13}
+\delta_D^{23}}{\bar{n}_g}\right) + \langle \delta_1 \delta_3\gamma_{t 4;3}\rangle\left(
1+\frac{\delta_D^{12}}{\bar{n}_g}\right) + \langle \delta_2 \delta_3\gamma_{t 4;3}\rangle
\right. \nn \\
& + & \left. \langle\delta_1 \delta_2\rangle \langle \delta_3 \gamma_{t 4;3}\rangle + 
\langle\delta_1 \delta_3\rangle \langle \delta_2 \gamma_{t 4;3}\rangle +
\langle\delta_2 \delta_3\rangle \langle \delta_1 \gamma_{t 4;3}\rangle \right. \nn \\
& + & \left. \langle\delta_1 \gamma_{t 4;3}\rangle \left(1+\frac{\delta_D^{12}
+\delta_D^{13}+\delta_D^{23}}{\bar{n}_g} + \frac{\delta_D^{12}\delta_D^{13}}{\bar{n}_g^2}\right)
+ \langle\delta_2 \gamma_{t 4;3}\rangle \left(1+\frac{\delta_D^{13}+\delta_D^{23}}{\bar{n}_g}
\right) + \langle\delta_3 \gamma_{t 4;3}\rangle \left(1+\frac{\delta_D^{12}}{\bar{n}_g}\right)
\right\}.
\label{eq:T1_final}
\ea
%%%%%%%%%%%%%%%%%%%%%%%%%%%%%%%%%%%%%%%%%%%%%%%%%%%%%%%
We now address the second term of Eq.~(\ref{eq:RH1_def}) and use the
fact that $\bar{n}_g=\alpha \bar{n}_R$:
%%%%%%%%%%%%%%%%%%%%%%%%%%%%%%%%%%%%%%%%%%%%%%%%%%%%%%%
\ba
\langle n_{g1} n_{R2} n_{g3} \gamma_{t 4;3}\rangle & = & \bar{n}_R\sum_{i,j}\langle\delta_D(\bz_1-\bx_i)
\delta_D(\bz_3-\bx_j)\gamma_t(\bz_4; \bz_3)\rangle=\frac{\bar{n}_g^3}{\alpha}\left\{
\langle\delta_1 \delta_3\gamma_{t4;3}\rangle + \langle\delta_1 \gamma_{t 4;3}\rangle 
\left(1+\frac{\delta_D^{13}}{\bar{n}_g}\right) + \langle\delta_3 \gamma_{t 4;3}\rangle \right\},
\nn \\
\langle n_{R1} n_{g2} n_{g3} \gamma_{t 4;3}\rangle & = & \frac{\bar{n}_g^3}{\alpha}\left\{
\langle\delta_2 \delta_3\gamma_{t4;3}\rangle + \langle\delta_2 \gamma_{t 4;3}\rangle 
\left(1+\frac{\delta_D^{23}}{\bar{n}_g}\right) + \langle\delta_3 \gamma_{t 4;3}\rangle \right\},
\nn \\
\langle n_{R1} n_{R2} n_{g3} \gamma_{t 4;3}\rangle & = & \frac{\bar{n}_g^3}{\alpha^2}
 \langle\delta_3 \gamma_{t 4;3}\rangle. 
\label{eq:T_234_final}
\ea
%%%%%%%%%%%%%%%%%%%%%%%%%%%%%%%%%%%%%%%%%%%%%%%%%%%%%%%
Combining Eqs.~(\ref{eq:T1_final}) and~(\ref{eq:T_234_final}), we
obtain a final expression for the first part of the cross-covariance,
i.e. Eq.~(\ref{eq:RH1_def}):
%%%%%%%%%%%%%%%%%%%%%%%%%%%%%%%%%%%%%%%%%%%%%%%%%%%%%%%
\ba
\langle \widehat{F}_1 \widehat{F}_2 n_3 \gamma_{t 4;3}\rangle & = &\frac{\bar{n}_g^3}{A_{2D}} 
\left\{\langle\delta_1 \delta_2 \delta_3 \gamma_{t 4;3}\rangle_c + 
\langle \delta_1 \delta_2\gamma_{t 4;3}\rangle\left(1+\frac{\delta_D^{13}
+\delta_D^{23}}{\bar{n}_g}\right) + \langle \delta_1 \delta_3\gamma_{t 4;3}\rangle
\frac{\delta_D^{12}}{\bar{n}_g} + \langle\delta_1 \delta_2\rangle 
\langle \delta_3 \gamma_{t 4;3}\rangle \right. \nn \\
& + & \left. \langle\delta_1 \delta_3\rangle \langle \delta_2 \gamma_{t 4;3}\rangle +
\langle\delta_2 \delta_3\rangle \langle \delta_1 \gamma_{t 4;3}\rangle +
\langle\delta_1 \gamma_{t 4;3}\rangle \left(\frac{\delta_D^{12}
+\delta_D^{23}}{\bar{n}_g} + \frac{\delta_D^{12}\delta_D^{13}}{\bar{n}_g^2}\right)
+ \langle\delta_2 \gamma_{t 4;3}\rangle \frac{\delta_D^{13}}{\bar{n}_g}
+ \langle\delta_3 \gamma_{t 4;3}\rangle \frac{\delta_D^{12}}{\bar{n}_g}\right\}.
\label{eq:RH1_v1}
\ea
%%%%%%%%%%%%%%%%%%%%%%%%%%%%%%%%%%%%%%%%%%%%%%%%%%%%%%%
We also need to compute the second part of the cross-covariance matrix
given by Eq.~(\ref{eq:def_cov_cross}). In completely similar way to
the calculation above, we write:
%%%%%%%%%%%%%%%%%%%%%%%%%%%%%%%%%%%%%%%%%%%%%%%%%%%%%%%
\ba
\langle \widehat{F}_g(\bz_1) \widehat{F}_g(\bz_2)\rangle\,\langle n_g(\bz_3) 
\gamma_t(\bz_4; \bz_3)\rangle & \equiv & \langle \widehat{F}_1 \widehat{F}_2 \rangle
\langle n_3 \gamma_{t 4;3}\rangle \nn \\
& = & \frac{1}{A_{2D}}\left[\langle n_{g1} n_{g2}\rangle -\alpha\left(\langle n_{R1} n_{g2}\rangle +
\langle n_{g1} n_{R2} \rangle \right) + \alpha^2\langle n_{R1} n_{R2} \rangle\right]
\langle n_{g3} \gamma_{t 4;3} \rangle \nn \\
& = & \frac{\bar{n}_g^3}{A_{2D}} \left[\langle\delta_3 \gamma_{t 4;3} \rangle \left(
\langle\delta_1 \delta_2 \rangle + \frac{\delta_D^{12}}{\bar{n}_g}\right)\right].
\label{eq:RH2_v1}
\ea
%%%%%%%%%%%%%%%%%%%%%%%%%%%%%%%%%%%%%%%%%%%%%%%%%%%%%%%
Finally, putting together Eqs.~(\ref{eq:RH1_v1}) and~(\ref{eq:RH2_v1}), we obtain the 
desired result:
%%%%%%%%%%%%%%%%%%%%%%%%%%%%%%%%%%%%%%%%%%%%%%%%%%%%%%%
\ba
 \langle \widehat{F}_1 \widehat{F}_2 n_3 \gamma_{t 4;3}\rangle - \langle 
\widehat{F}_1 \widehat{F}_2 \rangle\langle n_3 \gamma_{t 4;3}\rangle & = &
\frac{\bar{n}_g^3}{A_{2D}} \left\{\langle\delta_1 \delta_2 
\delta_3 \gamma_{t 4;3}\rangle_c + \langle \delta_1 \delta_2\gamma_{t 4;3}\rangle
\left(1+\frac{\delta_D^{13} +\delta_D^{23}}{\bar{n}_g}\right) 
+ \langle \delta_1 \delta_3\gamma_{t 4;3}\rangle \frac{\delta_D^{12}}{\bar{n}_g} 
+  \langle\delta_1 \delta_3\rangle \langle \delta_2 \gamma_{t 4;3}\rangle
\right.\nn\\
 & + & \left. \langle\delta_2 \delta_3\rangle \langle \delta_1 \gamma_{t 4;3}\rangle +
\langle\delta_1 \gamma_{t 4;3}\rangle \left(\frac{\delta_D^{12}
+\delta_D^{23}}{\bar{n}_g} + \frac{\delta_D^{12}\delta_D^{13}}{\bar{n}_g^2}\right)
+ \langle\delta_2 \gamma_{t 4;3}\rangle \frac{\delta_D^{13}}{\bar{n}_g}\right\}.
\label{eq:cross_cov_v1}
\ea
%%%%%%%%%%%%%%%%%%%%%%%%%%%%%%%%%%%%%%%%%%%%%%%%%%%%%%%
The above equation is general in the sense that it contains both
Gaussian and non-Gaussian terms, as well as contributions from the
0-lag correlations, i.e. all terms containing
$\delta_D^{12}$. Discarding the latter contributions, as well as the
non-Gaussian ones, we write a final and simplified expression:
%%%%%%%%%%%%%%%%%%%%%%%%%%%%%%%%%%%%%%%%%%%%%%%%%%%%%%%
\ba
\langle \widehat{F}_1 \widehat{F}_2 n_3 \gamma_{t 4;3}\rangle - \langle 
\widehat{F}_1 \widehat{F}_2 \rangle\langle n_3 \gamma_{t 4;3}\rangle & = &
\frac{\bar{n}_g^3}{A_{2D}} \left\{\langle\delta_1 \delta_3\rangle \langle 
\delta_2 \gamma_{t 4;3}\rangle + \langle\delta_2 \delta_3\rangle \langle 
\delta_1 \gamma_{t 4;3}\rangle + \langle\delta_1 \gamma_{t 4;3}\rangle \frac{
\delta_D^{23}}{\bar{n}_g} + \langle\delta_2 \gamma_{t 4;3}\rangle 
\frac{\delta_D^{13}}{\bar{n}_g}\right\}.
\label{eq:cross_cov_v2}
\ea
%%%%%%%%%%%%%%%%%%%%%%%%%%%%%%%%%%%%%%%%%%%%%%%%%%%%%%%
While Eq.~(\ref{eq:cross_cov_v1}) provides the general result,
Eq.~(\ref{eq:cross_cov_v2}) encompasses the approximations that we
have made throughout this paper.  Replacing this latter equation into
Eq.~(\ref{eq:cov_cross_v0}), we write:
%%%%%%%%%%%%%%%%%%%%%%%%%%%%%%%%%%%%%%%%%%%%%%%%%%%%%%%
\ba
{\rm Cov}[\widehat{w}_{\rm gg}(\btheta_1), \widehat{\gamma}^g_t(\btheta_2)] & = & 
\frac{\bar{n}_g^3}{N_g\,A_{2D}}\int_{\Omega_s} d^2x\, d^2y \left\{\hspace{-0.1cm}
\frac{ }{ }\langle\delta_g(\bx)\delta_g(\by)\rangle\langle\delta_g(\bx+\btheta_1)
\gamma_t(\btheta_2+\by;\by)\rangle + \langle\delta_g(\bx+\btheta_1)\delta_g(\by)\rangle
\langle \delta_g(\bx)\gamma_t(\btheta_2+\by; \by)\rangle \right. \nonumber\\
& + & \left. \langle \delta_g(\bx)\gamma_t(\btheta_2+\by; \by)\rangle 
\frac{\delta_D(\bx+\btheta_1-\by)}{\bar{n}_g} + \langle \delta_g(\bx+\btheta_1)
\gamma_t(\btheta_2+\by; \by)\rangle \frac{\delta_D(\bx-\by)}{\bar{n}_g} \right\}.
\label{eq:cov_cross_v3}
\ea
%%%%%%%%%%%%%%%%%%%%%%%%%%%%%%%%%%%%%%%%%%%%%%%%%%%%%%%
We shall detail only the computation of first term of the above
equation, since the steps are fairly similar to those taken in
sections \S\ref{app:ggl-cov} and \S\ref{app:gc-cov}. We label this
first term $C_1$ to shorten the notation. Using the definition of the
correlation function as the Fourier transform of the power spectrum,
and noting that $\bar{n}_g^3/N_g/A_{2D}=1/\Omega_s^2$, we write
%%%%%%%%%%%%%%%%%%%%%%%%%%%%%%%%%%%%%%%%%%%%%%%%%%%%%%%
\ba
C_1 & = & -\frac{1}{\Omega_s^2} \int_{\Omega_s} d^2x\,d^2y \int \frac{d^2l}{(2\pi)^2} 
e^{i\,\bl\cdot(\bx -\by)} \mathcal{C}_{\rm gg}(l)
\int \frac{d^2l'}{(2\pi)^2} e^{i\,\bl'\cdot(\bx+\btheta_1-\by-\btheta_2)}
\cos[2(\phi_{\btheta_2}-\phi_{\bl'})]\,\mathcal{C}_{g\kappa}(l')\nonumber \\
& = & -\frac{1}{\Omega_s^2}\int \frac{d^2 l}{(2\pi)^2} \frac{d^2 l'}{(2\pi)^2} \mathcal{C}_{\rm gg}(l)
\mathcal{C}_{g\kappa}(l')\cos[2(\phi_{\btheta_2}-\phi_{\bl'})]
e^{i\,\bl' \cdot (\btheta_1-\btheta_2)} \int_{\Omega_s} d^2x \, e^{i\,\bx \cdot(\bl+
\bl')}  \int_{\Omega_s} d^2y \,e^{-i\,\by \cdot(\bl+\bl')} \nonumber \\
& = & -\frac{1}{\Omega_s}\int \frac{d^2l}{(2\pi)^2} \mathcal{C}_{\rm gg}(l)
\mathcal{C}_{g\kappa}(l)\cos[2(\phi_{\btheta_2}-\phi_{\bl})] e^{i\,\bl \cdot (\btheta_2-\btheta_1)}.
\label{eq:C1}
\ea
%%%%%%%%%%%%%%%%%%%%%%%%%%%%%%%%%%%%%%%%%%%%%%%%%%%%%%%
The remaining three terms of Eq.~(\ref{eq:cov_cross_v3}) are computed
in the same way, leading to the following result for the
cross-covariance:
%%%%%%%%%%%%%%%%%%%%%%%%%%%%%%%%%%%%%%%%%%%%%%%%%%%%%%%
\ba
{\rm Cov}[\widehat{w}_{\rm gg}(\btheta_1), \widehat{\gamma}^g_t(\btheta_2)] & = &
-\frac{1}{\Omega_s}\int \frac{d^2l}{(2\pi)^2} \cos[2(\phi_{\btheta_2}-\phi_{\bl})] 
\left[e^{i\,\bl \cdot (\btheta_2-\btheta_1)} + e^{i\,\bl \cdot (\btheta_2+\btheta_1)}\right]  
\mathcal{C}_{g\kappa}(l)\left[\mathcal{C}_{\rm gg}(l) 
+\frac{1}{\bar{n}_g}\right].
\label{eq:cross_cov_angular}
\ea
%%%%%%%%%%%%%%%%%%%%%%%%%%%%%%%%%%%%%%%%%%%%%%%%%%%%%%%
%%%%%%%%%%%%%%%%%%%%%%%%%%%%%%%%%%%%%%%%%%%%%%%%%%%%%%%
\section{Galaxy catalogue comparison}
\label{App2}
\begin{figure*}
\centering {
\includegraphics[scale=0.6]{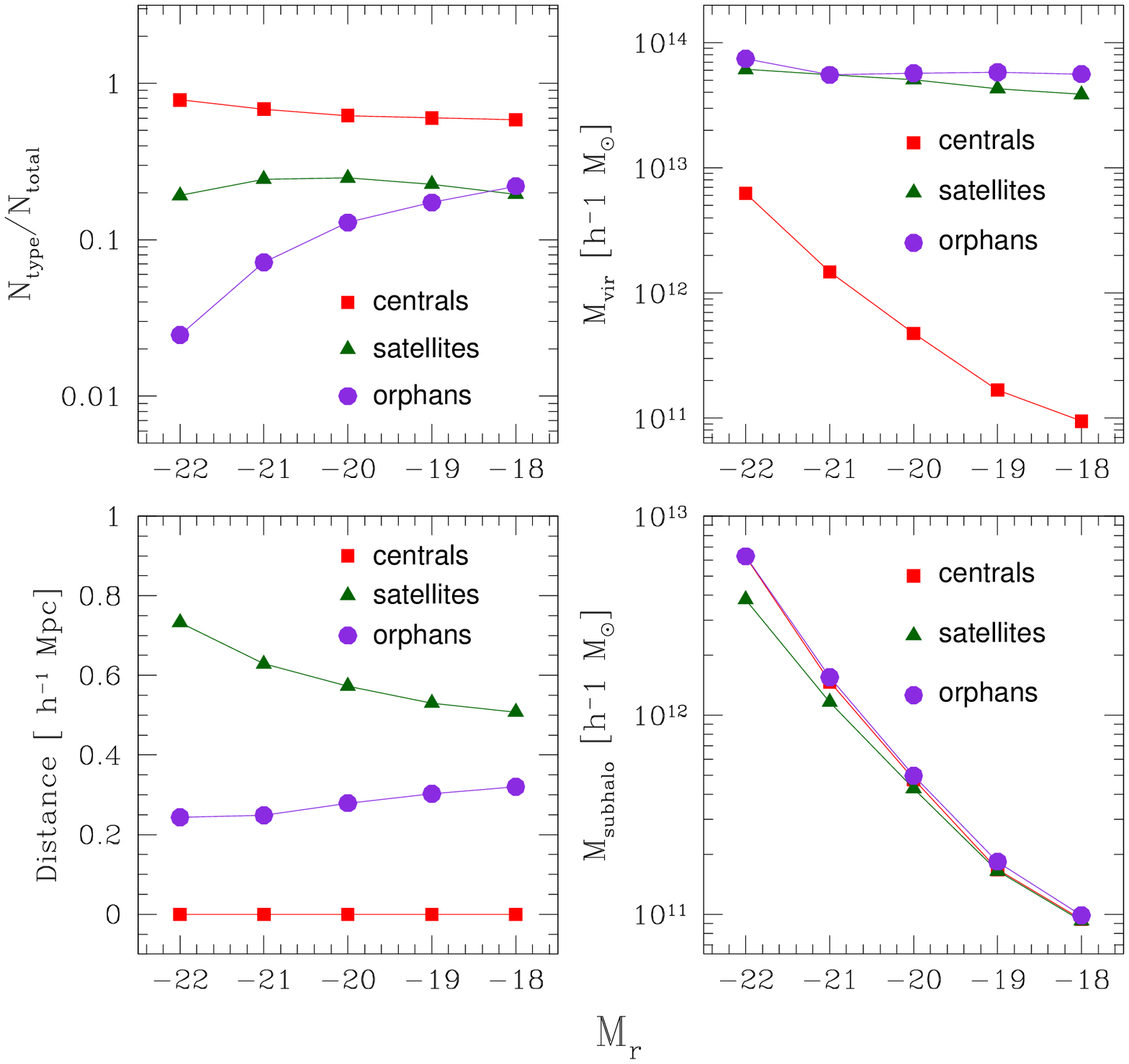}
\caption{The analogue of Figure~\ref{fig:gal_prop} for the Millennium
  simulation.  The galaxy catalogue is very similar to that described
  in \citet{Guoetal2011}. The redshift is 0.24.}
\label{fig:gal_prop_Bruno}}
\end{figure*}

%%%%%%%%%%%%%%%%%%%%%%%%%%%%%%%%%%%%%%%%%%%%%%%%%%%%%%%
Figure~\ref{fig:gal_prop_Bruno} presents various properties of
galaxies from the Millennium simulation as a function of $r$-band
magnitude, e.g. the fraction of each galaxy type, the host halo mass,
the distance to the central galaxy, and the subhalo mass. The galaxy
catalogue is very similar to the MXXL one, used throughout this
paper. The figure serves as a diagnostic for the impact of resolution
effects on the distribution of galaxies, and by comparing it to
Figure~\ref{fig:gal_prop} we selected only MXXL galaxies with $M_r <
-19$ as a `reliable' sample.
%%%%%%%%%%%%%%%%%%%%%%%%%%%%%%%%%%%%%%%%%%%%%%%%%%%%%%%
%%%%%%%%%%%%%%%%%%%%%%%%%%%%%%%%%%%%%%%%%%%%%%%%%%%%%%%
%%%%%%%%%%%%%%%%%%%%%%%%%%%%%%%%%%%%%%%%%%%%%%%%%%%%%%%
\section{The impact of shape noise}
\label{App3}
%%%%%%%%%%%%%%%%%%%%%%%%%%%%%%%%%%%%%%%%%%%%%%%%%%%%%%%
%%%%%%%%%%%%%%%%%%%%%%%%%%%%%%%%%%%%%%%%%%%%%%%%%%%%%%%
\begin{figure*}
\centering 
\begin{tabular}{cc}
  \includegraphics[scale=0.55]{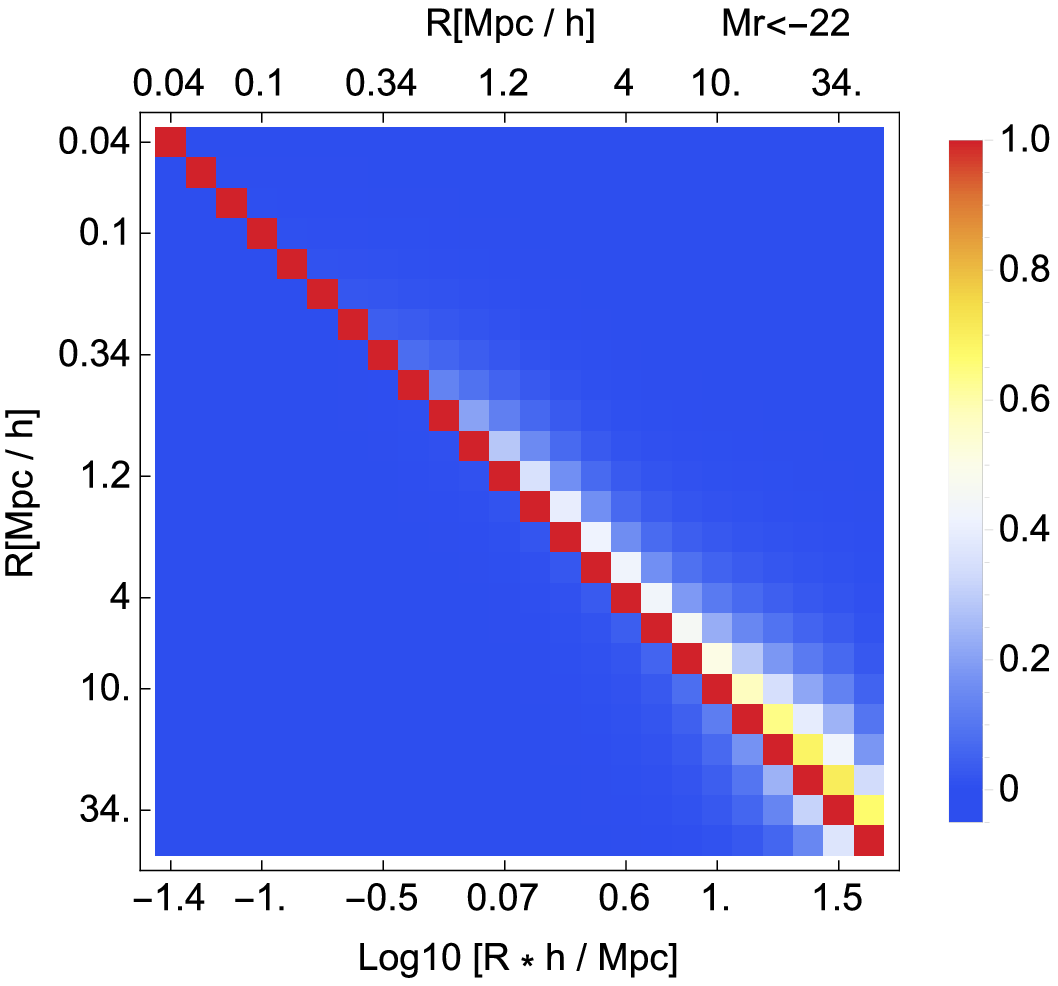}
  \includegraphics[scale=0.55]{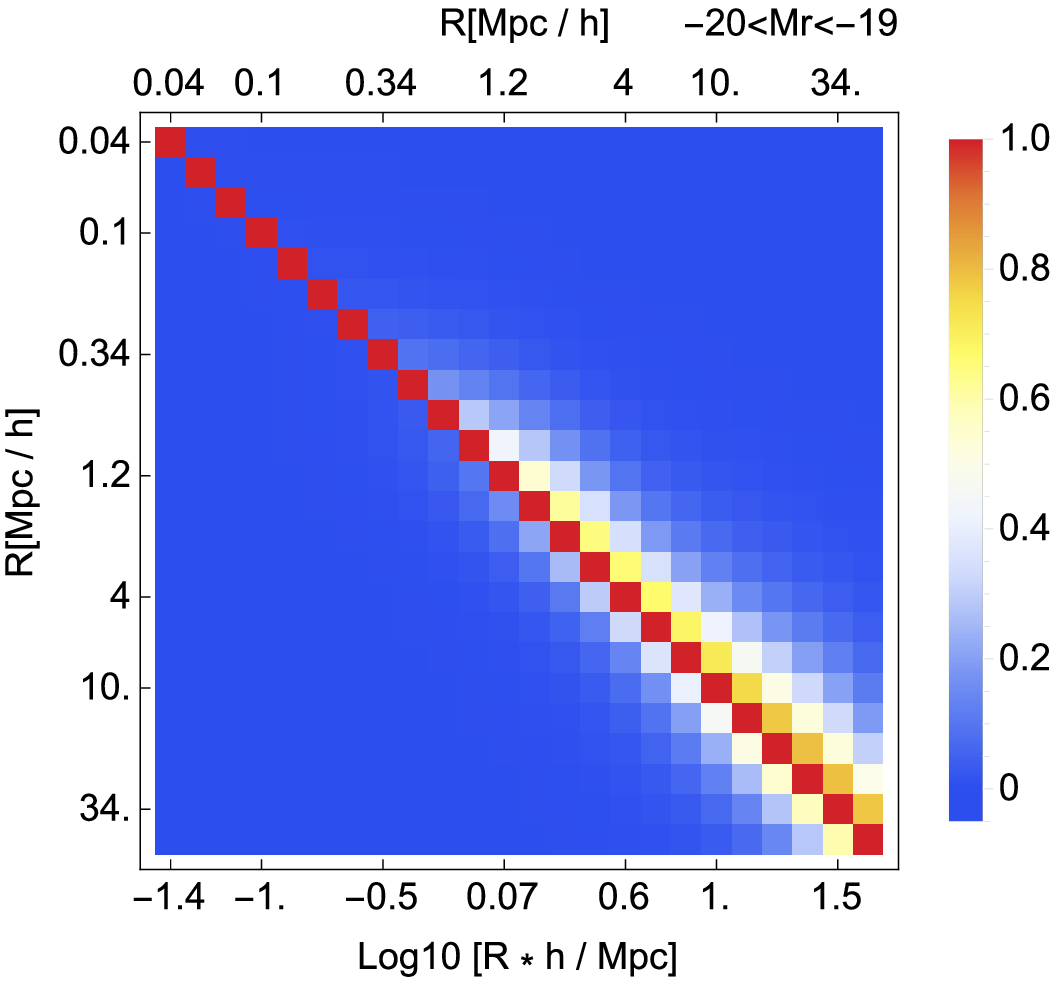}
\end{tabular}
%%%%%%%%%%%%%%%%%%%%%%%%%%%%%%%%%%%%%%%%%%%%%%%%%%%%%%%
\begin{tabular}{cc}
\includegraphics[scale=0.3]{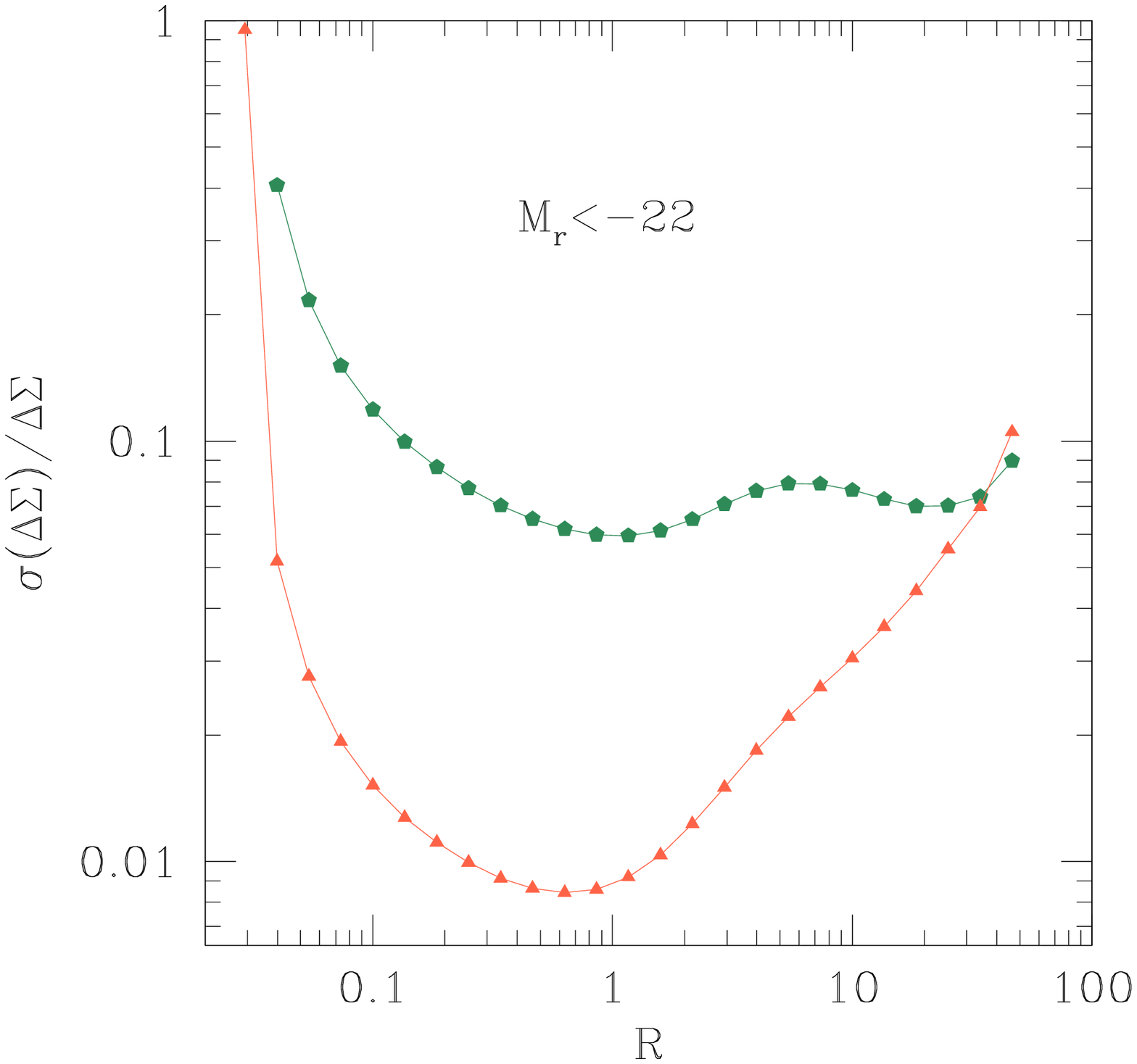}
\includegraphics[scale=0.3]{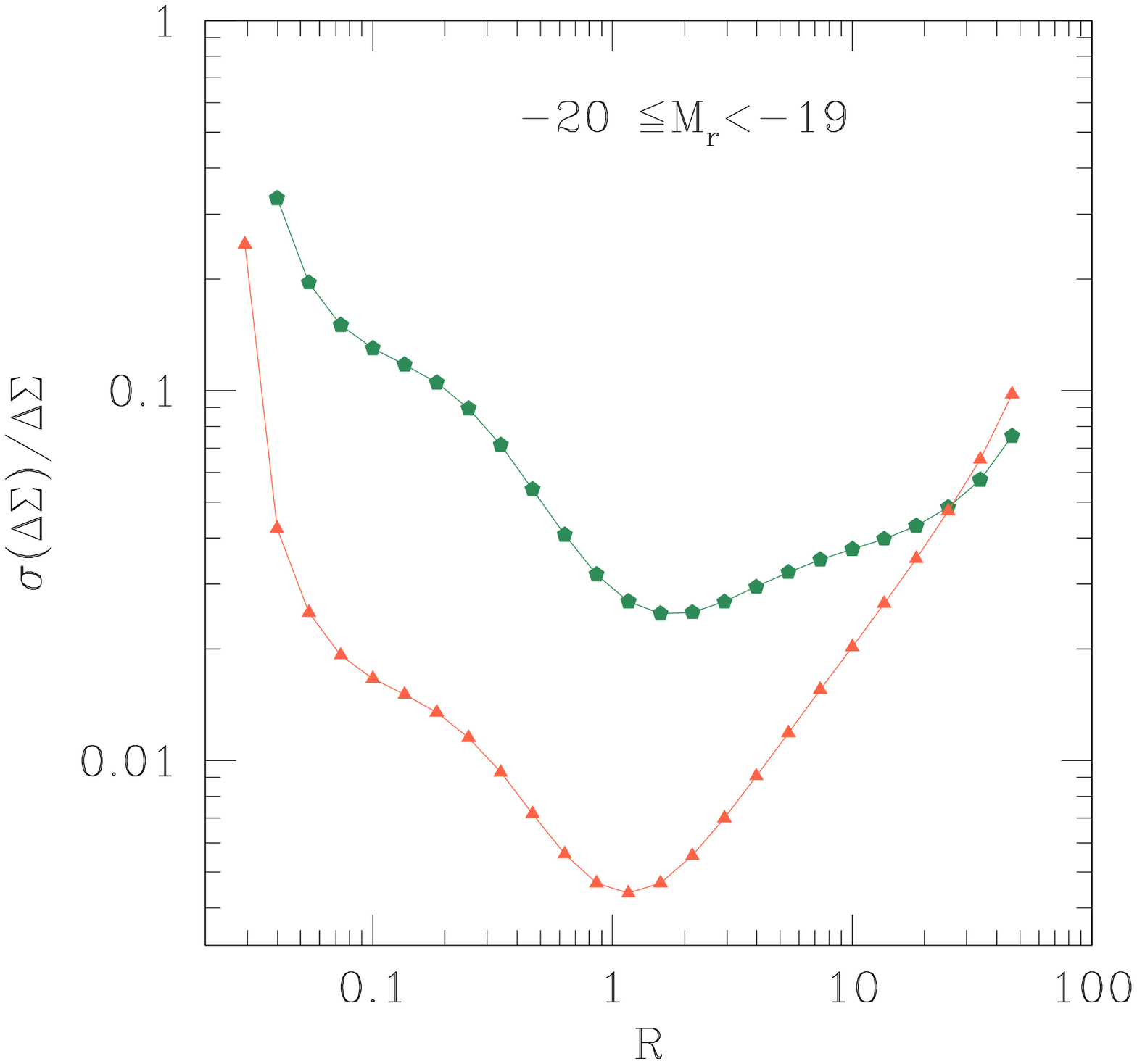}
\end{tabular}
\caption{Theoretical estimates of the GGL correlation matrix and
  variance, illustrating the impact of shape noise for our brightest
  and faintest galaxy bins. {\it Top panels:} For each magnitude bin,
  the upper right triangle corresponds to the level of noise from our
  MXXL measurements, while the bottom left triangle has a shape noise
  similar to the SDSS measurements of \citet{Mandelbaumetal2013}. {\it
    Bottom panels:} The noise-to-signal for the excess surface
  density. Green pentagons depict the prediction for the SDSS shape
  noise, and red triangles for the MXXL noise, i.e. the same as in
  Figure~\ref{fig:NS_GGL}. }
\label{fig:SDSS}
\end{figure*}
%%%%%%%%%%%%%%%%%%%%%%%%%%%%%%%%%%%%%%%%%%%%%%%%%%%%%%%
We present a theoretical calculation of the impact of shape noise on
the GGL results. Whilst we do not have shape noise in our simulations,
we have particle shot noise, which qualitatively acts in a similar way
to shape noise, but quantitatively is different. 

\noindent As an example, we use the shape noise values of
\cite{Mandelbaumetal2013}. We take $\sigma_{\gamma}^2/2=0.365^2$,
$\bar{n}_s=1.2$ galaxies/$\rm{arcmin}^2$. Our approach is to translate
the contribution that this shape noise level would have on our GGL
covariance into an effective shot noise contribution that we can just
replace into \Eqn{eq:cov_delta_sigma}. Matching the constants
multiplying the shape noise and shot noise terms in
\Eqns{eq:cov_shear4}{eq:cov_delta_sigma}, we write:
%%%%%%%%%%%%%%%%%%%%%%%%%%%%%%%%%%%%%%%%%%%%%%%%%%%%%%%
\be
\frac{\sigma_{\gamma}^2}{2 \bar{n}_s}\frac{1}{\nbar^{2D}_\g}=
\left(\frac{\rho_m^0}{\Sigma_{\rm crit}}\right)^2\frac{1}{D^4(z_l)}
\frac{1}{\nbar_p} \frac{1}{\nbar^{3D}_\g} \ ,
\ee
%%%%%%%%%%%%%%%%%%%%%%%%%%%%%%%%%%%%%%%%%%%%%%%%%%%%%%%
where we have differentiated between angular and volume densities;
$D(z_l)$ denotes the angular diameter distance to the lens
redshift. Assuming all lenses to be at the same redshift $z_l$, the
relationship between the angular and volume density of lens galaxies
is given by $\nbar^{2D}_\g=D^2(z_l) (2\chi_{\rm max}) \nbar^{3D}_\g$,
where $\chi_{\rm max}$ is the projection length used throughout this
work. Using this, the equation above can be expressed as:
%%%%%%%%%%%%%%%%%%%%%%%%%%%%%%%%%%%%%%%%%%%%%%%%%%%%%%%
\be
\frac{\sigma_{\gamma}^2}{2 \bar{n}_s}=\left(\frac{\rho_m^0}{\Sigma_{\rm crit}}\right)^2
\frac{2\chi_{\rm max}}{D^2(z_l)} \frac{1}{\nbar_p} \ .
\label{eq:shape_noise_conversion}
\ee
%%%%%%%%%%%%%%%%%%%%%%%%%%%%%%%%%%%%%%%%%%%%%%%%%%%%%%%
Evaluating \Eqn{eq:shape_noise_conversion} for a general source
redshift of $z_s=1$, we obtain the effective particle shot noise for
the SDSS measurements of \cite{Mandelbaumetal2013}: $\nbar^{\rm
  SDSS}_p=2.88\times 10^{-5}\, [\Mpc]^{-3}$. This is to be compared to
our shot noise value $\nbar^{\rm MXXL}_p=1.58 \times
10^{-2}\,[\Mpc]^{-3}$. Equivalently, the MXXL shot noise would
correspond to an effective source galaxy density of $\sim 660 \:{\rm
  galaxies/arcmin^2}$. For a survey with $\sigma^2_{\gamma}/2=0.3^2$,
the MXXL shot noise would be equivalent to $\nbar_s=440\,{\rm
  galaxies/arcmin^2}$. Note however that the SDSS survey volume is
significantly larger than that of our subcubes.

The top panels of Figure~\ref{fig:SDSS} illustrate the effect of shape
noise on the GGL correlation matrix, for the brightest and faintest
galaxy bins considered in this work. There are two consequences of
increased shape noise, or equivalently, particle shot noise. First,
the off-diagonal elements are present mostly on larger scales,
e.g. $\sim 5-10\,\Mpc$, compared to the lower MXXL noise, where they
stretch to $\sim 1\,\Mpc$. Second, the correlations are weaker. Since
our predictions are obtained using the measured bias, there is also a
dependence on magnitude, with the brighter galaxies being less
correlated than the fainter ones. The bottom panels of the figure show
the noise-to-signal, with the area of SDSS accounted for. As expected,
the SDSS predictions are higher.

%%%%%%%%%%%%%%%%%%%%%%%%%%%%%%%%%%%%%%%%%%%%%%%%%%%%%%%
%%%%%%%%%%%%%%%%%%%%%%%%%%%%%%%%%%%%%%%%%%%%%%%%%%%%%%%

\end{document}